\documentclass[11pt]{article}

\usepackage[table]{xcolor}

\usepackage{amsmath,amsfonts,amssymb,amsthm,epsfig,epstopdf,titling,url,array}
\usepackage{graphicx, psfrag}
\usepackage{dynkin-diagrams}
\usepackage{appendix}
\usepackage{cite}
\usepackage{longtable}
\usepackage{multirow}
\usepackage{float}
\usepackage{hyperref}
\usepackage[utf8]{inputenc}
\usepackage[english]{babel}
\usepackage{color}
\usepackage{booktabs,caption}
\usepackage[flushleft]{threeparttable}
\usepackage{lscape}
\usepackage{cancel}
\definecolor{Mygrey}{gray}{0.8}
\definecolor{Mywhite}{gray}{1.0}

\newcommand{\be}{\begin{equation}}
\newcommand{\ee}{\end{equation}}
\newcommand{\bea}{\begin{eqnarray}}
\newcommand{\eea}{\end{eqnarray}}

\newcolumntype{C}[1]{>{\centering\arraybackslash$}p{#1}<{$}}

\usepackage{colortbl}

\begin{document}

\begin{center}
\textbf{ \Large \bf Tensor Product CFTs and One-Character Extensions}
\end{center}

\vskip .6cm
\medskip

\vspace*{4.0ex}

\baselineskip=18pt

\centerline{\large \rm   Chethan N. Gowdigere$^{1\,ab}$, Sachin Kala$^{2\,ab}$, Jagannath Santara$^{3\,c}$}

\vspace*{4.0ex}

\centerline{\large \it  $^a$National Institute of Science Education and Research Bhubaneshwar,}

\centerline{\large \it  P.O. Jatni, Khurdha, 752050, Odisha, INDIA}

\vspace*{1.0ex}

\centerline{\large \it  $^b$Homi Bhabha National Institute, Training School Complex, }

\centerline{\large \it  Anushakti Nagar, Mumbai 400094, INDIA}

\vspace*{1.0ex}

\centerline{\large \it $^c$Department of Physics,}

\centerline{\large \it  Indian Institute of Technology Madras,}

\centerline{\large \it Chennai 600036, INDIA}

\vspace*{4.0ex}
\centerline{E-mail: $^1$chethan.gowdigere@niser.ac.in, $^2$sachin.kala@niser.ac.in, $^3$jagannath.santara@physics.iitm.ac.in}

\vspace*{5.0ex}
\centerline{\bf Abstract} \bigskip
We study one-character CFTs obtained as one-character extensions of the tensor products of a single CFT $\mathcal{C}$. The motivation comes from the fact that $28$ of the $71$ CFTs in the Schelleken’s list of $c = 24$ CFTs are such CFTs. We study for $\mathcal{C}$ : (i) any two-character WZW CFT with vanishing Wronskian index, (ii) the Ising CFT, (iii) the infinite class of $D_{r,1}$ CFTs and the $A_{4,1}$ CFT.  The characters being $S$-invariant homogenous polynomials of the characters of $\mathcal{C}$,  when organised in terms of a $S$-invariant basis, take compact forms allowing for closed form answers for high central charges. We find a $S$-invariant basis for  each of the CFTs studied. As an example, one can find an explicit expression for the character of the monster CFT as a degree-$48$ polynomial of the characters of the Ising CFT. In some CFTs, some of the $S$-invariant polynomials of characters compute, after using the $q$-series of the characters,  to a constant value. Hence, the characters of one-character extensions are more properly elements of the quotient ring of polynomials (of characters) with the ideal needed for the quotient, generated by $S$-invariant polynomials that compute to a constant.
In some cases, we are able to rule out the existence of one-character extension CFTs. In other cases, we predict their existence. We are able to conjecture a discrete set of six and four infinite series of one-character extension CFTs.

\vfill \eject

\baselineskip=18pt

\tableofcontents

\section{Introduction \label{1s}}

Two dimensional conformal field theories (CFTs) are a subject of great interest for a whole range of theoretical scientists \cite{Belavin:1984vu, Witten:1983ar, DiFrancesco:1997nk, Moore:1989vd, Fuchs:2009iz, Gaberdiel:1999mc}. They are studied intensely by theoretical physicists, of both the high-energy and condensed matter varieties, as well as mathematical physicists and physical mathematicians. They are studied by high-energy theorists, especially string theorists, because of it’s relevance to the world-sheet description of string theory and string perturbation theory. They are studied by condensed matter physicists because they describe two dimensional critical systems and the fixed point of the renormalisation group. They are also relevant for the Kondo problem, describe universality classes of quantum Hall systems and find applications in entanglement entropy as well as quantum computing. Two dimensional CFTs have contributed to many areas of pure mathematics viz. infinite dimensional Lie algebras, theory of modular forms, representation theory, etc. 

A classification of such an object,  which seems to possess  such a wide range of interest and application, should be desirable. But it seems that classifying all CFTs  may be quite hard and perhaps is at some distance in the future. A subclass of CFTs that one may attempt a classification for, are the so-called rational conformal field theories (RCFTs). These are two dimensional CFTs with some rationality properties : the central charge and the conformal dimensions of fields are restricted to be rational numbers. The rationality results in a finiteness property \cite{Anderson:1987ge} :  the partition function can be written in terms of a finite number of characters. 

A classification scheme for two dimensional rational conformal field theories was given by Mathur, Mukhi and Sen (MMS) in 1988.  The scheme is based on properties of the torus partition function of a RCFT. The torus partition function of a RCFT can be written in terms of a finite number of characters. One denotes this by $\mathbf{n}$ and is one of the two positive integers that forms a label for every RCFT.  The other positive integer (actually, non-negative) comes about as follows. The $\mathbf{n}$ characters of a RCFT, it turns out, constitute a set of linearly independent solutions for a certain $\mathbf{n}$’th order ordinary differential equation, known as the modular linear differential equation (MLDE) \cite{Eguchi:1987qd}. Thus there is a MLDE associated to every RCFT. Now, one considers the Wronskian of the characters, which turns out to be modular of weight $\mathbf{n} (\mathbf{n}-1)$. From the Riemann-Roch theorem a.k.a the valence formula for modular functions/forms, one obtains information about the location of the zeroes of the Wronskian. This zeroes' information  is encoded in what is termed as the Wronskian index or the Mukhi index\footnote{This important detail about a RCFT, which is part of the identification label of a RCFT in the MMS scheme, had not been named for most of it’s existence. It was given the name ``Wronskian index’’ by Prof. Sunil Mukhi in 2016. The name ``Mukhi index’’ has also been used by some workers in this field.} of a RCFT, denoted here by $\mathbf{l}$.  According to the MMS scheme, one should identify any RCFT by a label $\mathbf{[n, l]}$ consisting of the number of characters and the Wronskian/Mukhi index of that RCFT. 

A second crucial feature of the MMS classification program of RCFT is the fact that the $\mathbf{[n, l]}$ label is also a label for MLDEs. Given a $\mathbf{[n, l]}$, one can construct a unique MLDE, unique upto the presence of a finite number of parameters. A given  $\mathbf{[n, l]}$ CFT solves the $\mathbf{[n, l]}$ MLDE for some values of it’s parameters. Another $\mathbf{[n, l]}$ CFT solves the same $\mathbf{[n, l]}$ MLDE but for different values of the parameters. Thus to obtain \emph{all} $\mathbf{[n, l]}$ CFTs, one only needs to solve \emph{one} MLDE, over it's entire parameter space. In this way, the MMS classification program reduces the infinitely hard problem of classifying all RCFTs to an infinite number of (still hard  but hopefully only) finitely hard problems of classifying  RCFTs with a given $\mathbf{[n, l]}$. Historically, $\mathbf{[2, 0]}$ RCFTs were the  first to be classified  \cite{Mathur:1988na, Mathur:1988gt}. The classification of $\mathbf{[3,0]}$ CFTs was begun in \cite{Mathur:1988gt} and was completed in \cite{Das:2022uoe} ; this was achieved after the  classification of  $\mathbf{[3,0]}$ characters \cite{Kaidi:2021ent}, \cite{Das:2021uvd}, \cite{Bae:2021mej}. 

This paper largely concerns CFTs with $\mathbf{n} = 1$. They have a single character which can be thought of as an identity character, with a $q$-series : $\chi(q) = q^{-\frac{c}{24}} (1 + a_1 q + \ldots)$.  They have extremely simple modular transformation properties \cite{Schellekens:1992db}.  The identity is self-conjugate, and hence the charge conjugation matrix $C$ must be equal to $1$. Therefore $S = \pm1$ and the character satisfies  $\chi(-\frac{1}{\tau}) = \pm \chi(\tau)$. Choosing $\tau = i$ and from $\chi(i) \neq 0$, we conclude that $S = 1$. Thus the character of a one-character CFT is S-invariant.  Furthermore, since $(ST)^3 = C$, we find that $T = e^{\frac{2 k \pi i}{3}}, k \in \mathbf{Z}$. On the other hand we have that the action of $T$ on a character  is given in terms of the central charge as $T = e^{ (-\frac{c}{24})2 \pi i }$. It follows that the central charge $c$ must be a multiple of $8$. When $c$ is a multiple of $24$, we have $S = 1$ as well as $T = 1$ and hence the character itself is modular invariant. In this case, the character itself can serve as a modular invariant  partition function (as in \cite{Witten:2007kt}). For one-character CFTs with central charges given by other multiples of $8$, we have $S = 1$ (S-invariance), but $T \neq 1$.  The partition function of such CFTs  is simply $\chi\,\overline{\chi}$.

To put one-character CFTs within the purview of the MMS classification scheme :  their Wronskian/Mukhi indices, being  $\frac{c}{4}$, are even numbers.  One can readily write down the corresponding first-order MLDEs. But MLDE methods have not been that effective in studying one-character CFTs as they have been for multiple-character CFTs. Even in this paper, we will  develop methods that avoid MLDEs and are very algebraic, but at the end of the say, produce solutions of the MLDEs. 

The classification of one-character CFTs, available in the literature, is as follows. First,  $c = 8$ or $\mathbf{[1,2]}$ theories:  there is exactly one such CFT viz. the $E_{8,1}$ CFT. Then for $c = 16$ or $\mathbf{[1, 4]}$ theories : there are exactly two CFTs viz. $E_{8,1}\otimes E_{8,1}$ and the one-character extension of $D_{16,1}$, which we denote by $\mathcal{E}_1[D_{16,1}]$\footnote{Given a multiple-character CFT $\mathcal{C}$, if one can find a combination of characters which has the modular properties of a one-character CFT, one would have found a one-character extension  of  $\mathcal{C}$. The one-character CFT thus constructed is denoted by $\mathcal{E}_1[\mathcal{C}]$. Here $D_{16,1}$ is a three-character CFT. It has three characters $\chi_0, \chi_{\frac12}, \chi_2$ where the subscripts are the conformal dimensions. It turns out that $\chi_0 + \chi_2$ is $S$-invariant and $T = e^{\frac{4 \pi i}{3}}$ and hence defines a one-character CFT.}. For $c = 24$ or $\mathbf{[1, 6]}$ CFTs, there is the famous classification due to Schellekens \cite{Schellekens:1992db} where there are a total of $71$ one-character CFTs. For $c > 24$, we are very far from a classification; some relevant papers are \cite{King} and \cite{Das:2022slz}.

\begin{table}[h] 
\begin{center}
\begin{threeparttable}
\rowcolors{2}{Mywhite}{Mygrey}
\begin{tabular}{|c|c|c||c|c|c|}
\hline
\# & Schellekens No.  & One-character CFT & \# & Schellekens No.  & One-character CFT \\
\hline
1. & Sch. 1 &   $\mathcal{E}_1[U(1)^{\otimes 24}]$  & 15. & Sch. 38 &  $\mathcal{E}_1[C_{4,1}^{\otimes 4}]$ \\
2. & Sch. 2 &  $\mathcal{E}_1[A_{1,4}^{\otimes 12}]$ & 16. & Sch. 41 &  $\mathcal{E}_1[B_{6,2}^{\otimes 2}]$    \\
3. & Sch. 4 &  $\mathcal{E}_1[C_{4,10}]$ &  17. & Sch. 42 &  $\mathcal{E}_1[D_{4,1}^{\otimes 6}]$ \\
4. & Sch. 5 &  $\mathcal{E}_1[A_{1,2}^{\otimes 16}]$  & 18. & Sch. 46 &  $\mathcal{E}_1[A_{6,1}^{\otimes 4}]$  \\
5. & Sch. 6 &  $\mathcal{E}_1[A_{2,3}^{\otimes 6}]$  & 19. & Sch. 51 &  $\mathcal{E}_1[A_{8,1}^{\otimes 3}]$  \\
6. & Sch. 9 &  $\mathcal{E}_1[A_{4,5}^{\otimes 2}]$  & 20. & Sch. 54 &  $\mathcal{E}_1[D_{6,1}^{\otimes 4}]$ \\
7. & Sch. 11 &  $\mathcal{E}_1[A_{6,7}]$  & 21. & Sch. 57 &  $\mathcal{E}_1[B_{12, 2}]$  \\
8. & Sch. 12 &  $\mathcal{E}_1[C_{2,2}^{\otimes 6}]$ & 22. & Sch. 58 &  $\mathcal{E}_1[E_{6,1}^{\otimes 4}]$ \\
9. & Sch. 15 &  $\mathcal{E}_1[A_{1,1}^{\otimes 24}]$  & 23. & Sch. 60 &  $\mathcal{E}_1[A_{12,1}^{\otimes 2}]$   \\
10. & Sch. 23 &  $\mathcal{E}_1[B_{3,2}^{\otimes 4}]$ & 24. & Sch. 61 &  $\mathcal{E}_1[D_{8,1}^{\otimes 3}]$  \\
11. & Sch. 24 &  $\mathcal{E}_1[A_{2,1}^{\otimes 12}]$ & 25. & Sch. 66 &  $\mathcal{E}_1[D_{12,1}^{\otimes 2}]$  \\
12. & Sch. 29 &  $\mathcal{E}_1[B_{4,2}^{\otimes 3}]$ &   26. & Sch. 67 &  $\mathcal{E}_1[A_{24,1}]$\\
13. & Sch. 30 &  $\mathcal{E}_1[A_{3,1}^{\otimes 8}]$  & 27. & Sch. 68 &  $\mathcal{E}_1[E_{8,1}^{\otimes 3}]$ \\
14. & Sch. 37 &  $\mathcal{E}_1[A_{4,1}^{\otimes 6}]$ & 28. & Sch. 70 &  $\mathcal{E}_1[D_{24,1}]$ \\
\hline
\end{tabular} 
\end{threeparttable}
\end{center}
\caption{ One-character extensions of the tensor product of a single CFT, that are in the Schellekens list of $c = 24$ CFTs \cite{Schellekens:1992db}. }\label{t0}
\end{table}

We make the following observation about the one-character CFTs in the Schellekens classification. $28$ of the $71$ CFTs are one-character extensions of  tensor products of a single WZW CFT and we reproduce them here in table \ref{t0}. The second column contains  the numbering of the CFT in the original list of \cite{Schellekens:1992db}  and the third column contains the one-character CFT. 
Inspired by this observation, we are now motivated to search for one-character CFTs by considering tensor products of a single mutliple-character CFT. In this paper, as an initial study, we consider tensor products of $\mathbf{[2, 0]}$ and $\mathbf{[3, 0]}$ CFTs. We first have to take the tensor products followed by searching for one-character extensions. 

Let us denote the single CFT that we wish to tensor by $\mathcal{C}$ and let us take the tensor product of $N$ copies of $\mathcal{C}$. We will denote the tensor product CFT by $\mathcal{C}^{\otimes N}$. The $N$ will depend on the central charge of $\mathcal{C}$ as well as the central charge of the one-character extension we wish to construct. We will give details pertaining to $N$ in section \ref{2s}. $\mathcal{C}^{\otimes N}$ is a multiple-character CFT. We will first learn how to construct it’s characters. When $\mathcal{C}$ has two characters, say $\chi_0, \chi_1$, then the characters of $\mathcal{C}^{\otimes N}$ are the monomials of degree $N$ that can obtained from the homogenous polynomial $(\chi_0 + \chi_1)^N$. Thus, in this
case, $\mathcal{C}^{\otimes N}$ has $N+1$ characters. Similarly, when $\mathcal{C}$ has three characters, say $\chi_0, \chi_1, \chi_2$, then the characters of $\mathcal{C}^{\otimes N}$ are the monomials of degree $N$ that can obtained from the homogenous polynomial $(\chi_0 + \chi_1 + \chi_2)^N$. Thus in this case, $\mathcal{C}^{\otimes N}$ has $\frac{(N+1)(N+2)}{2}$ characters. The identity character of $\mathcal{C}^{\otimes N}$ is $\chi_0^N$ and hence $\mathcal{C}^{\otimes N}$ has a central charge equal to $N$ times the central charge of $\mathcal{C}$.

Now that we understand how to construct tensor product CFTs, we can now proceed to the next step of the process in this paper viz. constructing one-character extensions. The character of a one-character extension CFT,  $\mathcal{E}_1[\mathcal{C}^{\otimes N}]$ is a certain (homogenous) polynomial of the characters of the tensor product CFT. This polynomial includes the identity character $\chi_0^N$ and other characters that have a commensurate $q$-series with the identity character\footnote{Every character of   $\mathcal{C}^{\otimes N}$ has a $q$-series $q^{-\frac{c N}{24} + h} (a_0 + a_1 q + a_2 q^2 + \ldots)$ where $h$ is it’s conformal dimension; having a $q$-series commensurate with the identity character translates to $h$ being a positive integer.}.  Let us denote by $b$ the number of such characters. The candidate character of the one-character extension is a polynomial with $b+1$ terms (the $+1$  is for the identity character) and there are $b$ number of arbitrary co-efficients in it. This polynomial is a homogenous polynomial of degree $N$. Let us denote this polynomial by $P(\chi_0, \chi_1, \ldots)$. Now we need to impose the modular data of a one-character extension, which we reviewed above, on $P(\chi_0, \chi_1, \ldots)$. The choice of monomials already ensures the correct action of $T$ viz. by multiplication by a cube root of $1$. Thus we have to only impose $S$-invariance : 
\be \label{1}
S P (\chi_0, \chi_1, \ldots) ~\equiv~ P (S(\chi_0), S(\chi_1), \ldots). ~=~ P(\chi_0, \chi_1, \ldots)
\ee
In the above, the first $\equiv$ is simply the definition : the $S$-transformation of  $P(\chi_0, \chi_1, \ldots)$ is derived from the $S$-transformation of the $\chi_i$’s, that is the modular $S$-matrix of $\mathcal{C}$. The second equality is the imposition of $S$-invariance. This equality is an equality of two homogenous polynomials of degree $N$. Thus it results in as many equations as there are monomials of degree $N$, which are way larger than the number of undetermined coefficients in $P(\chi_0, \chi_1, \ldots)$. Thus we have an overdetermined linear system of equations for the $b$ undetermined coefficients in $P(\chi_0, \chi_1, \ldots)$.  This is our procedure. We solve simultaneously some finite number of linear equations  and we determine the polynomial $P(\chi_0, \chi_1, \ldots)$.  Our procedure is thus very simple technically : algebraic manipulations of homogenous polynomials in polynomial rings;  we do not solve any MLDEs here.

All these considerations are implemented in the subsequent sections. In section two, we begin by working out the systematics of the tensor product and the extension. Given a CFT $\mathcal{C}$, it is not always possible to obtain a one-character CFT with every central charge. For example, it is impossible to obtain a one-character CFT with central charge $8$  when $\mathcal{C} = G_{2,1}$. Hence, we begin in the next section by deriving, given $\mathcal{C}$, what are the allowed values of $N$ and what are the values of the central charges $8k$ of the one-character extension that can be achieved. 

Then, in section three, we apply the above delineated procedure when $\mathcal{C}$ is a $\mathbf{[2,0]}$ WZW CFT.  Recall that the list of $\mathbf{[2,0]}$ WZW CFTs  is in one-to-one correspondence with the Deligne series of exceptional Lie algebras viz. $\{ A_{1,1}, A_{2,1}, G_{2,1}, D_{4,1}, F_{4,1}, E_{6,1}, E_{7,1} \}$.  We describe in full detail  the computations for $A_{1,1}$. We are able to obtain one-character extensions $\mathcal{E}_1[A_{1,1}]$ with central charges upto $128$.  Then we give the results for $A_{2,1}$, $G_{2,1}$ and $D_{4,1}$.  The other WZW CFTs in the $\mathbf{[2,0]}$ classification are $F_{4,1}$, $E_{6,1}$  and $E_{7,1}$. It is well-known that they can be obtained  as a coset CFT : coseting $E_{8,1}$ by $G_{2,1}$, $A_{2,1}$,  and $A_{1,1}$ respectively.   Hence the modular S-matrices of $F_{4,1}$, $E_{6,1}$  and $E_{7,1}$ are identical to those of $G_{2,1}$, $A_{2,1}$,  and $A_{1,1}$ respectively. We can adapt the computations for the latter to obtain results for the former.

In section \ref{4s}, we apply the above described procedure for a few (including an infinite class) of $\mathbf{[3,0]}$ WZW CFTs. We study the $A_{4,1}$ CFT first. Then we study the Ising CFT. Finally we study the infinite set of $D_{r,1}$ CFTs for all $r$. It turns out there are four categories (i) $r$ is an odd number, (ii) $r$ is twice of an odd number, (iii) $r$ is four times and odd number, (iv) $r$ is a multiple of eight ; and there is a common formulation for all the solutions  in any category.

Before we proceed further in summarizing the contents of this paper, we should dwell on a major caveat. MLDE studies of CFTs  always proceed in two waves. In the first wave, one works only with characters and the emphasis is on scanning all of the MLDE parameter space and obtaining solutions with non-negative coefficients in the $q$-series. These solutions are referred to as admissible characters. One has to note that these solutions are far from being characters of CFTs. For example, every multiple of a non-identity character is also a solution, but clearly all multiples of all characters will not have the correct modular properties.  In the second wave of analysis, one worries about which of the admissible character solutions are characters of  CFTs. For example, for $\mathbf{[2,0]}$ CFTs, the first wave of analysis was done in the seminal paper \cite{Mathur:1988na}, and the second wave in \cite{Mathur:1988gt}. It is useful to note that  except one, all admissible character solutions of the $\mathbf{[2,0]}$ MLDE resulted in CFTs.  For $\mathbf{[3,0]}$ CFTs, the first wave of analysis was done in the papers \cite{Kaidi:2021ent}, \cite{Das:2021uvd}, \cite{Bae:2021mej} and the second wave in \cite{Das:2022uoe}.  Here the dichotomy between admissible character solutions (around $300$ in number) and actual CFTs (around $45$ in number) is a little more severe.  This dichotomy between admissible characters and actual CFTs is most severe for one-character CFTs.  For example,  for $c = 24$, there are an infinite number of admissible characters,  but only $71$ actual CFTs, the celebrated Schellekens list of CFTs \cite{Schellekens:1992db}. For $c > 24$, the first wave of analysis is readily performed and admissible characters are found to be infinite in number, for each central charge. But the second wave of analysis has not been done yet and a list of actual CFTs is not available. Now comes the caveat : in this paper, we are very much within the first wave of analysis. We are working with admissible characters only. Imposing modular data as we do via \eqref{1} followed by imposing positivity,  results in only admissible characters. We may use the expression ``one-character extension CFT'' but we are always only working with a ``character of a possible one-character extension CFT''.

In this paper, the way we have setup the problem of finding characters of possible one-character CFTs is algebraic. All our computations are within a  polynomial ring. Remember that our answers are homogenous polynomials that are invariant under the $S$-transformation. To start with we work with the natural basis of generators in which the problem is defined viz. the characters of the ${\cal C}$ CFT.  One of the major results of this paper is that this is a very inefficient basis of generators for the homogenous polynomial algebra. We find an alternate basis of generators, where the basis elements themselves are $S$-invariant. We discovered this basis from the answers of the first few cases. For computations of every two-character CFT that we perform in section \ref{2s}, we find two $S$-invariant polynomial basis $\{ P_1, P_2\}$. We express all our answers in terms of this basis and the expressions become compact. Similarly for the three-characer CFTs we study in section \ref{3s}, we find three $S$-invariant polynomial basis $\{ P_1, P_2, P_3\}$. Because of the efficient computations and compact forms, we are able to find explicit answers, for example to questions such as: what is the degree $48$ polynomial of characters of the Ising CFT that is the character of the monster CFT?

At the outset, it seems that the character of the one-character CFT $\mathcal{E}_1[\mathcal{C}^{\otimes N}]$ is an element of the polynomial  ring of characters of ${\cal C}$. But surprisingly we find that (for some CFTs) it is properly an element of a quotient ring of this  polynomial ring.  We find that for some  CFTs   there are polynomial relations  that exist between the characters. And then these polynomial relations generate ideals which should be used to obtain the quotient ring. This is the content of section \ref{5s}.

\section{ Systematics of Tensor Products and One-character Extensions\label{2s}}

In this section, we will describe the general theory and working method of this paper. We will first discuss the tensor product procedure and the characters of a tensor product. Then we discuss the procedure of obtaining one-character extensions. We will discuss how to obtain an ansatz for the character of a one-character extension  and impose the modular properties of one-character CFTs on the ansatz. We will learn to impose the correct $T$-transformation and then the methodology to impose $S$-invariance. We will also elucidate on the relationship between the number of tensor product factors, the central charge of the one-character extension and explicitly display this in tables for the CFTs we consider in this paper. 

The CFT which is being tensored is denoted by  $\mathcal{C}$. Let us denote it's central charge by  $c(\mathcal C)$ and let $n_{\mathcal C}$ denote the number of characters of $\mathcal{C}$. In this paper we have $n_{\mathcal C} = 2, 3$ and $\mathcal C$ is either a $\mathbf{[2,0]}$ or a $\mathbf{[3,0]}$ WZW CFT (except for the Ising CFT).  Let us denote the characters of $\cal C$ by
\be
\chi_0, \chi_1, \ldots \chi_{(n_{\mathcal C} - 1)}.
\ee
Here, $\chi_0$ is the identity character. The non-identity characters $\chi_i$ have  conformal dimensions $h_i, ~ i = 1, 2, \ldots (n_{\mathcal C} - 1)$. The numbering of characters is such that as $i$ increases, $h_i$ increases. For an unitary CFT, this is just a matter of naming the characters. For non-unitary CFTs (which we do not deal with, for the most part, in this paper), this can be done when one works with the unitary presentation of the non-unitary CFT.  The $q$-series of the characters take the form :
\be \label{3b}
\chi_i(q) = q^{-\frac{c(\mathcal C)}{24} + h_i}\,\left( p_0^{(i)} + p_1^{(i)}\,q + p_2^{(i)}\,q^2 + \ldots \right)
\ee
with the Fourier coefficients $p_k^{(i)}, k = 0, 1 \ldots$ all non-negative integers and $p_0^{(0)} = 1$. The quantity $\alpha \equiv -\frac{c(\cal C)}{24} + h_i$ is also referred to as the index of the $q$-series. We will refer to $q^\alpha$ as the ``index-part of the $q$-series.'' The other factor in \eqref{3b} viz.  $\left( p_0^{(i)} + p_1^{(i)}\,q + p_2^{(i)}\,q^2 + \ldots \right)$ will be referred to as the ``power-series part of the $q$-series.''

The action of $S$-transformation on the characters can be written as a $n_{\mathcal C} \times n_{\mathcal C}$ matrix and  this matrix can be extracted from the modular $S$-matrix of the CFT (which can be bigger in size, when multiple ``primary fields'' have the same character).  We will denote this  $n_{\mathcal C} \times n_{\mathcal C}$ matrix by
\be \label{4b}
S_{ij}, \qquad i, j = 0, 1, \ldots (n_{\mathcal C} -1 ).
\ee

Let us denote by $N$ the number of copies of the ${\cal C}$ CFT that we take in the tensor product : $\mathcal C^{\otimes N}$. The characters of this tensor product CFT are nothing but the monomials of the homogenous degree-$N$ polynomial 
\be
(\chi_0 + \chi_1 + \ldots \chi_{(n_{\mathcal C}-1)})^N
\ee
and they can be written explicitly : 
\be \label{6b}
\Pi_{i = 0}^{(n_{\mathcal C}-1)}~\chi_i^{d_i}, \qquad \qquad \sum_{i = 0}^{(n_{\mathcal C}-1)}\,d_i = N.
\ee
For $n_{\mathcal C} = 2$, the number characters of $\mathcal C^{\otimes N}$ is $(N+1)$ and  for $n_{\mathcal C} = 3$, the number characters of $\mathcal C^{\otimes N}$ is $\frac{(N + 1)(N + 2)}{2} $.  The index-part of the $q$-series of the identity character $\chi_0^N$ is  $q^{-\frac{c(\mathcal C)\,N}{24}}$. Hence, we find that the central charge of tensor product CFT, $\mathcal C^{\otimes N}$, is $N c(\mathcal C)$.

A one-character extension of a CFT $\mathcal D$ is a one-character CFT constructed from the CFT $\mathcal D$. The notation for this is $\mathcal E_1[\mathcal D]$. In this paper, we mostly study characters of one-character CFTs and not the CFTs themselves. We will use the same notation ($\mathcal E_1[\mathcal D]$) for the character as well.  The character of the one-character extension $\mathcal E_1[\mathcal D]$ has the extremely simple modular properties of one-character CFTs, which we have already detailed in the introduction. There is a $S$-invariance and $T$ acts on the single character by mutliplication of a cube root of unity, which results in the central charge of a one-character CFT to be a multiple of $8$.

Our  ultimate goal is to construct CFTs which are one-character extensions of the tensor product CFT $\mathcal C^{\otimes N}$ with $\mathcal C$ either a $\mathbf{[2,0]}$ or $\mathbf{[3,0]}$ WZW CFT. But in this paper, we will only address a smaller goal and achieve this to a large extent for the cases we consider. The smaller goal is the following :  using the characters of the tensor product CFT $\mathcal C^{\otimes N}$ and the accompanying modular data\footnote{The modular data for the $\mathcal C^{\otimes N}$ CFT can be derived from the modular data for $\cal C$.} as ingredients, construct a character with the modular properties of a one-character CFT.  A subsequent study (not in this paper) should then establish which of the characters of this paper will result in a genuine one-character CFT and this genuine one-character CFT should have the notation $\mathcal E_1[\mathcal C^{\otimes N}]$. For example, in table \ref{t0}, every $\mathcal E_1[\ldots]$ refers to a genuine one-character CFT (and not just a character with the required modular properties). But, with a slight abuse of notation, when we write $\mathcal E_1[\mathcal C^{\otimes N}]$, we mostly  mean only the character for a possibility for a one-character CFT.

The first step in constructing the character for a possible one-character CFT is to identify all characters of $\mathcal C^{\otimes N}$ that have a  $q$-series commensurate with the identity character.  By this we mean that it’s power-series part is similar to that of the identity character and when we compare the index-parts, we find that the indices differ by integers. 
Thus we need to identify every character whose index differs from the index of the identity character by an integer.  This happens for the character in \eqref{6b} when the $d_i$'s are such that 
\be \label{7b}
\sum_{i = 1}^{(n_{\mathcal C}-1)} ~ h_i \, d_i = M, \qquad \qquad M \in \mathbf Z^+.
\ee
We should  remember that the $d_i$'s need to satisfy the homogeneity constraint also:
\be \label{8b}
\sum_{i = 0}^{(n_{\mathcal C}-1)}\,d_i = N
\ee
Thus, to find all characters of $\mathcal C^{\otimes N}$ that have a $q$-series commensurate with the identity character $\chi_0^N$, we need to simultaneously solve, for the $d_i$'s, the two equations \eqref{7b} and \eqref{8b}. Note that $d_0$ does not occur in \eqref{7b}. Solutions for \eqref{7b} and \eqref{8b} exist for a finite number of values for $M$, only for  $M \leq N h_{(n_{\mathcal C}-1)}$. Let us denote the number of solutions by $b$ and the solutions themselves by 
\be \label{9b}
d_i^{(I)}, \qquad \qquad I = 1, 2, \ldots b.
\ee

Consider the following sum of characters : 
\be \label{10b}
\chi_0^N ~+  ~\sum_{I = 1}^b b_{I} \, \left( \Pi_{i = 0}^{(n_{\mathcal C}-1)} \,\chi_i^{d_i^{(I)}} \right) \equiv P(\chi_0, \chi_1, \ldots \chi_{(n_{\mathcal C}-1)}), 
\ee
where the $b_I$'s are arbitrary constants at this stage. Later, they will be fixed (partially, if not completely) by requiring $P(\chi_0, \chi_1, \ldots \chi_{(n_{\mathcal C}-1)})$ to satisfy the modular properties of a one-character CFT.

The $q$-series of each of the summands in \eqref{10b} begins\footnote{After factoring out this beginning, the rest of the $q$-series, the integral powers, starts with $q^{h_i \, d_i^{(I)}}$ for the summand in \eqref{10b} with coefficient $p_I$. For the identity character $\chi_0^N$, this rest of the $q$-series start with $q^0$. Thus the integral power part of the $q$-series of each of the summands of \eqref{10b} can be added like usual power series. This is the meaning of commensurate $q$-series. } with $q^{-\frac{c(\mathcal C)\,N}{24}}$. This means that the $q$-series of $P(\chi_0, \chi_1, \ldots \chi_{(n_{\mathcal C}-1)})$ begins with $q^{-\frac{c(\mathcal C)\,N}{24}}$.   Let us denote by $c_{(1)}$ the central charge of the one-character CFT which we want to construct, the one whose character is \eqref{10b}. We can thus conclude
\be \label{11b}
c_{(1)} = N c(\mathcal C).
\ee

Now we impose the modular properties of a one-character CFT.  First we impose the $T$-transformation: to be a cube root of unity, which means $c_{(1)}$ is multiple of $8$. Hence we have
\bea \label{12b}
c_{(1)} &=& 8\,k, \\
\label{13b}
N c(\mathcal C) &=& 8 \, k, \qquad k \in \mathbf Z^+
\eea

Note that the object $P(\chi_0, \chi_1, \ldots \chi_{(n_{\mathcal C}-1)})$ in \eqref{10b} with $N$ chosen according to \eqref{13b} is not yet the character of a one-character CFT. It incorporates the required $T$-transformation but not the $S$-invariance. It is in a sense a pre-character and we will refer to it as an ``ansatz character.'' It forms an ansatz for the next stage of the computation where we will impose $S$-invariance.  Before we proceed to that stage, we will note the consequences of the $T$-transformation.  
\eqref{13b} is the main equation  which will help us answer the questions :
\newline (1) Given a $\mathcal C$, what all $c_{(1)}$'s can be achieved? In other words what all central-charges of one-character CFTs can one hope to  achieve by the process of taking tensor products and one-character extensions?
\newline  (2) And for each of central charges of one-character CFT that is possible, what is the number of tensor product factors of $\mathcal C$ i.e. $N$  needed?

Since, $\mathcal C$ is a RCFT, it's central charge  $c(\mathcal C)$ is a rational number. Let it be $\frac{m}{n}$ with $m$ and $n$ being co-prime positive integers. We then have, following \eqref{13b},
\be \label{14b}
N = \frac{8\,k}{m}\,n
\ee
$N$ is a whole number. All the factors of $m$ need to be cancelled by factors of the numerator.  One of the factors of $m$ viz. $\gcd(m, 8)$ is cancelled by the $8$ of the numerator. The other factor of $m$ viz. $\frac{m}{\gcd(m, 8)}$ needs to be cancelled by $k$ in the numerator because $n$, being co-prime with $m$ cannot cancel it. Thus all possible values of $k$  are given by
\be \label{15b}
k = \text{multiples of}~ \frac{m}{\gcd(m, 8)} 
\ee
and the number of tensor product factors $N$ follows from \eqref{14b} and is given by
\be \label{16b}
N = \text{multiples of}~ \frac{8 n}{\gcd(m, 8)}.
\ee
The above answers the two questions we asked earlier. The answers are summarised for tensor products of $\bf [2,0]$ CFTs in table \ref{t2} and for $\bf [3,0]$ CFTs in \ref{t3}. For $A_{1,1}$, following entry 1 of table \ref{t2}, we find that the allowed tensor powers are all multiples of $8$ and the corresponding $c_{(1)}$'s are (the same) multiples of $8$. For, $A_{1,1}$, $A_{2,1}$ and $D_{4,1}$ CFTs, every possible $c_{(1)}$ can be achieved by some tensor power. For the other five $\bf [2,0]$ CFTs not all $c_{(1)}$'s can be achieved. We find that for $G_{2,1}$,  $c_{(1)} = 56, 112, 168, 224, \ldots$ can be achieved by taking tensor powers $N = 20, 40, 60, 80 \ldots$ respectively. For $F_{4,1}$, the one-character CFT with the smallest central charge that one can hope to obtain is $104$. From table \ref{t3}, entry 1 we see that one can hope to obtain one-character CFTs for every possible central charge by taking some number of tensor powers of the Ising CFT viz. ${\cal M}(4,3)$. An interesting case is the monster CFT which can be obtained by taking  tensor products of $48$ copies of the Ising CFT.  We will see this in section \ref{4s}  where we can give explicitly the character of the monster CFT in terms of the Ising characters. 

Another point to note is that there exist sets of CFTs with the same $N$ values. From table \ref{t2}, from rows 1 and 7, we get the pair $A_{1,1}$ and $E_{7,1}$. Similarly, from rows 3 and 5, we get the pair $G_{2,1}$ and $F_{4,1}$.  From table \ref{t3}, each of the rows 3, 4, 5, 6 and 7 consist of a infinite set of CFTs with the same $N$ values. One consequence of this is that all the CFTs in a set can be analysed simultaneously and eventually solved in one-shot. 

\begin{table}[h]
\begin{center}
\begin{threeparttable}
\rowcolors{2}{Mywhite}{Mygrey}
\resizebox{!}{6.5cm}{
   \renewcommand{\arraystretch}{1.3}
   $\begin{array}{|c|c|c|c|c|c|c|c|c|c|c|}
\hline
\# & \mathcal C  & & & & & & & & & \\
\hline
1. & A_{1,1} & c_{(1)} = 8 k & 8 & 16 & 24 & 32 & 40 & 48 & 56 & \ldots \\
&  & N & 8 & 16  & 24 & 32 & 40 & 48 & 56 & \ldots \\
2. & A_{2,1} & c_{(1)} = 8 k & 8 & 16 & 24 & 32 & 40 & 48 & 56 & \ldots \\
&  & N & 4 & 8  & 12 & 16 & 20 & 24 & 28 & \ldots \\
3. & G_{2,1} & c_{(1)} = 8 k & 56 & 112 & 168 & 224 & 280 & 336 & 392 & \ldots \\
&  & N & 20  & 40 & 60 & 80 & 100 & 120 & 140 & \ldots  \\
4. & D_{4,1} & c_{(1)} = 8 k & 8 & 16 & 24 & 32 & 40 & 48 & 56 & \ldots \\
&  & N & 2 & 4 & 6 & 8 & 10 & 12 & 14 & \ldots \\
5. & F_{4,1} & c_{(1)} = 8 k & 104 & 208 & 312 & 416 & 520 & 624 & 728 & \ldots \\
&  & N & 20  & 40 & 60 & 80 & 100 & 120 & 140 & \ldots \\
6. & E_{6,1} & c_{(1)} = 8 k & 24 & 48 & 72 & 96 & 120 & 144 & 168 & \ldots \\
&  & N & 4 & 8  & 12 & 16 & 20 & 24 & 28 & \ldots \\
7. & E_{7,1} & c_{(1)} = 8 k & 56 & 112 & 168 & 224 & 280 & 336 & 392 & \ldots \\
&  & N & 8 & 16  & 24 & 32 & 40 & 48 & 56 & \ldots \\
\hline
\end{array}$
}
\end{threeparttable}
\caption{ Allowed values of number of tensor product factors $N$ and central charge  $c_{(1)}$  of one-character extensions for ${\mathbf [2,0]}$ CFTs. }\label{t2}
\end{center}
\end{table}

\begin{table}[h]
\begin{center}
\begin{threeparttable}
\rowcolors{2}{Mywhite}{Mygrey}
\resizebox{\textwidth}{!}{
   \renewcommand{\arraystretch}{1.3}
   $\begin{array}{|c|c|c|c|c|c|c|c|c|c|c|}
\hline
\# & \mathcal C  & & & & & & & & & \\
\hline
1. & \mathcal M(4,3) & 8 k & 8 & 16 & 24 & 32 & 40 & 48 & 56 & \ldots \\
&  & N & 16 & 32  & 48 & 64 & 80 & 96 & 112 & \ldots \\
2. & A_{4,1} & 8 k & 8 & 16 & 24 & 32 & 40 & 48 & 56 & \ldots \\
&  & N & 2 & 4 & 6 & 8 & 10 & 12 & 14 & \ldots \\
3. & D_{r,1} & 8 k & 8(2p+1) & 16(2p+1) & 24(2p+1) & 32(2p+1) & 40(2p+1) & 48(2p+1) & 56(2p+1) & \ldots \\
& r = 2p+1  & N & 8 & 16  & 24 & 32 & 40 & 48 & 56 & \ldots \\
& p \in \mathbf Z^+ &  &  &   &  &  &  &  &  & \ldots \\
4. & D_{r,1} & 8 k & 8(2p+1) & 16(2p+1) & 24(2p+1) & 32(2p+1) & 40(2p+1) & 48(2p+1) & 56(2p+1) & \ldots \\
& r = 2(2p+1) & N & 4 & 8  & 12 & 16 & 20 & 24 & 28 & \ldots \\
& p \in \mathbf Z^+ &  &  &   &  &  &  &  &  & \ldots \\
5. & D_{r,1} & 8 k & 8(2p+1) & 16(2p+1) & 24(2p+1) & 32(2p+1) & 40(2p+1) & 48(2p+1) & 56(2p+1) & \ldots \\
& r = 4(2p+1) & N & 2 & 4  & 6 & 8 & 10 & 12 & 14 & \ldots \\
& p \in \mathbf Z^+ &  &  &   &  &  &  &  &  & \ldots \\
6. & D_{r,1} & 8 k & 8p' & 16p' & 24p' & 32p' & 40p' & 48p' & 56p' & \ldots \\
& r = 8p' & N & 1 & 2  & 3 & 4 & 5 & 6 & 7 & \ldots \\
& p' \in \mathbf Z^+ &  &  &   &  &  &  &  &  & \ldots \\
7. & B_{r,1} & 8 k & 8(2r+1) & 16(2r+1) & 24(2r+1) & 32(2r+1) & 40(2r+1) & 48(2r+1) & 56(2r+1) & \ldots \\
&  & N & 16 & 32 & 48 & 64 & 80 & 96 & 112 & \ldots \\
\hline
\end{array}$
}
\end{threeparttable}
\end{center}
\caption{Allowed values of number of tensor product factors $N$ and central charge  $c_{(1)}$  of one-character extensions for ${\mathbf [3,0]}$ CFTs. }\label{t3}
\end{table}

In \eqref{10b}, what we have is an ansatz for the character of the one-character extension $P(\chi_0, \ldots \chi_{(n_{\mathcal C}-1)})$. This ansatz  together with \eqref{13b} incorporates the correct $T$-transformation property. What is left to do is to impose $S$-invariance.  First we learn how to $S$-transform the ansatz character $P$ : the $S$-transformation of $P$ is derived from the $S$-transformation of the characters $\chi_i$, from \eqref{4b}.
\be \label{17b}
S(P(\chi_0, \chi_1, \ldots \chi_{(n_{\mathcal C}-1)})) \equiv P(S(\chi_0), S(\chi_1), \ldots S(\chi_{(n_{\mathcal C}-1)})), \qquad S(\chi_i) = \sum_j S_{ij}\,\chi_j.
\ee
Now we can impose $S$-invariance via 
\be \label{18b}
P(S(\chi_0), S(\chi_1), \ldots S(\chi_{(n_{\mathcal C}-1)})) = P(\chi_0, \chi_1, \ldots \chi_{(n_{\mathcal C}-1)}).
\ee
We know that the right hand side is a certain homogenous polynomial of degree $N$ with  the number of monomials being $b$  (and $b$ is smaller than the total number of monomials of degree $N$). But the left hand side, which is also a homogenous polynomial of degree $N$, may contain all the monomials. In anycase, we can describe \eqref{18b} as the vanishing of a homogenous polynomial of degree $N$. Hence, we have as many equations as there are monomials of degree $N$. Each of these equations is a linear equation for the $b$ unknown coefficients $b_I$ of the ansatz character $P$. We thus have an over-determined linear system of equations.  Thus, to make the ansatz character $S$-invariant, we need to choose the $b_I$'s  from a solution of these linear system of equations. We will use the notation $P^{\,\cal C}_{c_{(1)} = 8k}(\chi_0, \ldots)$ to denote the $S$-invariant character while we will reserve the notation $P(\chi_0, \ldots \chi_{(n_{\mathcal C}-1)})$ for the ansatz character before imposing $S$-invariance.

This is what we do in this paper, we pick a CFT $\mathcal{C}$ and a $N$ and a $k$, in accordance with the tables \ref{t2}, \ref{t3}. We set up the linear system of equations and solve them. It turns out that we always obtain solutions. Only very rarely, we get the rank of the coefficient matrix of the linear system of equations to be exactly $b$, giving unique solutions. For most of the cases, we find that the rank of the coefficient matrix is less than $b$ giving families of solutions. 

\section{ $\mathbf{[2,0]}$ CFTs : Tensor Products and One-character Extensions \label{3s}}

In this section, we study $\mathbf{[2,0]}$ CFTs. We implement the procedure described in section \ref{2s}  for each of the $\mathbf{[2,0]}$  WZW CFTs. We start with the smallest central charge viz. $A_{1,1}$ and go up the sequence in increasing central charges. For reasons that will become clear below, we report the results of  $A_{1,1}$ and $E_{7,1}$ together in section \ref{31s}, $A_{2,1}$ and $E_{6,1}$ together in section \ref{32s}, $G_{2,1}$ and $F_{4,1}$ together in section \ref{33s} and $D_{4,1}$ in section \ref{34s}. We will describe the $A_{1,1}$ case in some detail and be somewhat brief for the others.

\subsection{$\mathcal C = A_{1,1}$ and $E_{7,1}$ \label{31s}}

$A_{1,1}$ is a $\mathbf{[2,0]}$ CFT with central charge $1$. Let us denote it's two characters by $\chi_0, \chi_1$, the latter is the non-identity character with conformal dimension $h_1 = \frac14$. Let us note the $q$-series for the characters :
\begin{align}
     \chi_0& =q^{-\frac{1}{24}}(1+3 q+4 q^2+7 q^3+13 q^4+19 q^5+29 q^6+43 q^7+62 q^8+90 q^9+126 q^{10}+ \ldots) \nonumber \\
     \chi_1& =q^{\frac{5}{24}}(2+2q+6 q^2+8 q^3+14 q^4+20 q^5+34 q^6+46 q^7+70 q^8+96 q^9+138 q^{10}+ \ldots) \label{19b}
\end{align}
and the modular $S$-transformations for the characters :
\be \label{20b}
S(\chi_0) = \frac{\chi_0 + \chi_1}{\sqrt{2}}, \qquad \qquad S(\chi_1) = \frac{\chi_0 - \chi_1}{\sqrt{2}}.
\ee

The allowed values of $k$ and $N$ can be read from entry 1 of table \ref{t2}. Here, every possible $c_{(1)} = 8k$ is allowed and the corresponding number of tensor product factors is  $N = 8k$. 

We begin with \underline{$N = 8, c_{(1)} = 8$.} $A_{1,1}^{\otimes 8}$ has nine characters : $\chi_0^m\, \chi_1^n,  ~~m+n = 8$ with $m, n$ non-negative integers. The   conformal dimension of $\chi_0^m\, \chi_1^n$ is $\frac{n}{4}$. Solving \eqref{7b} and \eqref{8b} we obtain the  characters which have $q$-series commensurate with the identity : $\{ \chi_0^8, \, \chi_0^4\,\chi_1^4, \, \chi_1^8
 \}$. Hence, the ansatz character \eqref{10b} which incorporates the relevant $T$-transformation is 
\be
P(\chi_0, \chi_1) = \chi_0^8 + b_1 \, \chi_0^4\,\chi_1^4 + b_2 \, \chi_1^8,
\ee
which has two arbitrary coefficients.

Now, we impose $S$-invariance, \eqref{18b}, \eqref{20b}.  We get a system of  $9$ linear equations for the two variables $b_1$ and $b_2$. Even before we solve the equations, we expect that there is a unique solution with $b_1$ and $b_2$ taking some specific values. This expectation comes from the fact that characters for $c_{(1)} = 8$ do not contain any free parameters (see row 1 of table \ref{tapp} of the appendix \ref{app1}). And indeed we get the  following unique character for $N = 8, c_{(1)} = 8$ :
\be \label{22b}
P^{A_{1,1}}_{c_{(1)} = 8}(\chi_0, \chi_1) ~=~\chi_0^8 + 14 \, \chi_0^4\,\chi_1^4 +  \chi_1^8.
\ee
We can now compute the $q$-series of $P^{A_{1,1}}_{c_{(1)} = 8}(\chi_0, \chi_1)$ using \eqref{19b} and we get
$P^{A_{1,1}}_{c_{(1)} = 8}(\chi_0, \chi_1)(q) = j^{\frac13}$. 

We recall some well-known facts about characters and one-character CFTs with $c_{(1)} = 8$ : that there is a unique character viz. $j^{\frac13}$ and this corresponds to the CFT $E_{8,1}$. Hence, what we have derived is the fact that 
\be \label{23b}
{\cal E}_1[A_{1,1}^{\otimes 8}] = E_{8,1}
\ee
We then study the second entry of row 1 of table \ref{t2} viz.  \underline{$N = 16, c_{(1)} = 16$.} $A_{1,1}^{\otimes 16}$ has $17$ characters : $\chi_0^m\, \chi_1^n,  ~~m+n = 16$ with $m, n$ non-negative integers. The   conformal dimension of $\chi_0^m\, \chi_1^n$ is $\frac{n}{4}$. Solving \eqref{7b} and \eqref{8b} we obtain the  characters which have $q$-series commensurate with the identity : $\{ \chi_0^{16}, \, \chi_0^{12}\,\chi_1^4, \, \chi_0^{8}\,\chi_1^8, \, \chi_0^{4}\,\chi_1^{12},  \, \chi_1^{16}  \}$. Hence, the ansatz character \eqref{10b} which incorporates the relevant $T$-transformation is 
\be \label{24b}
P(\chi_0, \chi_1)= \chi_0^{16} + b_1 \, \chi_0^{12}\,\chi_1^4 + b_2 \, \chi_0^{8}\,\chi_1^8 + b_3 \, \chi_0^{4}\,\chi_1^{12}  +  b_4 \, \chi_1^{16},
\ee
which has four arbitrary coefficients.

Now, we impose $S$-invariance, \eqref{18b}, \eqref{20b}.  We get a system of  $17$ linear equations for the $4$ variables $b_1, b_2, b_3, b_4$. Even before we solve the equations, we expect that there is a unique solution with $b_1, b_2, b_3, b_4$ taking some specific values. This expectation comes from the fact that characters for $c_{(1)} = 16$ do not contain any free parameters (see row 2 of table \ref{tapp} of the appendix \ref{app1}). The result follows this expectation. We get the  following unique character for $N = 16, c_{(1)} = 16$ :
\be \label{25b}
P^{A_{1,1}}_{c_{(1)} = 16}(\chi_0, \chi_1) ~=~\chi_0^{16} + 28 \, \chi_0^{12}\,\chi_1^4 + 198 \, \chi_0^{8}\,\chi_1^8 + 28 \, \chi_0^{4}\,\chi_1^{12}  +    \chi_1^{16} 
\ee
We can now compute the $q$-series of $P^{A_{1,1}}_{c_{(1)} = 16}(\chi_0, \chi_1)$ using \eqref{19b} and we get
$P^{A_{1,1}}_{c_{(1)} = 16}(\chi_0, \chi_1)(q) = j^{\frac23}$. 

We recall some well-known facts about characters and one-character CFTs with $c_{(1)} = 16$ : that there is a unique character viz. $j^{\frac23}$ and there are two CFTs both with this character viz. $E_{8,1}^{\otimes 2}$ and ${\cal E}_1[D_{16,1}]$. Hence, what we have derived is the fact that 
\be \label{26b}
{\cal E}_1[A_{1,1}^{\otimes 16}] = E_{8,1}^{\otimes 2} \quad  \text{or} \quad {\cal E}_1[D_{16,1}].
\ee.
We further note that the $S$-invariant polynomial of degree $16$ in \eqref{25b} is nothing but the square of the $S$-invariant polynomial of degree $8$ in \eqref{22b}. Let us denote the latter by
\be
\label{27b}
P_1(\chi_0, \chi_1) \equiv \chi_0^8 + 14 \, \chi_0^4\,\chi_1^4 +  \chi_1^8
\ee
so that we can write both the results for $c_{(1)} = 8, 16$ in terms of this  $P_1$ :
\be \label{28b}
P^{A_{1,1}}_{c_{(1)} = 8}(\chi_0, \chi_1) = P_1(\chi_0, \chi_1), \qquad  P^{A_{1,1}}_{c_{(1)} = 16}(\chi_0, \chi_1) = P_1(\chi_0, \chi_1)^2
\ee
One of the major themes of this paper is the existence of a few $S$-invariant polynomials for every CFT ${\cal C}$.  Every character of  any one-character extension ${\cal E}_1[{\cal C}^{\otimes N}]$ can be written as a polynomial of these $S$-invariant polynomials. What we have discovered here in \eqref{27b} is the first $S$-invariant polynomial for $A_{1,1}$. And in \eqref{28b}, we have written the characters for ${\cal E}_1[{\cal A}_{1,1}^{\otimes 8}]$ and ${\cal E}_1[{\cal A}_{1,1}^{\otimes 16}]$ as polynomials of $P_1$. Note that as polynomials of the characters, they are of degree $8$ and $16$ respectively, but as polynomials of $P_1$, they are of degree $1$ and $2$ respectively. This reduction in degrees when one goes from the description in terms of characters and the description in terms of $S$-invariant polynomials makes for compact expressions and reporting results for large $N$.  We will  see that there is another $S$-invariant polynomial $P_2(\chi_0, \chi_1)$ for $A_{1,1}$. And all results for $A_{1,1}$ can be written as polynomials of $P_1$ and $P_2$. $P_2$ shows up first in the next case. 

We now study the third entry of row 1 of table \ref{t2} viz.  \underline{$N = 24, c_{(1)} = 24$.}  $A_{1,1}^{\otimes 24}$ has $25$ characters: $\chi_0^m\, \chi_1^n,  ~~m+n = 24$ with $m, n$ non-negative integers. The  conformal dimension of $\chi_0^m\, \chi_1^n$ is $\frac{n}{4}$. Solving \eqref{7b} and \eqref{8b} we obtain the  characters which have $q$-series commensurate with the identity : $\{ \chi_0^{24 - 4r} \chi_1^{4r} ~| ~r = 0, 1, \ldots 6\}$. Hence, the ansatz character \eqref{10b} which incorporates the relevant $T$-transformation is 
\be 
P(\chi_0, \chi_1) = \sum_{r = 0}^6 b_r~\chi_0^{24 - 4r} \chi_1^{4r}, ~\text{where} ~b_0 = 1
\ee
which has $6$ arbitrary coefficients.

Now, we impose $S$-invariance, \eqref{18b}, \eqref{20b}.  We get  a system of $25$ equations for $6$ variables $p_1, \ldots p_6$.  This time we do not get a unique solution. The coefficient matrix of the set of linear equations has rank $5$.  One  of the coefficients is undetermined and the rest are given in terms of this coefficient :
\begin{small}
\be \label{30b}
P^{A_{1,1}}_{c_{(1)} = 24}(\chi_0, \chi_1)  = \chi_0^{24} + b_1 \, \chi_0^{20} \chi_1^{4}  + (759 - 4 b_1) \, \chi_0^{16} \chi_1^{8}  + (2576  + 6 b_1) \, \chi_0^{12} \chi_1^{12}  + (759 - 4 b_1) \, \chi_0^{8} \chi_1^{16} + b_1 \, \chi_0^{4} \chi_1^{20} + \chi_1^{24}
\ee
\end{small}
This is in line with expectations. From row 3 of table \ref{tapp} of appendix \ref{app1}, we know that $c_{(1)} = 24$ characters come in a one-parameter family. Hence the character $P^{A_{1,1}}_{c_{(1)} = 24}(\chi_0, \chi_1) $ can have utmost one free parameter.

The polynomial in \eqref{30b} is $S$-invariant for all values of $b_1$. In particular setting  $b_1 = 0$ gives one of the summands of  $P^{A_{1,1}}_{c_{(1)} = 24}(\chi_0, \chi_1) $ which is separately $S$-invariant.  This then implies that the coefficient of $b_1$ is also separately $S$-invariant. Examining the coefficient of $b_1$,  we discover that it is the fourth power of the following $S$-invariant polynomial 
\be
\label{31b}
P_2(\chi_0, \chi_1) \equiv \chi_0^5\,\chi_1 -  \chi_0\,\chi_1^5.
\ee 
The final answer for is $c_{(1)} = 24$ can be written in terms of the two $S$-invariant polynomials viz. the $P_1$ of \eqref{27b} and the $P_2$ of \eqref{31b}.
\be \label{32b}
P^{A_{1,1}}_{c_{(1)} = 24}(\chi_0, \chi_1) = P_1^3 - 42 P_2^4 + b_1 P_2^4
\ee
We can now compute the $q$-series of $P^{A_{1,1}}_{c_{(1)} = 24}(\chi_0, \chi_1)$ using \eqref{19b} :
\be 
P^{A_{1,1}}_{c_{(1)} = 24}(\chi_0, \chi_1)(q) = j + 16 b_1 - 672 \nonumber 
\ee
To obtain an admissible character, we need to choose $b_1$ such that $744 + 16 b_1 - 672$ is a non-negative integer i.e. $b_1 = \frac{p - 72}{16}, p \geq 0, p \in \mathbf{Z}$. We thus have an infinite number of solutions for $N = 24, c_{(1)} = 24$ : 
\be \label{33b}
P^{A_{1,1}}_{c_{(1)} = 24}(\chi_0, \chi_1) = P_1^3 - 42 P_2^4 + b_1 P_2^4, \qquad b_1 = \frac{p - 72}{16}, ~p \geq 0, ~p \in \mathbf{Z}
\ee
This is in sharp contrast to the previous two cases, where we had unique solutions. Amongst the infinite solutions of \eqref{33b}, we wish to know which of them correspond to actual CFTs. We can answer this question  due to the availability of the classification of $c_{(1)} = 24$ CFTs from \cite{Schellekens:1992db}. In the table at the end of that paper entry no. 15 is the only CFT which is a one-character extension of a tensor product of  copies of $A_{1,1}$\footnote{The reader can find this CFT listed as entry no. 9 of table \ref{t0}.}, which has a character given by \eqref{33b} with $b_1 = 0$ ($p = 72$)\footnote{We need to emphasise that the methods of this paper which incorporate admissibility and the modular data are not sufficient to identify CFTs.  In fact, these methods give us an infinite number of solutions \eqref{33b}. We are able to definitively declare that only the $b_1 = 0$ member of this infinite set of solutions corresponds to a CFT only after invoking the existing classification of $c_{(1)} = 24$ CFTs from \cite{Schellekens:1992db}. For  $c_{(1)} > 24$, when no such classification exists, we will not be able to give a definitive list of CFTs. }. 

\vspace{5mm}
Let us note that there is a fundamental difference between the two  $S$-invariant polynomials \eqref{27b} and \eqref{31b}. When we evaluate the $q$-series for the latter, we obtain the  constant $2$.  We will refer to the latter as a constant $S$-invariant polynomial and discuss the significance of the same in section \ref{5s}.  The former has a non-trivial $q$-series and we will refer to it  as a non-constant $S$-invariant polynomial. Let us collect all the information about the two $S$-invariant polynomials of $A_{1,1}$ in one place :
\begin{equation}\label{34b}
\begin{array}{llr}
   P_1(\chi_0, \chi_1) &= \chi_0^8 + 14 \, \chi_0^4\,\chi_1^4 +  \chi_1^8 = j^{\frac13}\\
   P_2(\chi_0, \chi_1) &= \chi_0^5\,\chi_1 -  \chi_0\,\chi_1^5 = 2
\end{array}
\end{equation}
It turns out that the character $P^{A_{1,1}}_{c_{(1)} = 8k}(\chi_0, \chi_1)$ for every $k$ can be written as polynomials of $P_1$ and $P_2$. Again let us note the compactness of the expressions. In equation \eqref{32b} we have a degree-$4$ polynomial and when written in terms of the characters $\chi_0, \chi_1$ as in \eqref{30b} we have a degree-$24$ polynomial.

We have illustrated the computations for $c_{(1)} = 8, 16, 24$ in full detail above.  We follow the same procedure for \underline{$32 \leq N \leq 128$}, \underline{$4 \leq k \leq 16$}.  First, we solve \eqref{7b} and \eqref{8b} and find that there are a total of $b = 2k$ monomials commensurate with the identity character.  The ansatz character \eqref{10b} has $2k$ arbitrary coefficients. Then we impose $S$-invariance  \eqref{18b}, \eqref{20b}. This amounts to solving $8k + 1$ equations for $2k$ variables.  It turns out that the rank of the coefficient matrix is $2k - [ \frac{k}{3} ] $ where $[Q]$ is the integer part of the rational number $Q$. So that the solution  has $ [\frac{k}{3}] $ undetermined  coefficients. We will denote the undetermined coefficients by the $p_j$’s; these are linear combinations of the coefficients in the ansatz character, the $b_I$’s.  This means that the $S$-invariant character $P^{A_{1,1}}_{c_{(1)} = 8k}(\chi_0, \chi_1) $ can be written as a sum of of $[\frac{k}{3}] + 1$ different summands with each summand being separately $S$-invariant :
\be 
P^{A_{1,1}}_{c_{(1)} = 8k}(\chi_0, \chi_1)  = {\cal Z}_0  +  \sum_{j = 1}^{[\frac{k}{3}]}\,p_j \,{\cal Z}_j 
\ee
The number of $p$’s is  $[\frac{k}{3}]$.  The number of ${\cal Z}$'s is $[\frac{k}{3}] + 1$. We will give the results for $k = 16$ i.e. $c_{(1)} = 128$. 
\be \label{36c}
P^{A_{1,1}}_{c_{(1)} = 128}(\chi_0, \chi_1)  = {\cal Z}_0  +  p_1 \,{\cal Z}_1 +  p_2 \,{\cal Z}_2  +  p_3 \,{\cal Z}_3 +  p_4 \,{\cal Z}_4  +  p_5 \,{\cal Z}_5
\ee
with 
\begin{small}
\begin{align}
     {\cal Z}_0& = P_1^{16} - 224 P_1^{13} P_2^4 + 16336 P_1^{10} P_2^8 - 430656 P_1^7 P_2^{12} + 3196776 P_1^4 P_2^{16} - 2696256 P_1 P_2^{20} \nonumber \\
     {\cal Z}_1& = P_1^{13} P_2^4 -178 P_1^{10} P_2^8 + 8917 P_1^7 P_2^{12} - 116350 P_1^4 P_2^{16} + 158178 P_1 P_2^{20} \nonumber \\
     {\cal Z}_2& = P_1^{10} P_2^8 - 132 P_1^7 P_2^{12} + 3614 P_1^4 P_2^{16} - 10608 P_1 P_2^{20} \nonumber \\
     {\cal Z}_3& =   P_1^7 P_2^{12} - 86 P_1^4 P_2^{16} + 427 P_1 P_2^{20} \nonumber \\
     {\cal Z}_4& =  P_1^4 P_2^{16} - 40 P_1 P_2^{20} \nonumber \\
     {\cal Z}_5& =  P_1 P_2^{20}.
     \end{align}
\end{small}
It should be noted that the above expressions for the ${\cal Z}$’s in terms of the $P_1$ and $P_2$'s are extremely compact compared to the homogenous degree-$128$ polynomials of the characters $\chi_0, \chi_1$. In the following, we are able to rewrite the same in an even more compact manner :
\be \label{38c}
P^{A_{1,1}}_{c_{(1)} = 128}(\chi_0, \chi_1)  = \tilde{{\cal Z}}_0  +  \tilde{p}_1 \,\tilde{{\cal Z}}_1 +  \tilde{p}_2 \,\tilde{{\cal Z}}_2  +  \tilde{p}_3 \,\tilde{{\cal Z}}_3 +  \tilde{p}_4 \,\tilde{{\cal Z}}_4  +  \tilde{p}_5 \,\tilde{{\cal Z}}_5
\ee
with 
\begin{small}
\begin{align}
     \tilde{{\cal Z}}_0& = P_1^{16}, \quad \tilde{{\cal Z}}_1 = P_1^{13} P_2^4, \quad  \tilde{{\cal Z}}_2 =  P_1^{10} P_2^8, \quad  \tilde{{\cal Z}}_3 =  P_1^7 P_2^{12}, \quad  \tilde{{\cal Z}}_4 =  P_1^4 P_2^{16}, \quad   \tilde{{\cal Z}}_5 =  P_1 P_2^{20}  
     \end{align}
\end{small}
The relation between the constant coefficients of \eqref{36c} and \eqref{38c} is:
\begin{small}
\begin{align}
         \tilde{p}_1& = p_1 - 224, \quad  \tilde{p}_2 = p_2 - 178 p_1 + 16336, \quad \tilde{p}_3 = p_3 - 132 p_2 + 8917 p_1 - 430656, \nonumber \\
     \tilde{p}_4& =  p_4 - 86 p_3  + 3614 p_2 - 116350 p_1 + 3196776, \quad  \tilde{p}_5 =  p_5 - 40 p_4  + 427 p_3 - 10608 p_2 + 158178 p_1 - 2696256 \nonumber 
     \end{align}
\end{small}
Since, the most compact form is that of \eqref{38c}, we will report this form of the results.

In table \ref{t4}, we collect all the results for $k = 1, 2, \ldots 16$. Table \ref{t4} also contains results for $E_{7,1}$ for reasons that will be explained below. The way to read the results for $A_{1,1}$ from table \ref{t4} is as follows. The first column is the serial no. which is also the value of $k$, from which one can infer the value of $c_{(1)} = 8k$. The (top half of the) second column contains the number of tensor product factors $N$ (which equals $8k$).  The third column contains the character of the one-character extension $P^{A_{1,1}}_{c_{(1)} = 8k}(\chi_0, \chi_1)$.  The number of summands ranges from $1$, for $k = 1$, to $6$ for $k = 16$. For $k = 16$, the last entry of the table \ref{t4} we have six $S$-invariant summands $\tilde{{\cal Z}}_0, \tilde{{\cal Z}}_1, \ldots \tilde{{\cal Z}}_5$ (see \eqref{38c}). The expressions for the $\tilde{{\cal Z}}$'s when written in terms of the characters  $\chi_0, \chi_1$ are very voluminous. Each of them is a  degree $N$ polynomial.  For example for $N= 128$, the last entry of table \ref{t4}, we have  six homogenous degree-$128$ polynomials. But, when written in terms of the $S$-invariant polynomials, they are in fact polynomials of much smaller degree. For $N = 128$, the highest degree term that occurs is $21$.  This is one of the results to report in this paper. Solving the huge linear system of equations that follow from  $S$-invariance,  organising in terms of some number of $S$-invariant summands and writing each of  the summands in terms of $S$-invariant polynomials and thus achieving compact expressions.

Now we need to impose admissibility. For that, we first need to compute the $q$-series of $P^{A_{1,1}}_{c_{(1)} = 8k}$. We employ the explicit $q$-series of the characters available in \eqref{19b} and give the  $q$-series of $P^{A_{1,1}}_{c_{(1)} = 8k}$ in the top half of the fourth column. Then we need to compare this with the generic admissible character for that particular central charge, given in table \ref{tapp} of the appendix \ref{app1}. The generic admissible character for central charge $8k$ has $[\frac{k}{3}]$ parameters which are denoted by $N_1, N_2, \ldots$. And these $N$’s need to satisfy a set of nested inequalities for admissibility, which are available from table \ref{tapp} of the appendix \ref{app1}.  In the top half of the fourth column of table \ref{t4}, we give the relation between the $\tilde{p}$’s and the $N$’s. Then, the nested inequalities for the $N$’s translate to inequalities for the $\tilde{p}$’s. We always find an infinite number of admissible character solutions. 

\underline{Lessons learnt and a new method} :  All the results that we have obtained and reported in table \ref{t4} were by running the whole procedure as laid out  in section \ref{2s} . From all the computations that we did, we could learn some lessons and we are able to come up with the following procedure that allows us to write the final answer (the entries of column two of table \ref{t4}) quite quickly. 

We first note that $P_1$ is a homogenous polynomial of the characters of degree $8$ and that $P_2$ is a homogenous polynomial of the characters of degree $6$. A monomial of the form $P_1^a\,P_2^b$ is a homogenous polynomial of the characters of degree $8a + 6b$. And for this monomial to be present in $P^{A_{1,1}}_{c_{(1)} = 8k}$ it’s degree needs to be equal to $N$. Thus by finding all non-negative integral solutions to 
\be \label{40c}
8a + 6b = N = 8 k, 
\ee
we can obtain all the $\tilde{{\cal Z}}$'s  in $P^{A_{1,1}}_{c_{(1)} = 8k}$.  All except one of the $\tilde{{\cal Z}}$'s  viz. $\tilde{{\cal Z}}_0$ are multiplied by constants the $\tilde{p}$'s.  Which of the solutions of \eqref{40c} corresponds to $\tilde{{\cal Z}}_0$ is easily decided by seeing which of them contains the identity character $\chi_0^N$. Here it is the $a = k, b = 0$ solution. This quick computation of the solutions to \eqref{40c} gives us the same answer as the long procedure.  We need to mention one more caveat. We should only consider solutions of \eqref{40c} that give monomials of characters which are commensurate to the identity character. Here, for $A_{1,1}$, it turns out that every solution to \eqref{40c} gives monomials of characters that are  commensurate to the identity character.

\begin{table}[H]
\begin{center}
\begin{threeparttable}
\resizebox{\textwidth}{!}{
\renewcommand{\arraystretch}{1.3}
   $\begin{array}{|c|c|c|C{16cm}|}
   \hline
\# & N & P_{c_{(1)} = 8k}^{A_{1,1}} (\chi_0, \chi_1) ~\text{and} ~P_{c_{(1)} = 8k}^{E_{7,1}} (\chi_0, \chi_1)  &  q-\text{series}  \\
\hline
    &  \mathcal E_1[A_{1,1}^{~\otimes 8}] && j^{\frac{1}{3}}   \\
\cline{2-2}\cline{4-4}
  & && j^{\frac{1}{3}}(j^2+N_1j+N_2)   \\
    \multirow{-3}{*}{$1.$} & \multirow{-2}{*}{$\mathcal E_1[E_{7,1}^{~\otimes 8}]$}&\multirow{-3}{10em}{\centering $ {\cal \tilde{Z}}_0 = P_1 $}& N_{1}= -672,N_{2}= -145152\\
\hline
   &  \mathcal E_1[A_{1,1}^{~\otimes 16}] && j^{\frac{2}{3}}    \\
\cline{2-2}\cline{4-4}
    & && j^{\frac{2}{3}}(j^4+N_1j^3+N_2j^2+N_3j+N_4)    \\
  \multirow{-3}{*}{$ 2.$} & \multirow{-2}{*}{$\mathcal E_1[E_{7,1}^{~\otimes 16}]$}&\multirow{-3}{10em}{\centering $ {\cal \tilde{Z}}_0 = P_1^2$ }& N_{1}= -1344,N_{2}= 161280,N_{3}= 195084288,N_{4}= 21069103104\\
\hline
&  &&  j+N  \\
 &  \multirow{-2}{*}{$\mathcal E_1[A_{1,1}^{~\otimes 24}]$}&& N=16 \tilde{p}_1\\
\cline{2-2}\cline{4-4}
    & &&  j^7+N_1j^6+N_2j^5+N_3j^4+N_4j^3+N_5j^2+N_6j+N_7  \\
     \multirow{-3}{*}{$ 3.$} & \multirow{-1}{*}{$\mathcal E_1[E_{7,1}^{~\otimes 24}]$}&\multirow{-3}{10em}{\centering ${\cal \tilde{Z}}_0 = P_1^3,$\\ ${\cal \tilde{Z}}_1 = P_2^4$}&\ N_{1}=-2016,\  N_{2}=919296,\  N_{3}=200704 \left(49 \tilde{p}_1+1404\right),\\&&&\  N_{4}=-86704128 \left(28 \tilde{p}_1+1539\right),\  N_{5}=224737099776 \left(\tilde{p}_1-189\right),\ \ldots\\
\hline
 &&&   j^{\frac{1}{3}}(j+N) \\
   & \multirow{-2}{*}{$\mathcal E_1[A_{1,1}^{~\otimes 32}]$}&& N=16 \tilde{p}_1\\
\cline{2-2}\cline{4-4}
   &&&  j^{\frac{1}{3}} (j^9+N_{1} j^8+N_{2} j^7+N_{3} j^6+N_{4} j^5+N_{5} j^4+N_{6} j^3+N_{7} j^2+N_{8} j+N_{9})\\
   \multirow{-3}{*}{$ 4.$} & \multirow{-1}{*}{$\mathcal E_1[E_{7,1}^{~\otimes 32}]$}&\multirow{-3}{10em}{\centering ${\cal \tilde{Z}}_0 = P_1^4,$\\ ${\cal \tilde{Z}}_1 = P_1 P_2^4$}&
   N_{1}=-2688,\  N_{2}=2128896,\  N_{3}=200704 \left(49 \tilde{p}_1-216\right),\\&&&\  N_{4}=-19267584 \left(469 \tilde{p}_1+23679\right),\  N_{5}=4161798144 \left(103 \tilde{p}_1+1512\right),\ \ldots\\
\hline
 && &  j^{\frac{2}{3}}(j+N)  \\
 &  \multirow{-2}{*}{$\mathcal E_1[A_{1,1}^{~\otimes 40}]$}&&  N=16 \tilde{p}_1\\
\cline{2-2}\cline{4-4}
   &&& j^{\frac{2}{3}} (j^{11}+N_{1} j^{10}+N_{2} j^9+N_{3} j^8+N_{4} j^7+N_{5} j^6+N_{6} j^5+N_{7} j^4+N_{8} j^3+N_{9} j^2+N_{10} j+N_{11})\\
   \multirow{-4}{*}{$5.$} & \multirow{-2}{*}{$\mathcal E_1[E_{7,1}^{~\otimes 40}]$}&\multirow{-4}{10em}{\centering ${\cal \tilde{Z}}_0 = P_1^5,$\\ ${\cal \tilde{Z}}_1 = P_1^2 P_2^4$}&
   N_{1}=-3360,\  N_{2}=3790080,\  N_{3}=200704 \left(49 \tilde{p}_1-5400\right),\\&&&\  N_{4}=-19267584 \left(812 \tilde{p}_1+38205\right),\  N_{5}=1849688064 \left(2743 \tilde{p}_1+172557\right),\  \ldots\\
\hline
 &&&  j^2+N_1j+N_2  \\
 & \multirow{-2}{*}{$\mathcal E_1[A_{1,1}^{~\otimes 48}]$}&& N_{1}=16 \tilde{p}_1,\quad N_{2}=256 \tilde{p}_2\\
\cline{2-2}\cline{4-4}
    &&& j^{14}+N_{1} j^{13}+N_{2} j^{12}+N_{3} j^{11}+N_{4} j^{10}+N_{5} j^9+N_{6} j^8+N_{7} j^7+N_{8} j^6+N_{9} j^5\\&&&+N_{10} j^4+N_{11} j^3+N_{12} j^2+N_{13} j+N_{14}\\
  \multirow{-4}{*}{$6.$} & \multirow{-2}{*}{$\mathcal E_1[E_{7,1}^{~\otimes 48}]$}&\multirow{-4}{10em}{\centering ${\cal \tilde{Z}}_0 = P_1^6,$\\ ${\cal \tilde{Z}}_1 =  P_1^3 P_2^4,$\\
   ${\cal \tilde{Z}}_2 = P_2^8$}& N_{1}=-4032,\  N_{2}=5902848,\  N_{3}=200704 \left(49 \tilde{p}_1-15660\right),\\&&&\  N_{4}=-1589575680 \left(14 \tilde{p}_1+351\right),\ N_{5}=1387266048 \left(10207 \tilde{p}_1+700056\right),\ \ldots\\
\hline
 &&&  j^{\frac{1}{3}}(j^2+N_1j+N_2)  \\
& \multirow{-2}{*}{$\mathcal E_1[A_{1,1}^{~\otimes 56}]$}&&N_{1}=16 \tilde{p}_1,\quad N_{2}=256 \tilde{p}_2\\
\cline{2-2}\cline{4-4}
  & &&  j^{\frac{1}{3}} (j^{16}+N_{1} j^{15}+N_{2} j^{14}+N_{3} j^{13}+N_{4} j^{12}+N_{5} j^{11}+N_{6} j^{10}+N_{7} j^9+N_{8} j^8+N_{9} j^7+N_{10} j^6+N_{11} j^5\\&&&+N_{12} j^4+N_{13} j^3+N_{14} j^2+N_{15} j+N_{16})\\
\multirow{-4}{*}{$7.$} & \multirow{-2}{*}{$\mathcal E_1[E_{7,1}^{~\otimes 56}]$}&\multirow{-4}{10em}{\centering ${\cal \tilde{Z}}_0 = P_1^7,$\\ ${\cal \tilde{Z}}_1 =  P_1^4 P_2^4,$\\ ${\cal \tilde{Z}}_2 =  P_1 P_2^8$} & N_{1}=-4704,\  N_{2}=8467200,\  N_{3}=1404928 \left(7 \tilde{p}_1-4644\right),\\&&&\  N_{4}=-67436544 \left(428 \tilde{p}_1-10341\right),\  N_{5}=2774532096 \left(9979 \tilde{p}_1+649593\right),\ \ldots\\
\hline
& &&   j^{\frac{2}{3}}(j^2+N_1j+N_2) \\
& \multirow{-2}{*}{$\mathcal E_1[A_{1,1}^{~\otimes 64}]$}&&N_{1}=16 \tilde{p}_1,\quad N_{2}=256 \tilde{p}_2\\
\cline{2-2}\cline{4-4}
   & & &  j^{\frac{2}{3}} (j^{18}+N_{1} j^{17}+N_{2} j^{16}+N_{3} j^{15}+N_{4} j^{14}+N_{5} j^{13}+N_{6} j^{12}+N_{7} j^{11}+N_{8} j^{10}+N_{9} j^9+N_{10} j^8+N_{11} j^7\\&&&+N_{12} j^6+N_{13} j^5+N_{14} j^4+N_{15} j^3+N_{16} j^2+N_{17} j+N_{18}) \\
\multirow{-4}{*}{$8.$} & \multirow{-2}{*}{$\mathcal E_1[E_{7,1}^{~\otimes 64}]$}&\multirow{-4}{10em}{\centering ${\cal \tilde{Z}}_0 = P_1^8,$\\ ${\cal \tilde{Z}}_1 =  P_1^5 P_2^4,$\\ ${\cal \tilde{Z}}_2 =  P_1^2 P_2^8$} & N_{1}=-5376,\  N_{2}=11483136,\  N_{3}=1404928 \left(7 \tilde{p}_1-8208\right),\\&&&\  N_{4}=-134873088 \left(263 \tilde{p}_1-28566\right),\  N_{5}=462422016 \left(98731 \tilde{p}_1+4932144\right),\  \ldots\\
\hline
& &&  j^3+N_1j^2+N_2j+N_3  \\
& \multirow{-2}{*}{$\mathcal E_1[A_{1,1}^{~\otimes 72}]$}&&N_{1}=16 \tilde{p}_1,\quad N_{2}=256 \tilde{p}_2,\quad N_{3}=4096 \tilde{p}_3 \\
\cline{2-2}\cline{4-4}
   &&& j^{21}+N_{1} j^{20}+N_{2} j^{19}+N_{3} j^{18}+N_{4} j^{17}+N_{5} j^{16}+N_{6} j^{15}+N_{7} j^{14}+N_{8} j^{13}+N_{9}j^{12} \\
&&& +N_{10} j^{11}+N_{11} j^{10}+N_{12} j^9+N_{13} j^8+N_{14} j^7+N_{15} j^6+N_{16} j^5+N_{17} j^4+N_{18} j^3+N_{19} j^2+N_{20} j+N_{21} \\
\multirow{-4}{*}{$9.$} & \multirow{-2}{*}{$\mathcal E_1[E_{7,1}^{~\otimes 72}]$}&\multirow{-4}{10em}{\centering ${\cal \tilde{Z}}_0 = P_1^9,$\\ ${\cal \tilde{Z}}_1 =  P_1^6 P_2^4,$\\ ${\cal \tilde{Z}}_2 =  P_1^3 P_2^8,$\\ ${\cal \tilde{Z}}_3 =  P_2^{12}$} & N_{1}=-6048,\  N_{2}=14950656,\  N_{3}=200704 \left(49 \tilde{p}_1-92016\right),\\&&&\  N_{4}=-115605504 \left(364 \tilde{p}_1-85941\right),\  N_{5}=11098128384 \left(6133 \tilde{p}_1+123039\right),\  \ldots\\
\end{array}$
}
\end{threeparttable}
\end{center}
\end{table}

\begin{table}[H]
\begin{center}
\begin{threeparttable}
\resizebox{\textwidth}{!}{
\renewcommand{\arraystretch}{1.3}
$\begin{array}{|c|c|c|c|}
& &&  j^{\frac{1}{3}}(j^3+N_1j^2+N_2j+N_3)  \\
& \multirow{-2}{*}{$\mathcal E_1[A_{1,1}^{~\otimes 80}]$}&& N_{1}=16 \tilde{p}_1,\quad N_{2}=256 \tilde{p}_2,\quad N_{3}=4096 \tilde{p}_3\\
\cline{2-2}\cline{4-4}
   &&& j^{\frac{1}{3}} (j^{23}+N_{1} j^{22}+N_{2} j^{21}+N_{3} j^{20}+N_{4} j^{19}+N_{5} j^{18}+N_{6} j^{17}+N_{7} j^{16}+N_{8} j^{15}+N_{9} j^{14}+ N_{10}j^{13}+ N_{11}j^{12}\\
&&& +N_{12} j^{11}+N_{13} j^{10}+N_{14} j^9+N_{15} j^8+N_{16} j^7+N_{17} j^6+N_{18} j^5+N_{19} j^4+N_{20} j^3+N_{21} j^2+N_{22} j+N_{23})\\
\multirow{-4}{*}{$10.$} & \multirow{-2}{*}{$\mathcal E_1[E_{7,1}^{~\otimes 80}]$}&\multirow{-4}{10em}{\centering ${\cal \tilde{Z}}_0 = P_1^{10},$\\ ${\cal \tilde{Z}}_1 =  P_1^7 P_2^4,$\\ ${\cal \tilde{Z}}_2 = P_1^4 P_2^8,$\\ ${\cal \tilde{Z}}_3 =  P_1 P_2^{12}$} & N_{1}=-6720,\  N_{2}=18869760,\  N_{3}=200704 \left(49 \tilde{p}_1-137700\right),\\&&&\  N_{4}=-9633792 \left(5054 \tilde{p}_1-2094255\right),\  N_{5}=26358054912 \left(3601 \tilde{p}_1-99792\right),\ \ldots\\ 
\hline
&&&  j^{\frac{2}{3}}(j^3+N_1j^2+N_2j+N_3) \\
& \multirow{-2}{*}{$\mathcal E_1[A_{1,1}^{~\otimes 88}]$}&& N_{1}=16 \tilde{p}_1,\quad N_{2}=256 \tilde{p}_2,\quad N_{3}=4096 \tilde{p}_3\\
\cline{2-2}\cline{4-4}
   &&& j^{2/3} (j^{25}+N_{1} j^{24}+N_{2} j^{23}+N_{3} j^{22}+N_{4} j^{21}+N_{5} j^{20}+N_{6} j^{19}+N_{7} j^{18}+N_{8} j^{17}+N_{9} j^{16}+N_{10} j^{15}+N_{11} j^{14}+ N_{12}j^{13}\\&&&+ N_{13}j^{12}
+N_{14} j^{11}+N_{15} j^{10}+N_{16} j^9+N_{17} j^8+N_{18} j^7+N_{19} j^6+N_{20} j^5+N_{21} j^4+N_{22} j^3+N_{23} j^2+N_{24} j+N_{25})\\
\multirow{-4}{*}{$11.$} & \multirow{-2}{*}{$\mathcal E_1[E_{7,1}^{~\otimes 88}]$}&\multirow{-4}{10em}{\centering ${\cal \tilde{Z}}_0 =  P_1^{11},$\\ ${\cal \tilde{Z}}_1 = P_1^8 P_2^4,$\\ ${\cal \tilde{Z}}_2 =  P_1^5 P_2^8,$\\ ${\cal \tilde{Z}}_3 =   P_1^2 P_2^{12}$} & N_{1}=-7392,\  N_{2}=23240448,\  N_{3}=200704 \left(49 \tilde{p}_1-196020\right),\\&&&\  N_{4}=-48168960 \left(1148 \tilde{p}_1-747549\right),\  N_{5}=924844032 \left(136463 \tilde{p}_1-13166307\right),\ \ldots\\
\hline
&&&  j^4+N_1j^3+N_2j^2+N_3j+N_4  \\
& \multirow{-2}{*}{$\mathcal E_1[A_{1,1}^{~\otimes 96}]$}&&N_{1}=16 \tilde{p}_1,\quad N_{2}=256 \tilde{p}_2,\quad N_{3}=4096 \tilde{p}_3,\quad N_{4}=65536 \tilde{p}_4\\
\cline{2-2}\cline{4-4}
   &&& j^{28}+N_{1} j^{27}+N_{2} j^{26}+N_{3} j^{25}+N_{4} j^{24}+N_{5} j^{23}+N_{6} j^{22}+N_{7} j^{21}+N_{8} j^{20}+N_{9} j^{19}+N_{10} j^{18}+N_{11} j^{17}\\&&&+N_{12} j^{16}+ j^{15}N_{13}+ N_{14}j^{14} +N_{15} j^{13}+N_{16} j^{12}+N_{17} j^{11}+N_{18} j^{10}+N_{19} j^9+N_{20} j^8+N_{21} j^7+N_{22} j^6+N_{23} j^5\\&&&+N_{24} j^4+N_{25} j^3+N_{26} j^2+N_{27} j+N_{28}\\
\multirow{-5}{*}{$12.$} & \multirow{-3}{*}{$\mathcal E_1[E_{7,1}^{~\otimes 96}]$}&\multirow{-5}{10em}{\centering ${\cal \tilde{Z}}_0 =  P_1^{12},$\\ ${\cal \tilde{Z}}_1 = P_1^9 P_2^4,$\\ ${\cal \tilde{Z}}_2 = P_1^6 P_2^8,$\\ ${\cal \tilde{Z}}_3 =   P_1^3 P_2^{12},$\\ ${\cal \tilde{Z}}_4 =   P_2^{16}$} & N_{1}=-8064,\  N_{2}=28062720,\  N_{3}=200704 \left(49 \tilde{p}_1-268488\right),\\&&&\  N_{4}=-520224768 \left(119 \tilde{p}_1-113553\right),\  N_{5}=4161798144 \left(38911 \tilde{p}_1-7367976\right),\ \ldots\\
\hline
&&&  j^{\frac{1}{3}}(j^4+N_1j^3+N_2j^2+N_3j+N_4)  \\
& \multirow{-2}{*}{$\mathcal E_1[A_{1,1}^{~\otimes 104}]$}&& N_{1}=16 \tilde{p}_1,\quad N_{2}=256 \tilde{p}_2,\quad N_{3}=4096 \tilde{p}_3,\quad N_{4}=65536 \tilde{p}_4\\
\cline{2-2}\cline{4-4}
   &&&j^{\frac{1}{3}} (j^{30}+N_{1} j^{29}+N_{2} j^{28}+N_{3} j^{27}+N_{4} j^{26}+N_{5} j^{25}+N_{6} j^{24}+N_{7} j^{23}+N_{8} j^{22}+N_{9} j^{21}+N_{10} j^{20}+N_{11} j^{19}\\&&&+N_{12} j^{18}+N_{13} j^{17}+ N_{14}j^{16}+ N_{15}j^{15}+N_{16} j^{14}+N_{17} j^{13}+N_{18} j^{12}+N_{19} j^{11}+N_{20} j^{10}+N_{21} j^9+N_{22} j^8\\&&&+N_{23} j^7+N_{24} j^6+N_{25} j^5+N_{26} j^4+N_{27} j^3+N_{28} j^2+N_{29} j+N_{30})\\
\multirow{-5}{*}{$13.$} & \multirow{-3}{*}{$\mathcal E_1[E_{7,1}^{~\otimes 104}]$}&\multirow{-5}{10em}{\centering ${\cal \tilde{Z}}_0 =  P_1^{13},$\\ ${\cal \tilde{Z}}_1 = P_1^{10} P_2^4,$\\ ${\cal \tilde{Z}}_2 = P_1^7 P_2^8,$\\ ${\cal \tilde{Z}}_3 =   P_1^4 P_2^{12},$\\ ${\cal \tilde{Z}}_4 =  P_1 P_2^{16}$} & N_{1}=-8736,\  N_{2}=33336576,\  N_{3}=200704 \left(49 \tilde{p}_1-356616\right),\\&&&\  N_{4}=-19267584 \left(3556 \tilde{p}_1-4733937\right),\  N_{5}=149824733184 \left(1349 \tilde{p}_1-417417\right),\ \ldots\\ 
\hline
&&&  j^{\frac{2}{3}}(j^4+N_1j^3+N_2j^2+N_3j+N_4)  \\
& \multirow{-2}{*}{$\mathcal E_1[A_{1,1}^{~\otimes 112}]$}& & N_{1}=16 \tilde{p}_1,\quad N_{2}=256 \tilde{p}_2,\quad N_{3}=4096 \tilde{p}_3,\quad N_{4}=65536 \tilde{p}_4\\
\cline{2-2}\cline{4-4}
    &&& j^{\frac{2}{3}} (j^{32}+N_{1} j^{31}+N_{2} j^{30}+N_{3} j^{29}+N_{4} j^{28}+N_{5} j^{27}+N_{6} j^{26}+N_{7} j^{25}+N_{8} j^{24}+N_{9} j^{23}+N_{10} j^{22}+N_{11} j^{21}\\&&&+N_{12} j^{20}+N_{13} j^{19}+ N_{14}j^{18}+ N_{15}j^{17}+N_{16} j^{16}+N_{17} j^{15}+N_{18} j^{14}+N_{19} j^{13}+N_{20} j^{12}+N_{21} j^{11}+N_{22} j^{10}\\&&&+N_{23} j^9+N_{24} j^8+N_{25} j^7+N_{26} j^6+N_{27} j^5+N_{28} j^4+N_{29} j^3+N_{30} j^2+N_{31} j+N_{32})\\
\multirow{-5}{*}{$14.$} & \multirow{-3}{*}{$\mathcal E_1[E_{7,1}^{~\otimes 112}]$}&\multirow{-5}{10em}{\centering ${\cal \tilde{Z}}_0 =  P_1^{14},$\\ ${\cal \tilde{Z}}_1 = P_1^{11} P_2^4,$\\ ${\cal \tilde{Z}}_2 = P_1^8 P_2^8,$\\ ${\cal \tilde{Z}}_3 =   P_1^5 P_2^{12},$\\ ${\cal \tilde{Z}}_4 =  P_1^2 P_2^{16}$} & N_{1}=-9408,\  N_{2}=39062016,\  N_{3}=1404928 \left(7 \tilde{p}_1-65988\right),\\&&&\  N_{4}=-67436544 \left(1114 \tilde{p}_1-1994031\right),\  N_{5}=462422016 \left(533557 \tilde{p}_1-245326536\right),\ \ldots\\
\hline
&& & j^5+N_1j^4+N_2j^3+N_3j^2+N_4j+N_5  \\
& \multirow{-2}{*}{$\mathcal E_1[A_{1,1}^{~\otimes 120}]$}& &  N_{1}=16 \tilde{p}_1,\quad N_{2}=256 \tilde{p}_2,\quad N_{3}=4096 \tilde{p}_3,\quad N_{4}=65536 \tilde{p}_4,\quad N_{5}=1048576 \tilde{p}_5\\
\cline{2-2}\cline{4-4}
 &&&j^{35}+N_{1} j^{34}+N_{2} j^{33}+N_{3} j^{32}+N_{4} j^{31}+N_{5} j^{30}+N_{6} j^{29}+N_{7} j^{28}+N_{8} j^{27}+N_{9} j^{26}+N_{10} j^{25}+N_{11} j^{24}+N_{12} j^{23}\\&&&+N_{13} j^{22}+N_{14} j^{21}+N_{15} j^{20}+ N_{16}j^{19}+ N_{17}j^{18}  +N_{18} j^{17}+N_{19} j^{16}+N_{20} j^{15}+N_{21} j^{14}+N_{22} j^{13}+N_{23} j^{12}+N_{24} j^{11}\\&&&+N_{25} j^{10}+N_{26} j^9+N_{27} j^8+N_{28} j^7+N_{29} j^6+N_{30} j^5+N_{31} j^4+N_{32} j^3+N_{33} j^2+N_{34} j+N_{35}\\
\multirow{-5}{*}{$15.$} & \multirow{-3}{*}{$\mathcal E_1[E_{7,1}^{~\otimes 120}]$}&\multirow{-5}{10em}{\centering ${\cal \tilde{Z}}_0 =  P_1^{15},$\\ ${\cal \tilde{Z}}_1 = P_1^{12} P_2^4,$\\ ${\cal \tilde{Z}}_2 = P_1^9 P_2^8,$\\ ${\cal \tilde{Z}}_3 =   P_1^6 P_2^{12},$\\ ${\cal \tilde{Z}}_4 =  P_1^3 P_2^{16},$\\ ${\cal \tilde{Z}}_5 =  P_2^{20}$} & N_{1}=-10080,\  N_{2}=45239040,\  N_{3}=1404928 \left(7 \tilde{p}_1-83700\right),\\&&&\  N_{4}=-202309632 \left(404 \tilde{p}_1-944595\right),\  N_{5}=2774532096 \left(106607 \tilde{p}_1-68606811\right),\ \ldots\\
\hline
&& &  j^{\frac{1}{3}}(j^5+N_1j^4+N_2j^3+N_3j^2+N_4j+N_5) \\
& \multirow{-2}{*}{$\mathcal E_1[A_{1,1}^{~\otimes 128}]$}& & N_{1}=16 \tilde{p}_1,\quad N_{2}=256 \tilde{p}_2,\quad N_{3}=4096 \tilde{p}_3,\quad N_{4}=65536 \tilde{p}_4,\quad N_{5}=1048576 \tilde{p}_5\\
\cline{2-2}\cline{4-4}
 &&& j^{\frac{1}{3}} (j^{37}+N_{1} j^{36}+N_{2} j^{35}+N_{3} j^{34}+N_{4} j^{33}+N_{5} j^{32}+N_{6} j^{31}+N_{7} j^{30}+N_{8} j^{29}+N_{9} j^{28}+N_{10} j^{27}+N_{11} j^{26}\\&&&+N_{12} j^{25}+N_{13} j^{24}+N_{14} j^{23}+N_{15} j^{22}+N_{16} j^{21}+N_{17} j^{20}+ N_{18}j^{19}+N_{19} j^{18}+N_{20} j^{17}+N_{21} j^{16}+N_{22} j^{15}+N_{23} j^{14}\\&&&+N_{24} j^{13}+N_{25} j^{12}+N_{26} j^{11}+N_{27} j^{10}+N_{28} j^9+N_{29} j^8+N_{30} j^7+N_{31} j^6+N_{32} j^5+N_{33} j^4+N_{34} j^3+N_{35} j^2+N_{36} j+N_{37})\\
\multirow{-5}{*}{$16.$} & \multirow{-3}{*}{$\mathcal E_1[E_{7,1}^{~\otimes 128}]$} & \multirow{-5}{10em}{\centering ${\cal \tilde{Z}}_0 =  P_1^{16},$\\ ${\cal \tilde{Z}}_1 = P_1^{13} P_2^4,$\\ ${\cal \tilde{Z}}_2 = P_1^{10} P_2^8,$\\ ${\cal \tilde{Z}}_3 =   P_1^7 P_2^{12},$\\ ${\cal \tilde{Z}}_4 =  P_1^4 P_2^{16},$\\ ${\cal \tilde{Z}}_5 =  P_1 P_2^{20}$} & N_{1}=-10752,\  N_{2}=51867648,\  N_{3}=200704 \left(49 \tilde{p}_1-730080\right),\\&&&\  N_{4}=-96337920 \left(917 \tilde{p}_1-2735748\right),\  N_{5}=1387266048 \left(251777 \tilde{p}_1-217480032\right),\ \ldots\\
 \hline
\end{array}$
}
\end{threeparttable}
\end{center}
\caption{ Characters of one-character extensions of tensor products of $A_{1,1}$ (and $E_{7,1}$) written in a basis of $S$-invariant polynomials}
\label{t4}
\end{table}

This subsection also contains the study of the $E_{7,1}$ CFT.  The reasons to include this here will become clear. First let us note some details of the $E_{7,1}$ CFT. 

$E_{7,1}$ is a $\mathbf{[2,0]}$ CFT with central charge $7$. Let us denote it's two characters by $\chi_0, \chi_1$\footnote{We are deliberately using the symbols $\chi_0, \chi_1$ to denote  the characters of the $A_{1,1}$ CFT as well as the $E_{7,1}$ CFT.}, the latter is the non-identity character with conformal dimension $h_1 = \frac34$. Let us note the $q$-series for the characters :
\begin{small}
\begin{align}
     \chi_0& =q^{-\frac{7}{24}}( 1+133 q+1673 q^2+11914 q^3+63252 q^4+278313 q^5+1070006 q^6+3711557 q^7+ \ldots)  \nonumber \\
     \chi_1& =q^{\frac{11}{24}}(56+968 q+7504 q^2+42616 q^3+194768 q^4+772576 q^5+2742040 q^6+8933456 q^7 + \ldots) \label{51b}
\end{align}
\end{small}
and the modular $S$-transformations for the characters :
\be \label{52b}
S(\chi_0) = \frac{\chi_0 + \chi_1}{\sqrt{2}}, \qquad \qquad S(\chi_1) = \frac{\chi_0 - \chi_1}{\sqrt{2}}.
\ee

The allowed values of $k$ and $N$ can be read from entry 7 of table \ref{t2}. Here, only $c_{(1)} = 56k$ is possible and the corresponding number of tensor product factors is  $N = 8k$. 

Before we try to solve for the characters of one-character extensions of the tensor products of the $E_{7,1}$ CFT, we note from table \ref{t2}, from entries 1 and 7,  that the $N$ values for $E_{7,1}$ is identical to those of $A_{1,1}$. This is because they have the same value for $\frac{8 n}{\gcd(m,8)}$. Hence we are looking at the monomials of the same degree. Then we note that the $h_1$ for both these CFTs have $4$ in the denominator. What this means is that a monomial which solves the equations \eqref{7b} and \eqref{8b} for $A_{1,1}$ also solves the equations for $E_{7,1}$ (but maybe for different values of $M$). We thus conclude that the ansatz character \eqref{10b} for $E_{7,1}$ for a given $N$ is exactly the same as the ansatz character for $A_{1,1}$. As an illustration, consider $N = 16$. The ansatz character for $A_{1,1}$ has been worked out and given in \eqref{24b}. The only solutions to \eqref{7b} (for $A_{1,1}$)  are the five terms in \eqref{24b}, which  are solutions to \eqref{7b} for $M = 0, 1, 2, 3, 4$. When one studies the solutions to \eqref{7b} (for $E_{7,1}$), one finds that the only solutions are the five terms in \eqref{24b} (but this time, the $\chi_0, \chi_1$ stand for characters of $E_{7,1}$) for $M = 0, 3, 6, 9, 12$. Hence upto the point of computing the ansatz character, we only need to copy the results of $A_{1,1}$ with the understanding that 
the $\chi_0, \chi_1$ now stand for characters of $E_{7,1}$.

Now consider the $S$-invariance equations \eqref{18b}. Here, we meet the second important coincidence concerning the $A_{1,1}$ and $E_{7,1}$ CFTs; the first coincidence being that the $N$ values for both are identical. The $S$-matrix for both the $A_{1,1}$ and $E_{7,1}$ CFTs are identical \eqref{20b}, \eqref{52b}. What this means is that the system of linear equations we need to solve is exactly the same as that of $A_{1,1}$. As well as the solutions. We only need to remember that the $\chi_0, \chi_1$ now are characters of $E_{7,1}$.

The two $S$-invariant polynomials of $A_{1,1}$ are also $S$-invariant polynomials of $E_{7,1}$. But, they have  different $q$-series, which we compute  (using \eqref{51b}) :
\begin{equation}\label{53b}
\begin{array}{llr}
   P_1(\chi_0, \chi_1) &= \chi_0^8 + 14 \, \chi_0^4\,\chi_1^4 +  \chi_1^8 = j^{\frac{1}{3}}(j^2-672j-145152)\\\
   P_2(\chi_0, \chi_1) &=\chi_0^5\,\chi_1 -  \chi_0\,\chi_1^5 = 56j-3456
\end{array}
\end{equation}
We can see that both the $S$-invariant polynomials are non-constant $S$-invariant polynomials. In going from $A_{1,1}$ to $E_{7,1}$, $P_2$ even though having an identical expression in terms of characters,  has changed from being a constant $S$-invariant polynomial to a non-constant $S$-invariant polynomial. This is going to happen throughout. The answers in terms of characters are identical for $A_{1,1}$ and $E_{7,1}$ but in terms of $q$-series are very different. 

Now, we can report the results for $E_{7,1}$. Since the $A_{1,1}$ and $E_{7,1}$ solutions  have so much common information, we put everything in one table viz. table \ref{t4}. The way to read the  $E_{7,1}$ results from that table is as follows. In the first column, we have the serial number. $7$ times the serial number is the value of $k$ and the central charge is then $c_{(1)} = 8k$. In other words, $56$ times the value of the serial number is the central charge $c_{(1)}$. In the second column, the relevant part for $E_{7,1}$ is the bottom half. The third column does not have two halves. This is the common solution for both $A_{1,1}$ and $E_{7,1}$.  But we need to remember that the $\chi_0, \chi_1$ stand for characters of $E_{7,1}$ \eqref{51b} and thus what we have in third column is $P^{E_{7,1}}_{c_{(1)} = 8k}$. The bottom half of the fourth column contains the $q$-series for $P^{E_{7,1}}_{c_{(1)} = 8k}$, which is of course very different from the top half of the fourth column which contains the $q$-series for $P^{A_{1,1}}_{c_{(1)} = 8k}$.

Then we need to impose admissibility. We compare with the general admissible characters given in table \ref{tapp} of appendix \ref{app1}. We give the relation between the $N$’s and the $p$’s. The nested inequalities that the $N$’s satisfy (for admissibility) translates to nested inequalities for the $p$’s. For rows $3$ and below, we always find an infinite number of admissible character solutions.  In row $1$ and row $2$, we obtain unique admissible characters.

In row $1$, we have a unique solution, at central charge $c_{(1)} = 56$. The underlying reason we have a unique solution is that this is derived from the $c_{(1)} = 8$ solution for $A_{1,1}$.  In row $2$ also, we have a unique solution at $c_{(1)} = 112$. The underlying reason for this unique solution is that this is derived from the $c_{(1)} = 16$ solution for $A_{1,1}$.  After this, from row 3 onwards, we have infinite numbers of admissible characters and the methods of this paper cannot decide which of them correspond to genuine CFTs. We can make some guesses about some of them. 

Further considerations (beyond those in this paper) establish, in the case of $A_{1,1}$,  that (i) the single character obtained for $N = 8$ viz. $P_1$, (ii) the single  character obtained for $N = 16$ viz. $P_1^2$, (iii) the character obtained for $N =  24$  (\eqref{33b}) viz. $P_1^3 - 42 P_2^4$, all correspond to genuine CFT(s). We suspect that these further considerations would  establish that the same characters for $E_{7,1}$ also all correspond to genuine CFT(s). We thus are led to the following surmise/conjecture : \emph{For $E_{7,1}$, (i) there exist genuine one-character CFT(s) with central charge $56$ with character $P_1$, (ii) there exist genuine one-character CFT(s) with central charge $112$ with character $P_1^2$, (iii) there exist genuine one-character CFT(s) with central charge $168$ with character $P_1^3 - 42 P_2^4$. }

\subsection{$\mathcal C = A_{2,1}$ and $E_{6,1}$ \label{32s}}

$A_{2,1}$ is a $\mathbf{[2,0]}$ CFT with central charge $2$. Let us denote it's two characters by $\chi_0, \chi_1$, the latter is the non-identity character with conformal dimension $h_1 = \frac13$. Let us note the $q$-series for the characters :
\begin{small}
\begin{align}
     \chi_0& =q^{-\frac{1}{12}} \left(1+8 q+17 q^2+46 q^3+98 q^4+198 q^5+371 q^6+692 q^7+1205 q^8+2082 q^9+3463 q^{10}  + \ldots \right) \nonumber \\
     \chi_1& =q^{\frac{1}{4}}\left(3+9 q+27 q^2+57 q^3+126 q^4+243 q^5+465 q^6+828 q^7+1458 q^8+2463 q^9+4104 q^{10} + \ldots \right) \label{36b}
\end{align}
\end{small}
and the modular $S$-transformations for the characters :
\be \label{37b}
S(\chi_0) = \frac{\chi_0 + 2 \chi_1}{\sqrt{3}}, \qquad \qquad S(\chi_1) = \frac{\chi_0 - \chi_1}{\sqrt{3}}.\ee

The allowed values of $k$ and $N$ can be read from entry 2. of table \ref{t2}. Here, every possible $c_{(1)} = 8k$ is possible and the corresponding number of tensor product factors is  $N = 4k$. 

We will be brief in our description of the results here. The analysis proceeds very similar to $A_{1,1}$. We discover the first $S$-invariant polynomial when we are working with $N = 4, c_{(1)} = 8$. This is a degree-$4$ non-constant $S$-invariant polynomial. Just as for the $A_{1,1}$ case in section \ref{31s}, we discover the second $S$-invariant polynomial when working $ c_{(1)} = 24$, which happens when $N = 12$. This is a constant $S$-invariant polynomial of degree-$4$. The two  $S$-invariant polynomials for $A_{2,1}$ are :
\begin{equation}\label{38b}
   \begin{array}{llr}
   P_1(\chi_0, \chi_1) &=\chi_0^4+8 \chi_0 \chi_1^3 = j^{\frac{1}{3}}\\
   P_2(\chi_0, \chi_1) &=\chi_0^3 \chi_1-\chi_1^4 = 3
   \end{array}
\end{equation}
It is a peculiarity of $A_{2,1}$ that both the $S$-invariant polynomials are of the same degree. This then results in the fact that the characters of ${\cal E}_1[A_{2,1}^{\otimes N}]$ are homogenous polynomials of the $S$-invariant polynomials also. 

We have implemented the method outlined in section \ref{2s}  for $k = 1, 2, \ldots 16$.  For $c_{(1)} = 8k$ the character has $[\frac{k}{3}]$ arbitrary constants (the $p$’s) and there are $1 + [\frac{k}{3}]$ $S$-invariant summands the ${\cal Z}$'s. Each summand, when expressed in terms of the characters $\chi_0, \chi_1$ are polynomials of degree $4k$. But when expressed in terms of the $S$-invariant polynomials \eqref{38b}, they are polynomials of only degree $k$.  Because of this, we can handle in a compact manner,  computations upto $c_{(1)} = 128$ for which we have  polynomials of degree $16$ only. The result for $N = 64, c_{(1)} = 128, k = 16$ is :
\be \label{47c}
P^{A_{2,1}}_{c_{(1)} = 128}(\chi_0, \chi_1)  = {\cal Z}_0  +  p_1 \,{\cal Z}_1 +  p_2 \,{\cal Z}_2  +  p_3 \,{\cal Z}_3 +  p_4 \,{\cal Z}_4  +  p_5 \,{\cal Z}_5
\ee
with 
\begin{small}
\begin{align}
     {\cal Z}_0& = P_1^{16}-128 P_1^{13}P_2^3 +5248  P_1^{10}P_2^6-75648 P_1^7 P_2^9+292992 P_1^4 P_2^{12}-127488 P_1 P_2^{15} \nonumber \\
     {\cal Z}_1& = P_1^{13} P_2^3-101 P_1^{10} P_2^6+2791 P_1^7 P_2^9-19029 P_1^4 P_2^{12}+10788 P_1 P_2^{15}\nonumber \\
     {\cal Z}_2& =  P_1^{10} P_2^6-74 P_1^7 P_2^9+1063 P_1^4 P_2^{12}-1776 P_1 P_2^{15}\nonumber \\
     {\cal Z}_3& =  P_1^7 P_2^9-47 P_1^4 P_2^{12}+64 P_1 P_2^{15}  \nonumber \\
     {\cal Z}_4& =  P_1^4 P_2^{12}-20 P_1 P_2^{15} \nonumber \\
     {\cal Z}_5& = P_1 P_2^{15} .
     \end{align}
\end{small}
In the following, we are able to rewrite the same in an even more compact manner :
\be \label{49c}
P^{A_{2,1}}_{c_{(1)} = 128}(\chi_0, \chi_1)  = \tilde{{\cal Z}}_0  +  \tilde{p}_1 \,\tilde{{\cal Z}}_1 +  \tilde{p}_2 \,\tilde{{\cal Z}}_2  +  \tilde{p}_3 \,\tilde{{\cal Z}}_3 +  \tilde{p}_4 \,\tilde{{\cal Z}}_4  +  \tilde{p}_5 \,\tilde{{\cal Z}}_5
\ee
with 
\begin{small}
\begin{align}
     \tilde{{\cal Z}}_0& = P_1^{16}, \quad \tilde{{\cal Z}}_1 = P_1^{13}P_2^3, \quad  \tilde{{\cal Z}}_2 =  P_1^{10}P_2^6, \quad  \tilde{{\cal Z}}_3 =  P_1^7 P_2^9, \quad  \tilde{{\cal Z}}_4 =  P_1^4 P_2^{12}, \quad   \tilde{{\cal Z}}_5 =  P_1 P_2^{15}  
     \end{align}
\end{small}
The relation between the constant coefficients of \eqref{47c} and \eqref{49c} is:
\begin{small}
\begin{align}
         \tilde{p}_1& = p_1 - 128, \quad  \tilde{p}_2 = p_2 - 101 p_1 + 5248, \quad \tilde{p}_3 = p_3 -74 p_2 +2791 p_1 -75648, \nonumber \\
     \tilde{p}_4& =  p_4 -47 p_3  +1063 p_2 -19029 p_1 +292992, \quad  \tilde{p}_5 =  p_5 -20 p_4  + 64 p_3 -1776 p_2 +10788 p_1 -127488 \nonumber 
     \end{align}
\end{small}
Since, the most compact form is that of \eqref{49c}, we will report this form of the results in table \ref{t5}. This table also contains results for $E_{6,1}$ for reasons that will be explained below. The way to read the results for $A_{2,1}$ from table \ref{t5} is as follows.  The first column is the serial no. which is also the value of $k$, from which one can infer the value of $c_{(1)} = 8k$.  The (top half of the) second column contains the number of tensor product factors $N$ (which equals $4k$).  The third column contains the character $P^{A_{2,1}}_{c_{(1)} = 8k}(\chi_0, \chi_1)$.  The number of summands ranges from $1$, for $k = 1$, to $6$ for $k = 16$. For $k = 16$, the last entry of the table \ref{t5} we have six $S$-invariant summands $\tilde{{\cal Z}}_0, \tilde{{\cal Z}}_1, \ldots \tilde{{\cal Z}}_5$ (see \eqref{49c}).   The expressions for the $\tilde{{\cal Z}}$'s when written in terms of the characters  $\chi_0, \chi_1$ are very voluminous. Each of them is a  degree $N$ polynomial.  For example for $N= 64$, the last entry of table \ref{t5}, we have  six homogenous degree $64$ polynomials. But, when written in terms of the $S$-invariant polynomials \eqref{38b}, they are in fact polynomials of much smaller degree. For $N = 64$, the highest degree term that occurs is $16$.  This is one of the results to report in this paper. Solving the huge linear system of equations that follow from  $S$-invariance,  organising in terms of some number of $S$-invariant summands and writing each of  the summands in terms of $S$-invariant polynomials and thus achieving compact expressions.

Now we need to impose admissibility. For that, we first need to compute the $q$-series of $P^{A_{2,1}}_{c_{(1)} = 8k}$. We employ the explicit $q$-series of the characters available in \eqref{36b} and give the  $q$-series of $P^{A_{2,1}}_{c_{(1)} = 8k}$ in the top half of the fourth column. Then we need to compare this with the generic admissible character for that particular central charge, given in table \ref{tapp} of the appendix \ref{app1}. The generic admissible character for central charge $8k$ has $[\frac{k}{3}]$ parameters which are denoted by $N_1, N_2, \ldots$. And these $N$’s need to satisfy a set of nested inequalities for admissibility, which are available from table \ref{tapp} of the appendix \ref{app1}.  In the top half of the fourth column of table \ref{t5}, we give the relation between the $\tilde{p}$’s and the $N$’s. Then, the nested inequalities for the $N$’s translate to inequalities for the $\tilde{p}$’s. We always find an infinite number of admissible character solutions, except of course for $N = 4, c_{(1)} = 8$ and $N = 8, c_{(1)} = 16$.

Amongst all the admissible characters we have computed in table \ref{t5}, there are only three of them which are known to correspond to actual CFTs. The first one is in the first row, $N = 4, c_{(1)} = 8$, and this corresponds to the $E_{8,1}$ CFT.  The second one is the second row, $N = 8, c_{(1)} = 16$, and this corresponds to two CFTs viz. $E_{8,1}^{\otimes 2}$ and ${\cal E}_1[D_{16,1}]$. The third one is one of the infinite characters of the third row, $N = 12, c_{(1)} = 24$. Reading off from the top half of the fourth column of the third row, we have the character $j + 27 \tilde{p}_1$. Among these infinite characters, we can identify the one which corresponds to a CFT by invoking the classification of $c_{(1)} = 24$ CFTs of \cite{Schellekens:1992db}.  In the table at the end of that paper entry no. 24 is the only CFT which is a one-character extension of a tensor product of  copies of $A_{2,1}$\footnote{The reader can find this CFT listed as entry no. 11 of table \ref{t0}.}. This CFT has a character  $j - 648$. Thus amongst the infinite characters  $j + 27 \tilde{p}_1$ that we found, only one of them, the one with $\tilde{p}_1 = -24$ corresponds to an actual CFT. 

\underline{Lessons learnt and a new method} :  All the results that we have obtained and reported in table \ref{t5} were by running the whole procedure as laid out  in section \ref{2s} . Similar to what we did in section \ref{31s}, we have a simpler  procedure that allows us to write the final answer quickly.  We first note that $P_1$ is a homogenous polynomial of the characters of degree $4$ and that $P_2$ is a homogenous polynomial of the characters of degree $4$. A monomial of the form $P_1^a\,P_2^b$ is a homogenous polynomial of the characters of degree $4a + 4b$. And for this monomial to be present in $P^{A_{2,1}}_{c_{(1)} = 8k}$ it’s degree needs to be equal to $N$. Thus we look for  non-negative integral solutions to 
\be \label{51c}
4a + 4b = N = 4 k, 
\ee
we can obtain all the $\tilde{{\cal Z}}$'s  in $P^{A_{2,1}}_{c_{(1)} = 8k}$.  All except one of the $\tilde{{\cal Z}}$'s  viz. $\tilde{{\cal Z}}_0$ are multiplied by constants the $\tilde{p}$'s.  Which of the solutions of \eqref{51c} corresponds to $\tilde{{\cal Z}}_0$ is easily decided by seeing which of them contains the identity character $\chi_0^N$. Here it is the $a = k, b = 0$ solution. This quick computation of the solutions to \eqref{51c} gives us the same answer as the long procedure.  There is one caveat. We should only consider solutions of \eqref{51c} that give monomials of characters which are commensurate to the identity character. Consider, for example, $N = 4$, which has two solutions $a = 1, b = 0$ and $a = 0, b =1$ which correspond to $P_1$ and $P_2$ respectively. But we have to reject $P_2$ because the monomials of characters present in $P_2$ have conformal dimensions which are not integral and hence  are not commensurate with the identity character.

\begin{table}[H]
\begin{center}
\begin{threeparttable}
\resizebox{\textwidth}{!}{
\renewcommand{\arraystretch}{1.3}
   $\begin{array}{|c|c|c|C{10cm}|}
\hline
\# &    N & P_{c_{(1)} = 8k}^{A_{2,1}} (\chi_0, \chi_1) ~\text{and} ~P_{c_{(1)} = 8k}^{E_{6,1}} (\chi_0, \chi_1) &  q-\text{series}  \\
\hline
   & \mathcal E_1[A_{2,1}^{~\otimes 4}]&& j^{\frac{1}{3}}   \\
\cline{2-2}\cline{4-4}
 &&& j+N   \\
 \multirow{-3}{*}{$1.$}&\multirow{-2}{*}{$\mathcal E_1[E_{6,1}^{~\otimes 4}]$} &\multirow{-3}{10em}{\centering ${\cal \tilde{Z}}_0 = P_1 $}&N= -432\\
 \hline
& \mathcal E_1[A_{2,1}^{~\otimes 8}] & & j^{\frac{2}{3}}  \\
\cline{2-2}\cline{4-4}
&&& j^2+N_1 j+N_2  \\
 \multirow{-3}{*}{$2.$}&\multirow{-2}{*}{$\mathcal E_1[E_{6,1}^{~\otimes 8}]$} &\multirow{-3}{10em}{\centering ${\cal \tilde{Z}}_0 = P_1^2 $}&N_{1}= -864,N_{2}= 186624\\
\end{array}$
}
\end{threeparttable}
\end{center}
\end{table}

\begin{table}[H]
\begin{center}
\begin{threeparttable}
\resizebox{\textwidth}{!}{
\renewcommand{\arraystretch}{1.3}
   $\begin{array}{|c|c|c|C{15cm}|}
&&&  j+N  \\
&\multirow{-2}{*}{$  \mathcal E_1[A_{2,1}^{~\otimes 12}]$}&& N_{1}=27 \tilde{p}_1\\
\cline{2-2}\cline{4-4}
&&& j^3+N_{1} j^2+N_{2} j+N_{3}\\
 \multirow{-4}{*}{$3.$}&\multirow{-2}{*}{$\mathcal E_1[E_{6,1}^{~\otimes 12}]$} & \multirow{-4}{10em}{\centering ${\cal \tilde{Z}}_0 = P_1^3,$\\ ${\cal \tilde{Z}}_1 = P_2^3$}&N_{1}=-1296,\  N_{2}=2187 \left(9 \tilde{p}_1+256\right),\  N_{3}=-80621568\\
\hline
&&&   j^{\frac{1}{3}}(j+N) \\
&\multirow{-2}{*}{$  \mathcal E_1[A_{2,1}^{~\otimes 16}]$}&& N_{1}=27 \tilde{p}_1\\
\cline{2-2}\cline{4-4}
&&&  j^4+N_{1} j^3+N_{2} j^2+N_{3} j+N_{4} \\
 \multirow{-4}{*}{$4.$}&\multirow{-2}{*}{$\mathcal E_1[E_{6,1}^{~\otimes 16}]$} & \multirow{-4}{10em}{\centering ${\cal \tilde{Z}}_0 = P_1^4,$\\ ${\cal \tilde{Z}}_1 = P_1 P_2^3$}&N_{1}=-1728,\  N_{2}=2187 \left(9 \tilde{p}_1+512\right),\  N_{3}=-314928 \left(27 \tilde{p}_1+1024\right),\  N_{4}=34828517376\\
\hline
&&&   j^{\frac{2}{3}}(j+N) \\
&\multirow{-2}{*}{$  \mathcal E_1[A_{2,1}^{~\otimes 20}]$}&& N_{1}=27 \tilde{p}_1\\
\cline{2-2}\cline{4-4}
&&& j^5+N_{1} j^4+N_{2} j^3+N_{3} j^2+N_{4} j+N_{5}\\
 \multirow{-3}{*}{$5.$}&\multirow{-1}{*}{$\mathcal E_1[E_{6,1}^{~\otimes 20}]$} & \multirow{-3}{10em}{\centering ${\cal \tilde{Z}}_0 = P_1^5,$\\ ${\cal \tilde{Z}}_1 = P_1^2 P_2^3$}&N_{1}=-2160,\  N_{2}=729 \left(27 \tilde{p}_1+2560\right),\  N_{3}=-629856 \left(27 \tilde{p}_1+1280\right),\\&&&\  N_{4}=136048896 \left(27 \tilde{p}_1+1280\right),\  N_{5}=-15045919506432\\
\hline
&&&   j^2+N_1j+N_2   \\
&\multirow{-2}{*}{$  \mathcal E_1[A_{2,1}^{~\otimes 24}]$}&&N_{1}=27 \tilde{p}_1,\  N_{2}=729 \tilde{p}_2\\
\cline{2-2}\cline{4-4}
&&& j^6+N_{1} j^5+N_{2} j^4+N_{3} j^3+N_{4} j^2+N_{5} j+N_{6}\\
\multirow{-3}{*}{$6.$}&\multirow{-1}{*}{$\mathcal E_1[E_{6,1}^{~\otimes 24}]$} & \multirow{-3}{10em}{\centering ${\cal \tilde{Z}}_0 = P_1^6,$\\ ${\cal \tilde{Z}}_1 = P_1^3 P_2^3,$\\ ${\cal \tilde{Z}}_2 = P_2^6$}&\ N_{1}=-2592,\  N_{2}=2187 \left(9 \tilde{p}_1+1280\right),\  N_{3}=-314928 \left(81 \tilde{p}_1+5120\right),\\&&&\  N_{4}=1594323 \left(6912 \tilde{p}_1+243 \tilde{p}_2+327680\right),\ N_{5}=-176319369216 \left(9 \tilde{p}_1+512\right),\ \ldots\\
\hline
&&&   j^{\frac{1}{3}}(j^2+N_1j+N_2) \\
&\multirow{-2}{*}{$  \mathcal E_1[A_{2,1}^{~\otimes 28}]$}&& N_{1}=27 \tilde{p}_1,\  N_{2}=729 \tilde{p}_2\\
\cline{2-2}\cline{4-4}
&&&  j^7+N_{1} j^6+N_{2} j^5+N_{3} j^4+N_{4} j^3+N_{5} j^2+N_{6} j+N_{7}\\
\multirow{-3}{*}{$7.$}&\multirow{-1}{*}{$\mathcal E_1[E_{6,1}^{~\otimes 28}]$} &\multirow{-3}{10em}{\centering ${\cal \tilde{Z}}_0 = P_1^7,$\\ ${\cal \tilde{Z}}_1 =  P_1^4 P_2^3,$\\ ${\cal \tilde{Z}}_2 =  P_1 P_2^6$} & N_{1}=-3024,\  N_{2}=2187 \left(9 \tilde{p}_1+1792\right),\  N_{3}=-1259712 \left(27 \tilde{p}_1+2240\right),\\&&&\  N_{4}=531441 \left(512 \left(81 \tilde{p}_1+4480\right)+729 \tilde{p}_2\right),\ N_{5}=-688747536 \left(9216 \tilde{p}_1+243 \tilde{p}_2+458752\right),\ \ldots\\
\hline
&&&   j^{\frac{2}{3}}(j^2+N_1j+N_2) \\
&\multirow{-2}{*}{$  \mathcal E_1[A_{2,1}^{~\otimes 32}]$}&& N_{1}=27 \tilde{p}_1,\  N_{2}=729 \tilde{p}_2\\
\cline{2-2}\cline{4-4}
&&&  j^8+N_{1} j^7+N_{2} j^6+N_{3} j^5+N_{4} j^4+N_{5} j^3+N_{6} j^2+N_{7} j+N_{8}\\
\multirow{-2}{*}{$8.$} &\multirow{-1}{*}{$\mathcal E_1[E_{6,1}^{~\otimes 32}]$} &\multirow{-2}{10em}{\centering ${\cal \tilde{Z}}_0 = P_1^8,$\\ ${\cal \tilde{Z}}_1 = P_1^5 P_2^3,$\\ ${\cal \tilde{Z}}_2 =  P_1^2 P_2^6$} & N_{1}=-3456,\  N_{2}=729 \left(27 \tilde{p}_1+7168\right),\  N_{3}=-314928 \left(135 \tilde{p}_1+14336\right),\\&&&\  N_{4}=531441 \left(69120 \tilde{p}_1+729 \tilde{p}_2+4587520\right),\  N_{5}=-459165024 \left(34560 \tilde{p}_1+729 \tilde{p}_2+1835008\right),\ \ldots\\
\hline
&&&   j^3+N_1j^2+N_2j+N_3 \\
&\multirow{-2}{*}{$  \mathcal E_1[A_{2,1}^{~\otimes 36}]$}&&N_{1}=27 \tilde{p}_1,\  N_{2}=729 \tilde{p}_2,\  N_{3}=19683 \tilde{p}_3\\
\cline{2-2}\cline{4-4}
&&&  j^9+N_{1} j^8+N_{2} j^7+N_{3} j^6+N_{4} j^5+N_{5} j^4+N_{6} j^3+N_{7} j^2+N_{8} j+N_{9}\\
\multirow{-3}{*}{$9.$}&\multirow{-1}{*}{$\mathcal E_1[E_{6,1}^{~\otimes 36}]$} & \multirow{-3}{10em}{\centering ${\cal \tilde{Z}}_0 = P_1^9,$\\ ${\cal \tilde{Z}}_1 =  P_1^6 P_2^3,$\\ ${\cal \tilde{Z}}_2 =  P_1^3 P_2^6,$\\ ${\cal \tilde{Z}}_3 =  P_2^9$} & N_{1}=-3888,\  N_{2}=6561 \left(3 \tilde{p}_1+1024\right),\  N_{3}=-1889568 \left(27 \tilde{p}_1+3584\right),\\&&&\  N_{4}=4782969 \left(11520 \tilde{p}_1+81 \tilde{p}_2+917504\right),\  N_{5}=-2066242608 \left(15360 \tilde{p}_1+243 \tilde{p}_2+917504\right),\ \ldots\\
\hline
&&&   j^{\frac{1}{3}}(j^3+N_1j^2+N_2j+N_3) \\
&\multirow{-2}{*}{$  \mathcal E_1[A_{2,1}^{~\otimes 40}]$}&&  N_{1}=27 \tilde{p}_1,\  N_{2}=729 \tilde{p}_2,\  N_{3}=19683 \tilde{p}_3\\
\cline{2-2}\cline{4-4}
&&&  j^{10}+N_{1} j^9+N_{2} j^8+N_{3} j^7+N_{4} j^6+N_{5} j^5+N_{6} j^4+N_{7} j^3+N_{8} j^2+N_{9} j+N_{10}\\
\multirow{-3}{*}{$10.$} &\multirow{-1}{*}{$\mathcal E_1[E_{6,1}^{~\otimes 40}]$} &\multirow{-3}{10em}{\centering ${\cal \tilde{Z}}_0 = P_1^{10},$\\ ${\cal \tilde{Z}}_1 = P_1^7 P_2^3,$\\ ${\cal \tilde{Z}}_2 =  P_1^4 P_2^6,$\\ ${\cal \tilde{Z}}_3 =  P_1 P_2^9$} & N_{1}=-4320,\  N_{2}=6561 \left(3 \tilde{p}_1+1280\right),\  N_{3}=-944784 \left(63 \tilde{p}_1+10240\right),\\&&&\  N_{4}=1594323 \left(48384 \tilde{p}_1+243 \tilde{p}_2+4587520\right),\  N_{5}=-8264970432 \left(6720 \tilde{p}_1+81 \tilde{p}_2+458752\right),\ \ldots\\
\end{array}$
}
\end{threeparttable}
\end{center}
\end{table}

\begin{table}[H]
\begin{center}
\begin{threeparttable}
\resizebox{\textwidth}{!}{
\renewcommand{\arraystretch}{1.3}
$\begin{array}{|c|c|c|c|}
&&&   j^{\frac{2}{3}}(j^3+N_1j^2+N_2j+N_3) \\
&\multirow{-2}{*}{$  \mathcal E_1[A_{2,1}^{~\otimes 44}]$}&& N_{1}=27 \tilde{p}_1,\  N_{2}=729 \tilde{p}_2,\  N_{3}=19683 \tilde{p}_3\\
\cline{2-2}\cline{4-4}
&&& j^{11}+N_{1} j^{10}+N_{2} j^9+N_{3} j^8+N_{4} j^7+N_{5} j^6+N_{6} j^5+N_{7} j^4+N_{8} j^3+N_{9} j^2+N_{10} j+N_{11}\\
\multirow{-3}{*}{$11.$}&\multirow{-1}{*}{$\mathcal E_1[E_{6,1}^{~\otimes 44}]$} &\multirow{-3}{10em}{\centering ${\cal \tilde{Z}}_0 =  P_1^{11},$\\ ${\cal \tilde{Z}}_1 = P_1^8 P_2^3,$\\ ${\cal \tilde{Z}}_2 =  P_1^5 P_2^6,$\\ ${\cal \tilde{Z}}_3 =   P_1^2 P_2^9$} & N_{1}=-4752,\  N_{2}=729 \left(27 \tilde{p}_1+14080\right),\  N_{3}=-7558272 \left(9 \tilde{p}_1+1760\right),\\&&&\  N_{4}=1594323 \left(64512 \tilde{p}_1+243 \tilde{p}_2+7208960\right),\  N_{5}=-688747536 \left(14336 \left(9 \tilde{p}_1+704\right)+1215 \tilde{p}_2\right),\ \ldots\\
\hline
&&&  j^4+N_1j^3+N_2j^2+N_3j+N_4  \\
&\multirow{-2}{*}{$  \mathcal E_1[A_{2,1}^{~\otimes 48}]$}& &N_{1}=27 \tilde{p}_1,\  N_{2}=729 \tilde{p}_2,\  N_{3}=19683 \tilde{p}_3,\  N_{4}=531441 \tilde{p}_4\\
\cline{2-2}\cline{4-4}
&&& j^{12}+N_{1} j^{11}+N_{2} j^{10}+N_{3} j^9+N_{4} j^8+N_{5} j^7+N_{6} j^6+N_{7} j^5+N_{8} j^4+N_{9} j^3+N_{10} j^2+N_{11} j+N_{12}\\
\multirow{-3}{*}{$12.$}&\multirow{-1}{*}{$\mathcal E_1[E_{6,1}^{~\otimes 48}]$} & \multirow{-3}{10em}{\centering ${\cal \tilde{Z}}_0 =  P_1^{12},$\\ ${\cal \tilde{Z}}_1 = P_1^9 P_2^3,$\\ ${\cal \tilde{Z}}_2 = P_1^6 P_2^6,$\\ ${\cal \tilde{Z}}_3 =   P_1^3 P_2^9,$\\ ${\cal \tilde{Z}}_4 =   P_2^{12}$} & N_{1}=-5184,\  N_{2}=2187 \left(9 \tilde{p}_1+5632\right),\  N_{3}=-314928 \left(243 \tilde{p}_1+56320\right),\\&&&\  N_{4}=4782969 \left(27648 \tilde{p}_1+81 \tilde{p}_2+3604480\right),\  N_{5}=-4132485216 \left(512 \left(63 \tilde{p}_1+5632\right)+243 \tilde{p}_2\right),\ \ldots\\ 
\hline
&&&  j^{\frac{1}{3}}(j^4+N_1j^3+N_2j^2+N_3j+N_4)  \\
&\multirow{-2}{*}{$  \mathcal E_1[A_{2,1}^{~\otimes 52}]$}&& N_{1}=27 \tilde{p}_1,\  N_{2}=729 \tilde{p}_2,\  N_{3}=19683 \tilde{p}_3,\  N_{4}=531441 \tilde{p}_4\\
\cline{2-2}\cline{4-4}
&&& j^{13}+N_{1} j^{12}+N_{2} j^{11}+N_{3} j^{10}+N_{4} j^9+N_{5} j^8+N_{6} j^7+N_{7} j^6+N_{8} j^5+N_{9} j^4+N_{10} j^3+N_{11} j^2\\&&&+N_{12} j+N_{13}\\
\multirow{-4}{*}{$13.$}&\multirow{-2}{*}{$\mathcal E_1[E_{6,1}^{~\otimes 52}]$} &\multirow{-4}{10em}{\centering ${\cal \tilde{Z}}_0 =  P_1^{13},$\\ ${\cal \tilde{Z}}_1 = P_1^{10} P_2^3,$\\ ${\cal \tilde{Z}}_2 = P_1^7 P_2^6,$\\ ${\cal \tilde{Z}}_3 =   P_1^4 P_2^9,$\\ ${\cal \tilde{Z}}_4 =  P_1 P_2^{12}$} & N_{1}=-5616,\  N_{2}=2187 \left(9 \tilde{p}_1+6656\right),\  N_{3}=-629856 \left(135 \tilde{p}_1+36608\right),\\&&&\  N_{4}=531441 \left(311040 \tilde{p}_1+729 \tilde{p}_2+46858240\right),\  N_{5}=-2066242608 \left(92160 \tilde{p}_1+567 \tilde{p}_2+9371648\right),\ \ldots\\
\hline
&&&  j^{\frac{2}{3}}(j^4+N_1j^3+N_2j^2+N_3j+N_4)  \\
&\multirow{-2}{*}{$  \mathcal E_1[A_{2,1}^{~\otimes 56}]$}&& N_{1}=27 \tilde{p}_1,\  N_{2}=729 \tilde{p}_2,\  N_{3}=19683 \tilde{p}_3,\  N_{4}=531441 \tilde{p}_4\\
\cline{2-2}\cline{4-4}
&&& j^{14}+N_{1} j^{13}+N_{2} j^{12}+N_{3} j^{11}+N_{4} j^{10}+N_{5} j^9+N_{6} j^8+N_{7} j^7+N_{8} j^6+N_{9} j^5+N_{10} j^4+N_{11} j^3\\&&&+N_{12} j^2+N_{13} j+N_{14}\\
\multirow{-4}{*}{$14.$}&\multirow{-2}{*}{$\mathcal E_1[E_{6,1}^{~\otimes 56}]$} &\multirow{-4}{10em}{\centering ${\cal \tilde{Z}}_0 =  P_1^{14},$\\ ${\cal \tilde{Z}}_1 = P_1^{11} P_2^3,$\\ ${\cal \tilde{Z}}_2 = P_1^8 P_2^6,$\\ ${\cal \tilde{Z}}_3 =   P_1^5 P_2^9,$\\ ${\cal \tilde{Z}}_4 =  P_1^2 P_2^{12}$} & N_{1}=-6048,\  N_{2}=729 \left(27 \tilde{p}_1+23296\right),\  N_{3}=-314928 \left(297 \tilde{p}_1+93184\right),\\&&&\  N_{4}=531441 \left(380160 \tilde{p}_1+729 \tilde{p}_2+65601536\right),\  N_{5}=-1836660096 \left(352 \left(405 \tilde{p}_1+46592\right)+729 \tilde{p}_2\right),\ \ldots\\
\hline
&&&  j^5+N_1j^4+N_2j^3+N_3j^2+N_4j+N_5 \\
&\multirow{-2}{*}{$  \mathcal E_1[A_{2,1}^{~\otimes 60}]$}&& N_{1}=27 \tilde{p}_1,\  N_{2}=729 \tilde{p}_2,\  N_{3}=19683 \tilde{p}_3,\  N_{4}=531441 \tilde{p}_4,\  N_{5}=14348907 \tilde{p}_5\\
\cline{2-2}\cline{4-4}
&&&j^{15}+N_{1} j^{14}+N_{2} j^{13}+N_{3} j^{12}+N_{4} j^{11}+N_{5} j^{10}+N_{6} j^9+N_{7} j^8+N_{8} j^7+N_{9} j^6+N_{10} j^5+N_{11} j^4\\&&&+N_{12} j^3+N_{13} j^2+N_{14} j+N_{15}\\
\multirow{-4}{*}{$15.$}&\multirow{-2}{*}{$\mathcal E_1[E_{6,1}^{~\otimes 60}]$} &\multirow{-4}{10em}{\centering ${\cal \tilde{Z}}_0 =  P_1^{15},$\\ ${\cal \tilde{Z}}_1 = P_1^{12}
   P_2^3,$\\ ${\cal \tilde{Z}}_2 = P_1^9 P_2^6,$\\ ${\cal \tilde{Z}}_3 =   P_1^6 P_2^9,$\\ ${\cal \tilde{Z}}_4 =  P_1^3 P_2^{12},$\\ ${\cal \tilde{Z}}_5 =  P_2^{15}$} & N_{1}=-6480,\  N_{2}=2187 \left(9 \tilde{p}_1+8960\right),\  N_{3}=-1259712 \left(81 \tilde{p}_1+29120\right),\\&&&\  N_{4}=1594323 \left(512 \left(297 \tilde{p}_1+58240\right)+243 \tilde{p}_2\right),\ N_{5}=-688747536 \left(11264 \left(45 \tilde{p}_1+5824\right)+2187 \tilde{p}_2\right),\ \ldots\\
\hline
&&&  j^{\frac{1}{3}}(j^5+N_1j^4+N_2j^3+N_3j^2+N_4j+N_5) \\
&\multirow{-2}{*}{$  \mathcal E_1[A_{2,1}^{~\otimes 64}]$}&&  N_{1}=27 \tilde{p}_1,\  N_{2}=729 \tilde{p}_2,\  N_{3}=19683 \tilde{p}_3,\  N_{4}=531441 \tilde{p}_4,\  N_{5}=14348907 \tilde{p}_5\\
\cline{2-2}\cline{4-4}
&&& j^{16}+N_{1} j^{15}+N_{2} j^{14}+N_{3} j^{13}+N_{4} j^{12}+N_{5} j^{11}+N_{6} j^{10}+N_{7} j^9+N_{8} j^8+N_{9} j^7+N_{10} j^6+N_{11} j^5\\&&&+N_{12} j^4+N_{13} j^3+N_{14} j^2+N_{15} j+N_{16}\\
\multirow{-4}{*}{$16.$}&\multirow{-2}{*}{$\mathcal E_1[E_{6,1}^{~\otimes 64}]$} &\multirow{-4}{10em}{\centering ${\cal \tilde{Z}}_0 = P_1^{16},$\\ ${\cal \tilde{Z}}_1 = P_1^{13} P_2^3,$\\ ${\cal \tilde{Z}}_2 = P_1^{10} P_2^6,$\\ ${\cal \tilde{Z}}_3 = P_1^7 P_2^9,$\\ ${\cal \tilde{Z}}_4 = P_1^4 P_2^{12}$, ${\cal \tilde{Z}}_5 =  P_1 P_2^{15}$} & N_{1}=-6912,\  N_{2}=2187 \left(9 \tilde{p}_1+10240\right),\  N_{3}=-314928 \left(351 \tilde{p}_1+143360\right),\\&&&\  N_{4}=531441 \left(6656 \left(81 \tilde{p}_1+17920\right)+729 \tilde{p}_2\right),\  N_{5}=-1377495072 \left(3328 \left(99 \tilde{p}_1+14336\right)+1215 \tilde{p}_2\right),\ \ldots\\
\hline
\end{array}$
}
\end{threeparttable}
\end{center}
\caption{Characters of one-character extensions of tensor products of $A_{2,1}$ (and $E_{6,1}$) written in a basis of $S$-invariant polynomials}
\label{t5}
\end{table}

This subsection also contains the study of the $E_{6,1}$ CFT.  The reasons to include this here will become clear. First let us note some details of the $E_{6,1}$ CFT.

$E_{6,1}$ is a $\mathbf{[2,0]}$ CFT with central charge $6$. Let us denote it's two characters by $\chi_0, \chi_1$, the latter is the non-identity character with conformal dimension $h_1 = \frac23$. Let us note the $q$-series for the characters :
\begin{small}
\begin{align}
     \chi_0& =q^{-\frac{1}{4}} \left(1+78 q+729 q^2+4382 q^3+19917 q^4+77274 q^5+264644 q^6+827388 q^7+2400597 q^8 + \ldots \right) \nonumber \\
     \chi_1& =q^{\frac{5}{12}} \left( 27+378 q+2484 q^2+12312 q^3+49896 q^4+177660 q^5+570807 q^6+1695762 q^7 + \ldots \right) \label{48b}
\end{align}
\end{small}
and the modular $S$-transformations for the characters :
\be \label{49b}
S(\chi_0) = \frac{\chi_0 + 2 \chi_1}{\sqrt{3}}, \qquad \qquad S(\chi_1) = \frac{\chi_0 - \chi_1}{\sqrt{3}}.
\ee

The allowed values of $k$ and $N$ can be read from entry 6. of table \ref{t2}. Here,  only $c_{(1)} = 24 k$ is possible and the corresponding number of tensor product factors is  $N = 4k$. 

Before we try to solve for the characters of one-character extensions of the tensor products of the $E_{6,1}$ CFT, we note from table \ref{t2}, from entries 2 and 6,  that the $N$ values for $E_{6,1}$ is identical to those of $A_{2,1}$. This is because they have the same value for $\frac{8 n}{\gcd(m,8)}$. Hence we are looking at the monomials of the same degree. Then we note that the $h_1$ for both these CFTs have $3$ in the denominator. What this means is that a monomial which solves the equations \eqref{7b} and \eqref{8b} for $A_{2,1}$ also solves the equations for $E_{6,1}$ (but maybe for different values of $M$). We thus conclude that the ansatz character \eqref{10b} for $E_{6,1}$ for a given $N$ is exactly the same as the ansatz character for $A_{2,1}$. 

Now consider the $S$-invariance equations \eqref{18b}. Here, we meet the second important coincidence concerning the $A_{2,1}$ and $E_{6,1}$ CFTs; the first coincidence being that the $N$ values for both are identical. The $S$-matrix for both the $A_{2,1}$ and $E_{6,1}$ CFTs are identical \eqref{37b}, \eqref{49b}. What this means is that the system of linear equations we need to solve is exactly the same as that of $A_{2,1}$. As well as the solutions. We only need to remember that the $\chi_0, \chi_1$ now are characters of $E_{6,1}$.

The two $S$-invariant polynomials of $A_{2,1}$ are also $S$-invariant polynomials of $E_{6,1}$. But, they have  different $q$-series, which we compute  (using \eqref{48b}) :
\begin{equation}\label{50b}
\begin{array}{llr}
   P_1(\chi_0, \chi_1) &= \chi_0^4+8 \chi_0 \,\chi_1^3 = j-432 \\\
   P_2(\chi_0, \chi_1) &=\chi_0^3 \,\chi_1-\chi_1^4 = 27j^{\frac{1}{3}}
\end{array}
\end{equation}
We can see that both the $S$-invariant polynomials are non-constant $S$-invariant polynomials. In going from $A_{2,1}$ to $E_{6,1}$, $P_2$ even though having an identical expression in terms of characters, has changed from being a constant $S$-invariant polynomial to a non-constant $S$-invariant polynomial. This is going to happen throughout. The answers in terms of characters are identical for $A_{2,1}$ and $E_{6,1}$ but in terms of $q$-series are very different. 

Now, we can report the results for $E_{6,1}$. Since the $A_{2,1}$ and $E_{6,1}$ solutions  have so much common information, we put everything in one table viz. table \ref{t5}. The way to read the  $E_{6,1}$ results from that table is as follows. In the first column, we have the serial number. $3$ times the serial number is the value of $k$ and the central charge is then $c_{(1)} = 8k$. In other words, $24$ times the value of the serial number is the central charge $c_{(1)}$. In the second column, the relevant part for $E_{6,1}$ is the bottom half. The third column does not have two halves. This is the common solution for both $A_{2,1}$ and $E_{6,1}$.  But we need to remember that the $\chi_0, \chi_1$ stand for characters of $E_{6,1}$ \eqref{48b} and thus what we have in third column is $P^{E_{6,1}}_{c_{(1)} = 8k}$. The bottom half of the fourth column contains the $q$-series for $P^{E_{6,1}}_{c_{(1)} = 8k}$, which is of course very different from the top half of the fourth column which contains the $q$-series for $P^{A_{2,1}}_{c_{(1)} = 8k}$.

Then we need to impose admissibility. We compare with the general admissible characters given in table \ref{tapp} of appendix \ref{app1}. We give the relation between the $N$’s and the $\tilde{p}$’s. The nested inequalities that the $N$’s satisfy (for admissibility) translates to nested inequalities for the $\tilde{p}$’s. For rows $3$ and below, we always find an infinite number of admissible character solutions.  In row $1$ and row $2$, we obtain unique admissible characters. 

Amongst all the admissible characters we have computed in table \ref{t5} for $E_{6,1}$, there is only one of them which is known to correspond to actual CFT. This is in the first row, $N = 4, c_{(1)} = 24$. The character here is $j - 432$. We again invoke the classification of $c_{(1)} = 24$ CFTs of \cite{Schellekens:1992db}. In the table at the end of that paper entry no. 58 is the only CFT which is a one-character extension of a tensor product of  copies of $E_{6,1}$\footnote{The reader can find this CFT listed as entry no. 22 of table \ref{t0}.}. This CFT has a character  $j - 432$.  Thus, amongst the infinite number of characters that we have in table \ref{t5}, for $E_{6,1}$, we can be sure that we have a CFT only in exactly one case.  The question for the others will require further analysis. We can make some guesses about some of them. 

Further considerations (beyond those in this paper) establish, in the case of $A_{2,1}$,  that (i) the single character obtained for $N = 4$ viz. $P_1$, (ii) the single  character obtained for $N = 8$ viz. $P_1^2$, (iii) the character obtained for $N =  12$   viz. $P_1^3 - 27 P_2^3$, all correspond to genuine CFT(s). We suspect that these further considerations would  establish that the same characters for $E_{6,1}$ also all correspond to genuine CFT(s). We thus are led to the following surmise/conjecture : \emph{For $E_{6,1}$, (i) there exist genuine one-character CFT(s) with central charge $24$ with character $P_1$, (ii) there exist genuine one-character CFT(s) with central charge $48$ with character $P_1^2$, (iii) there exist genuine one-character CFT(s) with central charge $72$ with character $P_1^3 - 27 P_2^3$. } Part (i) of the surmise is true \cite{Schellekens:1992db} (see previous paragraph).

\subsection{$\mathcal C = G_{2,1}$ and $F_{4,1}$ \label{33s}}
 
 $G_{2,1}$ is a $\mathbf{[2,0]}$ CFT with central charge $\frac{14}{5}$ . Let us denote it's two characters by $\chi_0, \chi_1$, the latter is the non-identity character with conformal dimension $h_1 = \frac25$. Let us note the $q$-series for the characters :
\begin{small}
\begin{align}
     \chi_0& = q^{-\frac{7}{60}} \left( 1+14 q+42 q^2+140 q^3+350 q^4+840 q^5+1827 q^6+3858 q^7+7637 q^8+14756 q^9 + \ldots \right) \nonumber \\
     \chi_1& = q^{\frac{17}{60}} \left( 7+34 q+119 q^2+322 q^3+819 q^4+1862 q^5+4025 q^6+8218 q^7+16177 q^8  + \ldots \right) \label{39b}
\end{align}
\end{small}
and the modular $S$-transformations for the characters :
\be \label{40b}
   S(\chi_0)=\frac{ \chi_0 + \varphi \chi_1}{\sqrt{2 + \varphi }},\qquad S(\chi_1)=\frac{ \varphi \chi_0 -  \chi_1}{\sqrt{2 + \varphi}}\quad\text{where}\quad\varphi = \frac{1+\sqrt{5}}{2}
\ee

The allowed values of $k$ and $N$ can be read from entry 3. of table \ref{t2}. The allowed tensor powers $N$ are multiples of $20$: $N = 20 r$ and the corresponding $c_{(1)}$'s are multiples of $56$: $c_{(1)} = 56 r$ with $r = 1, 2, \ldots$
 
The problem of finding one-character extensions for tensor products of $G_{2,1}$ is a little bit harder when compared to $A_{1,1}$ and $A_{2,1}$. For $A_{1,1}$, to study $16$ different $c_{(1)}$’s, we needed to work with $N = 128$ tensor powers and for $A_{2,1}$, we needed to work with $N = 64$ tensor powers. To achieve the same, i.e. to study $16$ different $c_{(1)}$’s for $G_{2,1}$, we need to work with $N = 320$ tensor powers, which is substantially much harder. Here, we use the long procedure (the ones outlined in section \ref{2s}) for small $N$ and for higher $N$ we employ the new method that we discovered in sections \ref{31s} and \ref{32s}.

We start the analysis with the smallest number of tensor powers i.e. the  $N = 20, c_{(1)} = 56$ case. We discover the first $S$-invariant polynomial $P_1$ here. It is a non-constant $S$-invariant polynomial which we call $P_1$. Then we work on the next higher number of tensor powers, the $N = 40, c_{(1)} = 112$ case. For this, we find that the character is  $P_1^2$.  The second $S$-invariant polynomial, which is also a non-constant $S$-invariant polynomial is discovered when working on the $N = 60, c_{(1)} = 168$ case. It is a degree-$12$ polynomial of the characters. We collect both the $S$-invariant polynomials for $G_{2,1}$ and their $q$-series here:
\begin{equation}\label{41b}
\begin{array}{llr}
   P_1(\chi_0, \chi_1) &=\chi_0^{20}-228 \chi_0^{15} \chi_1^5+494 \chi_0^{10} \chi_1^{10}+228 \chi_0^5 \chi_1^{15}+\chi_1^{20} =j^{\frac{1}{3}}(j^2-1456j-3670016) \\\
   P_2(\chi_0, \chi_1) &=\chi_0^{11} \chi_1+11 \chi_0^6 \chi_1^6 - \chi_0 \chi_1^{11} = 7j-4096 
\end{array}
\end{equation}

There is a similarity between all the three CFTs we have studied so far viz. $A_{1,1}$, $A_{2,1}$ and $G_{2,1}$. In each of them, one discovers $P_1$ for the smallest $c_{(1)}$ and $P_2$ while working with three times the smallest $c_{(1)}$.  

For $N \geq 80$, we will employ the method that we have described in the ``Lessons learnt and a new method’’ parts of sections \ref{31s} and \ref{32s}. First we note that $P_1$ is a polynomial of the characters of degree $20$ and $P_2$ has degree $12$.  The monomial $P_1^a\,P_2^b$ is a degree-$20 a + 12 b$ polynomial of the characters. To obtain all such monomials that will contribute to $P_{c_{(1)} = 56r}^{G_{2,1}} (\chi_0, \chi_1)$, we need to find non-negative integral solutions for : 
\be \label{58c}
20 a + 12 b = N = 20 r, \qquad r = 1, 2, \ldots
\ee
For $r = 1$, the equation $20 a + 12 b = 20$ has only one solution $(a, b) = (1,0)$ i.e. $P_1$ which is in agreement with what we found by the long method. For $r = 2$, the equation $20 a + 12 b = 40$ has only one solution $(a, b) = (2,0)$ i.e. $P_1^2$ which is in agreement with what we found by the long method. For $r = 3$, the equation $20 a + 12 b = 60$ has  two solutions $(a, b) = (3,0), (0,5)$ i.e. $P_1^3, P_2^5$ which is in agreement with what we found by the long method.  We also note that all the solutions we have obtained so far viz. $P_1, P_1^2, P_1^3, P_2^5$ all contain monomials of characters that are commensurate with the identity character. This will happen for a $P_1^a\,P_2^b$ with $a$ taking any value but $b$ should be a multiple of $5$. We have solved the \eqref{58c} for upto $N = 160$ and it turns out that every non-negative integral solution gives only monomials of characters that are commensurate with the identity character i.e. $b$ is always a multiple of $5$.  We can see this, directly from \eqref{58c}, which allows us to write $12 b = 20 r - 20 a$ ; since $5$ divides the right hand side, it follows that $5$ must divide $b$.

In table \ref{t6}, we collect all the results for $r = 1, 2, \ldots 8$.  Table \ref{t6} also contains results for $F_{4,1}$ for reasons that will be explained below.  The way to read the results for $G_{2,1}$ from table \ref{t6} is as follows.  The first column is the serial no. which is also the value of $r$, from which one can infer the value of $c_{(1)} = 56r$.  The (top half of the) second column contains the number of tensor product factors $N$ (which equals $20r$).   The third column contains the character of the one-character extension $P^{G_{2,1}}_{c_{(1)} = 56 r}(\chi_0, \chi_1)$.  The number of summands ranges from $1$, for $r = 1$, to $3$ for $r = 8$.  For $r = 8$, the last entry of the table \ref{t6} we have three $S$-invariant summands $\tilde{{\cal Z}}_0, \tilde{{\cal Z}}_1, \tilde{{\cal Z}}_2$.  The expressions for the $\tilde{{\cal Z}}$'s when written in terms of the characters  $\chi_0, \chi_1$ are very voluminous. Each of them is a  degree $N$ polynomial.  For example for $N= 160$, the last entry of table \ref{t6}, we have  three homogenous degree-$160$ polynomials. But, when written in terms of the $S$-invariant polynomials, they are in fact polynomials of much smaller degree. For $N = 160$, the highest degree term that occurs is $12$.  

Now we study admissibility. For that, we first need to compute the $q$-series of $P^{G_{2,1}}_{c_{(1)} = 56r}$. We employ the explicit $q$-series of the characters available in \eqref{39b} and give the  $q$-series of $P^{G_{2,1}}_{c_{(1)} = 56 r}$ in the top half of the fourth column. 

\underline{$N = 20, ~c_{(1)} = 56$} : Here, we have a unique solution, whose $q$-series is :
\begin{small}
\begin{align*}
   P^{G_{2,1}}_{c_{(1)} = 56}& = j^{\frac{1}{3}}(j^2 - 1456 j - 3670016)  = q^{-7/3} \left( 1 + 280 q - 3793916 q^2 -894427520 q^3 + \ldots  \right),
\end{align*}
\end{small}
is clearly not an admissible character and hence cannot be a character of a CFT.  We conclude that there are \emph{no one-character CFTs possible at central charge $56$} that are one-character extensions of tensor products of the $G_{2,1}$ CFT (see footnote \ref{foot}). Thus, the methods of this paper, while being too primitive to be able to conclude for the existence of CFTs, are good enough to conclude for the non-existence of CFTs. 

\underline{$N = 40, ~c_{(1)} = 112$} : Here, we have a unique solution, whose $q$-series is :
\begin{small}
\begin{align*}
   P^{G_{2,1}}_{c_{(1)} = 112}& = j^{\frac{2}{3}}(j^4 - 2912 j^3 - 5220096 j^2 + 10687086592 j + 13469017440256) \nonumber \\
   & = q^{-\frac{14}{3}} \left( 1 + 560 q -  7509432 q^2 -3913448000 q^3 + \ldots  \right),
\end{align*}
\end{small}
is clearly not an admissible character and hence cannot be a character of a CFT.  We conclude that there are \emph{no one-character CFTs possible at central charge $112$} that are one-character extensions of tensor products of the $G_{2,1}$ CFT (see footnote \ref{foot}). 

\underline{$N = 60, ~c_{(1)} = 168$} : To discuss the solutions, let us introduce the following notation :
\begin{small}
\be \label{59c}
P^{G_{2,1}}_{c_{(1)} = 56 r} = q^{-\frac{7 r}{3}} \left(1 + a_1^{(r)}\,q + a_2^{(r)}\,q^2 + a_3^{(r)}\,q^3 + \ldots \right).
\ee
\end{small}
From table \ref{t6}, the solution is :
\be \label{60c}
P^{G_{2,1}}_{c_{(1)} = 168} = P_1^3 + \tilde{p}_1 P_2^5.
\ee
Computing, we find for example 
\be
a_2^{(3)} = 7 (2401 \tilde{p}_1-1592364),  \qquad a_3^{(3)} = 280 (47677 \tilde{p}_1-32268248).
\ee
Requiring $a_n^{(3)}$ to be a non-negative integer gives us multiple inequalities for $\tilde{p}$. We solved all these inequalities for  $n \leq 500$ and found the following solution :
\be \label{62c}
\tilde{p}_1 \geq 7587.
\ee
We seem to have obtained an infinite number of admissible characters in \eqref{60c} and \eqref{62c}. We are not sure if any of them correspond to genuine CFTs\footnote{\label{foot}We were informed by Brandon Rayhaun that none of these admissible characters correspond to genuine CFTs. We were directed to the research \cite{BookerDavydov}, \cite{Moller:2024xtt} ;  “it is actually impossible to construct a meromorphic extension of $G_{2,1}^N$ or $F_{4,1}^N$ for any N; even if it is possible to find healthy looking characters, it turns out that more subtle CFT data breaks down…” We are very grateful to Brandon Rayhaun for his comments. His comments directed us to a more careful investigation of the $G_{2,1}$ and $F_{4,1}$ cases.}.

\underline{$N = 80, ~c_{(1)} = 224$} : From table \ref{t6}, the solution is :
\be \label{63c}
P^{G_{2,1}}_{c_{(1)} = 224} = P_1^4 + \tilde{p}_1 P_1 P_2^5.
\ee
We found that requiring $a_1^{(4)}, \ldots  a_5^{(4)}$ to be non-negative and integral gives $\tilde{p}_1 \geq 900$ while requiring $a_6^{(4)}$ to be non-negative and integral gives $\tilde{p}_1 \leq - 2436$. Hence, \eqref{63c} is not an admissible character for any value of $\tilde{p}_1$ and thus cannot be a character of a CFT.  We conclude that there are \emph{no one-character CFTs possible at central charge $224$} that are one-character extensions of tensor products of the $G_{2,1}$ CFT (see footnote \ref{foot}). 

\underline{$N = 100, ~c_{(1)} = 280$} : From table \ref{t6}, the solution is :
\be \label{64c}
P^{G_{2,1}}_{c_{(1)} = 280} = P_1^5 + \tilde{p}_1 P_1^2 P_2^5.
\ee
We found that requiring $a_1^{(5)}, \ldots  a_5^{(5)}$ to be non-negative and integral gives $\tilde{p}_1 \leq 1310$ while requiring $a_6^{(5)}$ to be non-negative and integral gives $\tilde{p}_1 \geq  10237$. Hence, \eqref{64c} is not an admissible character for any value of $\tilde{p}_1$ and thus cannot be a character of a CFT.  We conclude that there are \emph{no one-character CFTs possible at central charge $280$} that are one-character extensions of tensor products of the $G_{2,1}$ CFT (see footnote \ref{foot}). 

\underline{$N = 120, ~c_{(1)} = 336$} : From table \ref{t6}, the solution is :
\be \label{65c}
P^{G_{2,1}}_{c_{(1)} = 336} = P_1^6 + \tilde{p}_1 P_1^3 P_2^5 + \tilde{p}_2  P_2^{10}.
\ee
Requiring the non-negativity and integrality of each $a_n^{(6)}$ gives us some region of the positive quadrant of the $\tilde{p}_1-\tilde{p}_2$ plane. We studied upto $n = 500$ and we found multiple regions that are solutions to all the constraints; thus we seem to have obtained an infinite number of admissible characters. We are not sure if any of them correspond to genuine CFTs (see footnote \ref{foot}).

\underline{$N = 140, ~c_{(1)} = 392$} : From table \ref{t6}, the solution is :
\be \label{66c}
P^{G_{2,1}}_{c_{(1)} = 392} = P_1^7 + \tilde{p}_1 P_1^4 P_2^5 + \tilde{p}_2  P_1 P_2^{10}.
\ee
In this case, we analysed  the constraints coming from requiring the non-negativity and integrality of $a_n^{(7)}$ for $n \leq 8$. We found no common solutions to all the constraints. Hence we conclude that \eqref{66c} is not an admissible character for any value of $\tilde{p}_1, \tilde{p}_2 $ and thus cannot be a character of a CFT. We conclude that there are \emph{no one-character CFTs possible at central charge $392$} that are one-character extensions of tensor products of the $G_{2,1}$ CFT (see footnote \ref{foot}). 

\underline{$N = 160, ~c_{(1)} = 448$} : From table \ref{t6}, the solution is :
\be \label{67c}
P^{G_{2,1}}_{c_{(1)} = 448} = P_1^8 + \tilde{p}_1 P_1^5 P_2^5 + \tilde{p}_2  P_1^2 P_2^{10}.
\ee
In this case, we analysed  the constraints coming from requiring the non-negativity and integrality of $a_n^{(8)}$ for $n \leq 10$. We found no common solutions to all the constraints. Hence we conclude that \eqref{67c} is not an admissible character for any value of $\tilde{p}_1, \tilde{p}_2 $ and thus cannot be a character of a CFT. We conclude that there are \emph{no one-character CFTs possible at central charge $448$} that are one-character extensions of tensor products of the $G_{2,1}$ CFT (see footnote \ref{foot}). 

\begin{table}[H]
\begin{center}
\begin{threeparttable}
\resizebox{\textwidth}{!}{
\renewcommand{\arraystretch}{1.3}
$\begin{array}{|c|c|c|c|}
\hline
  \# &  N & P_{c_{(1)} = 56r}^{G_{2,1}} (\chi_0, \chi_1)~\text{and}~P_{c_{(1)} = 104r}^{F_{4,1}} (\chi_0, \chi_1)  &  q\text{-series}  \\
\hline
  &  & & j^{\frac{1}{3}}(j^2+N_1j+N_2)\\
   &  \multirow{-2}{*}{$\mathcal E_1[G_{2,1}^{~\otimes 20}]$}&& N_1 = -1456,\ N_2 = -3670016   \\
\cline{2-2}\cline{4-4}
 \multirow{-2}{*}{$1.$} && & j^{\frac{1}{3}}(j^4 +N_1j^3 +N_2j^2 +N_3j +N_4) \\
      &  \multirow{-2}{*}{$\mathcal E_1[F_{4,1}^{~\otimes 20}]$}&  \multirow{-4}{*}{${\cal \tilde{Z}}_0 = P_1$}& N_{1}= -2184,\quad N_{2}= 1086696,\quad N_{3}= -2815351344,\quad N_{4}= 402462867456\\
\hline
&   && j^{\frac{2}{3}}(j^4+N_1j^3+N_2j^2+N_3j+N_4)\\
& \multirow{-2}{*}{$\mathcal E_1[G_{2,1}^{~\otimes 40}]$} && N_1 = -2912,\quad N_2 = -5220096,\ N_3 = 10687086592,\ N_4 = 13469017440256\\
\cline{2-2}\cline{4-4}
& && j^{\frac{2}{3}} (j^8+N_1j^7+N_2j^6+N_3j^5+N_4j^4+N_5j^3+N_6j^2+N_7j+N_8)\\
  \multirow{-3}{*}{$2.$} &  \multirow{-1}{*}{$\mathcal E_1[F_{4,1}^{~\otimes 40}]$}&  \multirow{-3}{*}{${\cal \tilde{Z}}_0 = P_1^2$}& N_{1}= -4368,\quad N_{2}= 6943248,\quad N_{3}= -10377390816,\\&&&\quad N_{4}= 14283288601920,\quad N_{5}= -7876819893286656,\ \ldots\\
\hline
&&&  j^7+N_1j^6+N_2j^5+N_3j^4+N_4j^3+N_5j^2+N_6j+N_7  \\
   & \multirow{-1}{*}{$\mathcal E_1[G_{2,1}^{~\otimes 60}]$}&&N_{1}=-4368,\  N_{2}+4650240=16807 \tilde{p}_1,\  N_{3}=-1003520 (49 \tilde{p}_1-28873),\\&&&\  N_{4}=8220835840 (7 \tilde{p}_1+2076),\ N_{5}=-6734508720128 (5 \tilde{p}_1+8736),\ \ldots\\ 
\cline{2-2}\cline{4-4}
&  && j^{13}+N_{1}j^{12}+N_{2}j^{11}+N_{3}j^{10}+N_{4}j^9+N_{5}j^8+N_{6}j^7+N_{7}j^6+N_{8}j^5 +N_{9}j^4+N_{10}j^3+N_{11}j^2+N_{12}j+N_{13}\\
 \multirow{-4}{*}{$3.$}   &  \multirow{-1}{*}{$\mathcal E_1[F_{4,1}^{~\otimes 60}]$} &\multirow{-4}{10em}{\centering$ {\cal \tilde{Z}}_0 =P_1^3,$\\$ {\cal \tilde{Z}}_1 = P_2^5 $}& N_{1}= -6552,\quad N_{2}= 17569656,\quad N_{3}= 208 (57122 \tilde{p}_1-120079917),\\&&&\ N_{4}= -28080 (1913587 \tilde{p}_1-727880976),\quad N_{5}= 17740944 (5094505 \tilde{p}_1-554277348), \ldots\\
\end{array}$
}
\end{threeparttable}
\end{center}
\end{table}

\begin{table}[H]
\begin{center}
\begin{threeparttable}
\resizebox{\textwidth}{!}{
\renewcommand{\arraystretch}{1.3}
$\begin{array}{|c|c|c|c|}
   &&&  j^{\frac{1}{3}} (j^9+N_{1} j^8+N_{2} j^7+N_{3} j^6+N_{4} j^5+N_{5} j^4+N_{6} j^3+N_{7} j^2+N_{8} j+N_{9})\\
   & \multirow{-1}{*}{$\mathcal E_1[G_{2,1}^{~\otimes 80}]$}&&N_{1}=-5824,\  N_{2}+1960448=16807 \tilde{p}_1,\  N_{3}+73643472 \tilde{p}_1=51776012288,\\&&&\  N_{4}=3211264 (21007 \tilde{p}_1-2508095),\  N_{5}=26306674688 (2395 \tilde{p}_1-7223216),\ \ldots\\
\cline{2-2}\cline{4-4}
   &  &&j^{\frac{1}{3}} (j^{17}+N_{1} j^{16}+N_{2} j^{15}+N_{3} j^{14}+N_{4} j^{13}+N_{5} j^{12}+N_{6} j^{11}+N_{7} j^{10}+N_{8} j^9+N_{9} j^8+N_{10} j^7+N_{11} j^6\\&&&+N_{12} j^5+N_{13} j^4+N_{14} j^3+N_{15} j^2+N_{16} j+N_{17})\\
   \multirow{-5}{*}{$4.$}   &  \multirow{-2}{*}{$\mathcal E_1[F_{4,1}^{~\otimes 80}]$} &\multirow{-5}{10em}{\centering$ {\cal \tilde{Z}}_0 =P_1^4,$\\$ {\cal \tilde{Z}}_1 = P_1P_2^5 $}&N_{1}=-8736,\  N_{2}=32965920,\  N_{3}+81410996160=11881376 \tilde{p}_1,\\&&&\  N_{4}+79682448144 \tilde{p}_1=167432156161920,\  N_{5}=219024 (1007409169 \tilde{p}_1-1299574681632),\ \ldots\\
\hline
   &&& j^{\frac{2}{3}} (j^{11}+N_{1} j^{10}+N_{2} j^9+N_{3} j^8+N_{4} j^7+N_{5} j^6+N_{6} j^5+N_{7} j^4+N_{8} j^3+N_{9} j^2+N_{10} j+N_{11})\\
   & \multirow{-1}{*}{$\mathcal E_1[G_{2,1}^{~\otimes 100}]$}&&N_{1}=-7280,\  N_{2}=16807 \tilde{p}_1+2849280,\  N_{3}+98114464 \tilde{p}_1=76004597760,\\&&&\  N_{4}=12544 (9008447 \tilde{p}_1-6078216960),\  N_{5}=51380224 (4574851 \tilde{p}_1-7168336539),\ \ldots\\
\cline{2-2}\cline{4-4}
   &  &&j^{\frac{2}{3}} (j^{21}+N_{1} j^{20}+N_{2} j^{19}+N_{3} j^{18}+N_{4} j^{17}+N_{5} j^{16}+N_{6} j^{15}+N_{7} j^{14}+N_{8} j^{13}+N_{9} j^{12}+N_{10} j^{11}+N_{11} j^{10}\\&&&+N_{12} j^9+N_{13} j^8+N_{14} j^7+N_{15} j^6+N_{16} j^5+N_{17} j^4+N_{18} j^3+N_{19} j^2+N_{20} j+N_{21})\\
 \multirow{-5}{*}{$5.$}   &  \multirow{-2}{*}{$\mathcal E_1[F_{4,1}^{~\otimes 100}]$} &\multirow{-5}{10em}{\centering$ {\cal \tilde{Z}}_0 =P_1^5,$\\$ {\cal \tilde{Z}}_1 = P_1^2P_2^5 $}&N_{1}=-10920,\  N_{2}=53132040,\  N_{3}=208 (57122 \tilde{p}_1-796717755),\\&&&\  N_{4}+105631373328 \tilde{p}_1=406055077384320,\  N_{5}=24336 (16748220593 \tilde{p}_1-34315641058464),\ \ldots\\
\hline
&&& j^{14}+N_{1} j^{13}+N_{2} j^{12}+N_{3} j^{11}+N_{4} j^{10}+N_{5} j^9+N_{6} j^8+N_{7} j^7+N_{8} j^6+N_{9} j^5+N_{10} j^4+N_{11} j^3\\&&&+N_{12} j^2+N_{13} j+N_{14}\\
   & \multirow{-2}{*}{$\mathcal E_1[G_{2,1}^{~\otimes 120}]$}&&N_{1}=-8736,\  N_{2}=16807 \tilde{p}_1+9778944,\  N_{3}+122585456 \tilde{p}_1=98573762560,\\&&&\  N_{4}=49 (3962748160 \tilde{p}_1+5764801 \tilde{p}_2-4027852062720),\  N_{5}=200704 (2145486235 \tilde{p}_1-8235430 \tilde{p}_2-2671774769664),\ \ldots\\
\cline{2-2}\cline{4-4}
&  && j^{26}+N_{1} j^{25}+N_{2} j^{24}+N_{3} j^{23}+N_{4} j^{22}+N_{5} j^{21}+N_{6} j^{20}+N_{7} j^{19}+N_{8} j^{18}+N_{9} j^{17}+N_{10} j^{16}+N_{11} j^{15}\\&&&+N_{12} j^{14}+N_{13} j^{13}+N_{14} j^{12}+N_{15} j^{11}+N_{16} j^{10}+N_{17} j^9+N_{18} j^8+N_{19} j^7+N_{20} j^6+N_{21} j^5+N_{22} j^4\\&&&+N_{23} j^3+N_{24} j^2+N_{25} j+N_{26}\\
   \multirow{-7}{*}{$6.$}   &  \multirow{-3}{*}{$\mathcal E_1[F_{4,1}^{~\otimes 120}]$} &\multirow{-7}{10em}{\centering$ {\cal \tilde{Z}}_0 =P_1^6,$\\$ {\cal \tilde{Z}}_1 = P_1^3P_2^5 ,$\\$ {\cal \tilde{Z}}_2 = P_2^{10} $}&N_{1}=-13104,\  N_{2}=78068016,\  N_{3}=416 (28561 \tilde{p}_1-712595529),\\&&&\  N_{4}+131580298512 \tilde{p}_1=856866120267456,\  N_{5}=73008 (8919502787 \tilde{p}_1-28161219373488),\ \ldots\\
\hline
   &&& j^{\frac{1}{3}} (j^{16}+N_{1} j^{15}+N_{2} j^{14}+N_{3} j^{13}+N_{4} j^{12}+N_{5} j^{11}+N_{6} j^{10}+N_{7} j^9+N_{8} j^8+N_{9} j^7+N_{10} j^6+N_{11} j^5\\&&&+N_{12} j^4+N_{13} j^3+N_{14} j^2+N_{15} j+N_{16})\\
   & \multirow{-2}{*}{$\mathcal E_1[G_{2,1}^{~\otimes 140}]$}&&N_{1}=-10192,\  N_{2}=49 (343 \tilde{p}_1+384256),\  N_{3}=-21952 (6699 \tilde{p}_1-5302336),\\&&&\  N_{4}=343 (906638848 \tilde{p}_1+823543 \tilde{p}_2-1098475307008),\  N_{5}=784 (762474425344 \tilde{p}_1-1127 (2336173 \tilde{p}_2+691107069952)),\ \ldots\\
\cline{2-2}\cline{4-4}
   &  && j^{\frac{1}{3}} (j^{30}+N_{1} j^{29}+N_{2} j^{28}+N_{3} j^{27}+N_{4} j^{26}+N_{5} j^{25}+N_{6} j^{24}+N_{7} j^{23}+N_{8} j^{22}+N_{9} j^{21}+N_{10} j^{20}+N_{11} j^{19}\\&&&+N_{12} j^{18}+N_{13} j^{17}+N_{14} j^{16}+N_{15} j^{15}+N_{16} j^{14}+N_{17} j^{13}+N_{18} j^{12}+N_{19} j^{11}+N_{20} j^{10}+N_{21} j^9+N_{22} j^8\\&&&+N_{23} j^7+N_{24} j^6+N_{25} j^5+N_{26} j^4+N_{27} j^3+N_{28} j^2+N_{29} j+N_{30})\\
   \multirow{-7}{*}{$7.$}   &  \multirow{-3}{*}{$\mathcal E_1[F_{4,1}^{~\otimes 140}]$} &\multirow{-7}{10em}{\centering$ {\cal \tilde{Z}}_0 =P_1^7,$\\$ {\cal \tilde{Z}}_1 = P_1^4P_2^5 ,$\\$ {\cal \tilde{Z}}_2 = P_1P_2^{10} $}&   N_{1}=-15288,\  N_{2}=107773848,\  N_{3}+483995702736=11881376 \tilde{p}_1,\\&&&\  N_{4}+157529223696 \tilde{p}_1=1626421540161600,\  N_{5}=365040 (2606503055 \tilde{p}_1-12257814370032),\ \ldots\\
\hline
   &&& j^{\frac{2}{3}} (j^{18}+N_{1} j^{17}+N_{2} j^{16}+N_{3} j^{15}+N_{4} j^{14}+N_{5} j^{13}+N_{6} j^{12}+N_{7} j^{11}+N_{8} j^{10}+N_{9} j^9+N_{10} j^8+N_{11} j^7\\&&&+N_{12} j^6+N_{13} j^5+N_{14} j^4+N_{15} j^3+N_{16} j^2+N_{17} j+N_{18})\\
   & \multirow{-2}{*}{$\mathcal E_1[G_{2,1}^{~\otimes 160}]$}&&N_{1}=-11648,\  N_{2}=16807 \tilde{p}_1+29998080,\  N_{3}=-27440 (6251 \tilde{p}_1-4605952),\\&&&\  N_{4}=343 (2560 (527753 \tilde{p}_1-700792576)+823543 \tilde{p}_2),\  N_{5}=14112 (48518762240 \tilde{p}_1-147 (1193297 \tilde{p}_2+235834966016)),\ \ldots\\
\cline{2-2}\cline{4-4}
&  && j^{\frac{2}{3}} (j^{34}+N_{1} j^{33}+N_{2} j^{32}+N_{3} j^{31}+N_{4} j^{30}+N_{5} j^{29}+N_{6} j^{28}+N_{7} j^{27}+N_{8} j^{26}+N_{9} j^{25}+N_{10} j^{24}+N_{11} j^{23}\\&&&+N_{12} j^{22}+N_{13} j^{21}+N_{14} j^{20}+N_{15} j^{19}+N_{16} j^{18}+N_{17} j^{17}+N_{18} j^{16}+N_{19} j^{15}+N_{20} j^{14}+N_{21} j^{13}+N_{22} j^{12}\\&&&+N_{23} j^{11}+N_{24} j^{10}+N_{25} j^9+N_{26} j^8+N_{27} j^7+N_{28} j^6+N_{29} j^5+N_{30} j^4+N_{31} j^3+N_{32} j^2+N_{33} j+N_{34})\\
   \multirow{-7}{*}{$8.$}   &  \multirow{-3}{*}{$\mathcal E_1[F_{4,1}^{~\otimes 160}]$} &\multirow{-7}{10em}{\centering$ {\cal \tilde{Z}}_0 =P_1^8,$\\$ {\cal \tilde{Z}}_1 = P_1^5P_2^5 ,$\\$ {\cal \tilde{Z}}_2 = P_1^2P_2^{10} $}&N_{1}=-17472,\  N_{2}=142249536,\  N_{3}+738802546560=11881376 \tilde{p}_1,\\&&&\  N_{4}=-3120 (58807099 \tilde{p}_1-911547794448),\  N_{5}=170352 (7680761855 \tilde{p}_1-52023031697088),\ \ldots\\
\hline
\end{array}$
}
\end{threeparttable}
\end{center}
\caption{Characters of one-character extensions of tensor products of $G_{2,1}$  (and $F_{4,1}$) written in a basis of $S$-invariant polynomials}\label{t6}
\end{table}

This subsection also contains the study of the $F_{4,1}$ CFT.  The reasons to include this here will become clear. First let us note some details of the $F_{4,1}$ CFT. 

 $F_{4,1}$ is a $\mathbf{[2,0]}$ CFT with central charge $\frac{26}{5}$ . Let us denote it's two characters by $\chi_0, \chi_1$, the latter is the non-identity character with conformal dimension $h_1 = \frac35$. Let us note the $q$-series for the characters :
\begin{small}
\begin{align}
     \chi_0& = q^{-\frac{13}{60}} \left(1+52 q+377 q^2+1976 q^3+7852 q^4+27404 q^5+84981 q^6+243230 q^7+649454 q^8  + \ldots \right) \nonumber \\
     \chi_1& = q^{\frac{23}{60}} \left( 26+299 q+1702 q^2+7475 q^3+27300 q^4+88452 q^5+260650 q^6+714727 q^7 + \ldots \right) \label{45b}
\end{align}
\end{small}
and the modular $S$-transformations for the characters :
\be \label{46b}
   S(\chi_0)=\frac{ \chi_0 + \varphi \chi_1}{\sqrt{2 + \varphi }},\qquad S(\chi_1)=\frac{ \varphi \chi_0 -  \chi_1}{\sqrt{2 + \varphi}}\quad\text{where}\quad\varphi = \frac{1+\sqrt{5}}{2}
\ee

The allowed values of $k$ and $N$ can be read from entry 5. of table \ref{t2}. The allowed tensor powers $N$ are multiples of $20$: $N = 20 r$ and the corresponding $c_{(1)}$'s are multiples of $104$: $c_{(1)} = 104 \,r$ with $r = 1, 2, \ldots$

Before we try to solve for the characters of one-character extensions of the tensor products of the $F_{4,1}$ CFT, we note from table \ref{t2}, from entries 3 and 5,  that the $N$ values for $F_{4,1}$ is identical to those of $G_{2,1}$. This is because they have the same value for $\frac{8 n}{\gcd(m,8)}$. So we are looking at the monomials of the same degree. Then we note that the $h_1$ for both these CFT have $5$ in the denominator. What this means is that a monomial which solves the equations \eqref{7b} and \eqref{8b} for $G_{2,1}$ also solves the equations for $F_{2,1}$ (but for different values of $M$). We thus conclude that the ansatz character \eqref{10b} for $F_{4,1}$ for a given $N$ is exactly the same as the ansatz character for $G_{2,1}$.

Now consider the $S$-invariance equations \eqref{18b}. Since the $S$-matrix for both the $F_{4,1}$ and $G_{2,1}$ CFTs are identical \eqref{40b}, \eqref{46b}, we get the same set of linear equations and hence the same solutions as we had obtained for $G_{2,1}$. We have the same two $S$-invariant polynomials but they compute (using \eqref{45b}) to different $q$-series :
\begin{equation}\label{47b}
\begin{array}{llr}
   P_1(\chi_0, \chi_1) &=\chi_0^{20}-228 \chi_0^{15} \chi_1^5+494 \chi_0^{10} \chi_1^{10}+228 \chi_0^5 \chi_1^{15}+\chi_1^{20} \\&=j^{\frac{1}{3}}(j^4-2184 j^3+1086696 j^2-2815351344 j+402462867456)\\
   P_2(\chi_0, \chi_1) &=\chi_0^{11} \chi_1+11 \chi_0^6 \chi_1^6 - \chi_0 \chi_1^{11} = 26j^2-23517j-2985984
\end{array}
\end{equation}

Now, we can report the results for $F_{4,1}$. Since the $G_{2,1}$ and $F_{4,1}$ solutions  have so much common information, we put everything in one table viz. table \ref{t6}.  The way to read the results for $F_{4,1}$ from table \ref{t6} is as follows.  The first column is the serial no. which is also the value of $r$, from which one can infer the value of $c_{(1)} = 104 r$.  The (top half of the) second column contains the number of tensor product factors $N$ (which equals $20r$).   The third column contains the character of the one-character extension $P^{F_{4,1}}_{c_{(1)} = 104 r}(\chi_0, \chi_1)$.  This is the common solution for both $G_{2,1}$ and $F_{4,1}$.  But we need to remember that the $\chi_0, \chi_1$ stand for characters of $F_{4,1}$ \eqref{45b} and thus what we have in third column is $P^{F_{4,1}}_{c_{(1)} = 104 r}$.  The bottom half of the fourth column contains the $q$-series for $P^{F_{4,1}}_{c_{(1)} = 104 r}$, which is of course very different from the top half of the fourth column which contains the $q$-series for $P^{G_{2,1}}_{c_{(1)} = 56 r}$.

Now we study admissibility. To discuss the solutions, let us introduce the following notation :
\begin{small}
\be \label{71c}
P^{F_{4,1}}_{c_{(1)} = 104 r} = q^{-\frac{13 r}{3}} \left(1 + \tilde{a}_1^{(r)}\,q + \tilde{a}_2^{(r)}\,q^2 + \tilde{a}_3^{(r)}\,q^3 + \ldots \right).
\ee
\end{small}

\underline{$N = 20, ~c_{(1)} = 104$} : Here, we have a unique solution, whose $q$-series features
\begin{small}
\begin{align*}
\tilde{a}_3^{(1)} = - 2541171568, \quad \tilde{a}_4^{(1)} = - 2229771385010, \ldots.
\end{align*}
\end{small}
This is clearly not an admissible character and hence cannot be a character of a CFT.  We conclude that there are \emph{no one-character CFTs possible at central charge $104$} that are one-character extensions of tensor products of the $F_{4,1}$ CFT (see footnote \ref{foot}). 

\underline{$N = 40, ~c_{(1)} = 208$} : Here, we have a unique solution, whose $q$-series features
\begin{small}
\begin{align*}
\tilde{a}_3^{(1)} = - 3998039136, \quad \tilde{a}_4^{(1)} = - 9473425941460, \ldots.
\end{align*}
\end{small}
This is clearly not an admissible character and hence cannot be a character of a CFT.  We conclude that there are \emph{no one-character CFTs possible at central charge $208$} that are one-character extensions of tensor products of the $F_{4,1}$ CFT (see footnote \ref{foot}). 

\underline{$N = 60, ~c_{(1)} = 312$} :  From table \ref{t6}, the solution is :
\be \label{72c}
P^{F_{4,1}}_{c_{(1)} = 312} = P_1^3 + \tilde{p}_1 P_2^5.
\ee
Computing, we find for example 
\be
\tilde{a}_3^{(3)} = 208 (57122 \tilde{p}_1-15604513),  \qquad \tilde{a}_4^{(3)} = 390 (88881832 \tilde{p}_1-51383203665), \ldots
\ee
Requiring $\tilde{a}_n^{(3)}$ to be a non-negative integer gives us multiple inequalities for $\tilde{p}$. We solved all these inequalities for  $n \leq 500$ and found the following solution :
\be \label{73c}
\tilde{p}_1 \geq 4958.
\ee
We seem to have obtained an infinite number of admissible characters in \eqref{72c} and \eqref{73c}. We are not sure if any of them correspond to genuine CFTs (see footnote \ref{foot}).

\underline{$N = 80, ~c_{(1)} = 416$} : From table \ref{t6}, the solution is :
\be \label{74c}
P^{F_{4,1}}_{c_{(1)} = 416} = P_1^4 + \tilde{p}_1 P_1 P_2^5.
\ee
We found that requiring $\tilde{a}_1^{(4)}, \ldots  \tilde{a}_9^{(4)}$ to be non-negative and integral gives $\tilde{p}_1 \geq 810$ while requiring $\tilde{a}_{10}^{(4)}$ to be non-negative and integral gives $\tilde{p}_1 \leq 128$. Hence, \eqref{74c} is not an admissible character for any value of $\tilde{p}_1$ and thus cannot be a character of a CFT.  We conclude that there are \emph{no one-character CFTs possible at central charge $416$} that are one-character extensions of tensor products of the $F_{4,1}$ CFT (see footnote \ref{foot}).

\underline{$N = 100, ~c_{(1)} = 520$} : From table \ref{t6}, the solution is :
\be \label{75c}
P^{F_{4,1}}_{c_{(1)} = 520} = P_1^5 + \tilde{p}_1 P_1^2 P_2^5.
\ee
We found that requiring $\tilde{a}_1^{(5)}, \ldots  \tilde{a}_9^{(5)}$ to be non-negative and integral gives $\tilde{p}_1 \geq 958$ while requiring $\tilde{a}_{10}^{(5)}$ to be non-negative and integral gives $\tilde{p}_1 \leq  549$. Hence, \eqref{75c} is not an admissible character for any value of $\tilde{p}_1$ and  thus cannot be a character of a CFT.  We conclude that there are \emph{no one-character CFTs possible at central charge $520$} that are one-character extensions of tensor products of the $F_{4,1}$ CFT (see footnote \ref{foot}). 

\underline{$N = 120, ~c_{(1)} = 624$} : From table \ref{t6}, the solution is :
\be \label{76c}
P^{F_{4,1}}_{c_{(1)} = 624} = P_1^6 + \tilde{p}_1 P_1^3 P_2^5 + \tilde{p}_2  P_2^{10}.
\ee
Requiring the non-negativity and integrality of each $\tilde{a}_n^{(6)}$ gives us some region of the positive quadrant of the $\tilde{p}_1-\tilde{p}_2$ plane. We studied upto $n = 500$ and we found multiple regions that are solutions to all the constraints; thus we seem to have obtained an infinite number of admissible characters. We are not sure if any of them correspond to genuine CFTs (see footnote \ref{foot}).

\underline{$N = 140, ~c_{(1)} = 728$} : From table \ref{t6}, the solution is :
\be \label{77c}
P^{F_{4,1}}_{c_{(1)} = 728} = P_1^7 + \tilde{p}_1 P_1^4 P_2^5 + \tilde{p}_2  P_1 P_2^{10}.
\ee
In this case, we analysed  the constraints coming from requiring the non-negativity and integrality of $\tilde{a}_n^{(7)}$ for $n \leq 17$. We found no common solutions to all the constraints. Hence we conclude that \eqref{77c} is not an admissible character for any value of $\tilde{p}_1, \tilde{p}_2 $ and thus cannot be a character of a CFT. We conclude that there are \emph{no one-character CFTs possible at central charge $728$} that are one-character extensions of tensor products of the $F_{4,1}$ CFT (see footnote \ref{foot}). 

\underline{$N = 160, ~c_{(1)} = 832$} : From table \ref{t6}, the solution is :
\be \label{78c}
P^{G_{2,1}}_{c_{(1)} = 448} = P_1^8 + \tilde{p}_1 P_1^5 P_2^5 + \tilde{p}_2  P_1^2 P_2^{10}.
\ee
In this case, we analysed  the constraints coming from requiring the non-negativity and integrality of $a_n^{(8)}$ for $n \leq 17$. We found no common solutions to all the constraints. Hence we conclude that \eqref{78c} is not an admissible character for any value of $\tilde{p}_1, \tilde{p}_2 $ and thus cannot be a character of a CFT. We conclude that there are \emph{no one-character CFTs possible at central charge $832$} that are one-character extensions of tensor products of the $F_{4,1}$ CFT (see footnote \ref{foot}).

\subsection{$\mathcal C = D_{4,1}$ \label{34s}}

$D_{4,1}$ is a $\mathbf{[2,0]}$ CFT with central charge $4$. Let us denote it's two characters by $\chi_0, \chi_1$, the latter is the non-identity character with conformal dimension $h_1 = \frac12$. Let us note the $q$-series for the characters :
\begin{small}
\begin{align}
     \chi_0& =q^{-\frac{1}{6}} \left(1+28 q+134 q^2+568 q^3+1809 q^4+5316 q^5+13990 q^6+34696 q^7+80724 q^8  + \ldots \right) \nonumber \\
     \chi_1& =q^{\frac{1}{3}} \left( 8+64 q+288 q^2+1024 q^3+3152 q^4+8704 q^5+22208 q^6+53248 q^7 + \ldots \right) \label{42b}
\end{align}
\end{small}
and the modular $S$-transformations for the characters :
\be \label{43b}
S(\chi_0) = \frac{\chi_0 + 3 \chi_1}{2}, \qquad \qquad S(\chi_1) = \frac{\chi_0 - \chi_1}{2}.
\ee

The allowed values of $k$ and $N$ can be read from entry 4 of table \ref{t2}. Here, every possible $c_{(1)} = 8k$ is possible and the corresponding number of tensor product factors is  $N = 2k$.

We will be brief in our description of the results here. The analysis proceeds very similar to $A_{1,1}$. We discover the first $S$-invariant polynomial when we are working with $N = 2, c_{(1)} = 8$. This is a quadratic non-constant $S$-invariant polynomial. Just as for the $A_{1,1}$ case in section \ref{31s}, we discover the second $S$-invariant polynomial when working at $c_{(1)} = 24$, which happens when $N = 6$.  This is a constant $S$-invariant polynomial of degree-$3$. The two  $S$-invariant polynomials for $D_{4,1}$ are:
\begin{equation}\label{44b}
\begin{array}{llr}
   P_1(\chi_0, \chi_1) &=\chi_0^2+3 \chi_1^2 = j^{\frac{1}{3}}\\
   P_2(\chi_0, \chi_1) &=\chi_0^2 \chi_1-\chi_1^3 = 8
\end{array}
\end{equation}

We have implemented the method outlined in section \ref{2s}  for $k = 1, 2, \ldots 16$.  For $c_{(1)} = 8k$ the character has $[\frac{k}{3}]$ arbitrary constants (the $p$’s) and there are $1 + [\frac{k}{3}]$ $S$-invariant summands the ${\cal Z}$'s. Each summand, when expressed in terms of the characters $\chi_0, \chi_1$ are polynomials of degree $2k$. But when expressed in terms of the $S$-invariant polynomials \eqref{44b}, they are polynomials of much smaller degree.  Because of this, we can handle in a compact manner,  computations upto $c_{(1)} = 128$ for which we have  polynomials of degree $16$ only.  The result for $N = 32, c_{(1)} = 128, k = 16$ is :
\be \label{83c}
P^{D_{4,1}}_{c_{(1)} = 128}(\chi_0, \chi_1)  = {\cal Z}_0  +  p_1 \,{\cal Z}_1 +  p_2 \,{\cal Z}_2  +  p_3 \,{\cal Z}_3 +  p_4 \,{\cal Z}_4  +  p_5 \,{\cal Z}_5
\ee
with 
\begin{small}
\begin{align}
     {\cal Z}_0& = P_1^{16}-48 P_1^{13} P_2^2+696 P_1^{10} P_2^4-3216 P_1^7 P_2^6+3420 P_1^4 P_2^8-1968 P_1 P_2^{10}  \nonumber \\
     {\cal Z}_1& = P_1^{13} P_2^2-37 P_1^{10} P_2^4+337 P_1^7 P_2^6-645 P_1^4 P_2^8-427 P_1 P_2^{10}\nonumber \\
     {\cal Z}_2& =  P_1^{10} P_2^4-26 P_1^7 P_2^6+99 P_1^4 P_2^8-164 P_1 P_2^{10} \nonumber \\
     {\cal Z}_3& =  P_1^7 P_2^6-15 P_1^4 P_2^8-18 P_1 P_2^{10} \nonumber \\
     {\cal Z}_4& =  P_1^4 P_2^8-4 P_1 P_2^{10}  \nonumber \\
     {\cal Z}_5& = P_1 P_2^{10}  .
     \end{align}
\end{small}
In the following, we are able to rewrite the same in an even more compact manner :
\be \label{85c}
P^{D_{4,1}}_{c_{(1)} = 128}(\chi_0, \chi_1)  = \tilde{{\cal Z}}_0  +  \tilde{p}_1 \,\tilde{{\cal Z}}_1 +  \tilde{p}_2 \,\tilde{{\cal Z}}_2  +  \tilde{p}_3 \,\tilde{{\cal Z}}_3 +  \tilde{p}_4 \,\tilde{{\cal Z}}_4  +  \tilde{p}_5 \,\tilde{{\cal Z}}_5
\ee
with 
\begin{small}
\begin{align}
     \tilde{{\cal Z}}_0& = P_1^{16}, \quad \tilde{{\cal Z}}_1 = P_1^{13}P_2^2, \quad  \tilde{{\cal Z}}_2 =  P_1^{10}P_2^4, \quad  \tilde{{\cal Z}}_3 =  P_1^7 P_2^6, \quad  \tilde{{\cal Z}}_4 =  P_1^4 P_2^{8}, \quad   \tilde{{\cal Z}}_5 =  P_1 P_2^{10}  
     \end{align}
\end{small}
The relation between the constant coefficients of \eqref{83c} and \eqref{85c} is:
\begin{small}
\begin{align}
         \tilde{p}_1& = p_1 - 48, \quad  \tilde{p}_2 = p_2 - 37 p_1 + 696, \quad \tilde{p}_3 = p_3 - 26 p_2 + 337 p_1 - 3216, \nonumber \\
     \tilde{p}_4& =  p_4 - 15 p_3  + 99 p_2 - 645 p_1 + 3420, \quad  \tilde{p}_5 =  p_5 - 4 p_4  - 18 p_3 -164 p_2  - 427 p_1 - 1968. \nonumber 
     \end{align}
\end{small}
Since, the most compact form is that of \eqref{85c}, we will report this form of the results in table \ref{t7}.  The way to read the results is as follows.   The first column is the serial no. which is also the value of $k$, from which one can infer the value of $c_{(1)} = 8k$.   The second column contains the number of tensor product factors $N$ (which equals $2k$).  The third column contains the character $P^{D_{4,1}}_{c_{(1)} = 8k}(\chi_0, \chi_1)$.  The number of summands ranges from $1$, for $k = 1$, to $6$ for $k = 16$. For $k = 16$, the last entry of the table \ref{t7} we have six $S$-invariant summands $\tilde{{\cal Z}}_0, \tilde{{\cal Z}}_1, \ldots \tilde{{\cal Z}}_5$ (see \eqref{85c}).   The expressions for the $\tilde{{\cal Z}}$'s when written in terms of the characters  $\chi_0, \chi_1$ are very voluminous. Each of them is a  degree $N$ polynomial.  For example for $N= 32$, the last entry of table \ref{t7}, we have  six homogenous degree $32$ polynomials. But, when written in terms of the $S$-invariant polynomials \eqref{44b}, they are in fact polynomials of much smaller degree. For $N = 32$, the highest degree term that occurs is $16$.  This is one of the results to report in this paper. Solving the huge linear system of equations that follow from  $S$-invariance,  organising in terms of some number of $S$-invariant summands and writing each of  the summands in terms of $S$-invariant polynomials and thus achieving compact expressions.

Now we need to impose admissibility. For that, we first need to compute the $q$-series of $P^{D_{4,1}}_{c_{(1)} = 8k}$. We employ the explicit $q$-series of the characters available in \eqref{42b} and give the  $q$-series of $P^{D_{4,1}}_{c_{(1)} = 8k}$ in the  fourth column of table \ref{t7}. Then we need to compare this with the generic admissible character for that particular central charge, given in table \ref{tapp} of the appendix \ref{app1}. The generic admissible character for central charge $8k$ has $[\frac{k}{3}]$ parameters which are denoted by $N_1, N_2, \ldots$. And these $N$’s need to satisfy a set of nested inequalities for admissibility, which are available from table \ref{tapp} of the appendix \ref{app1}.  In the fourth column of table \ref{t7}, we give the relation between the $\tilde{p}$’s and the $N$’s. Then, the nested inequalities for the $N$’s translate to inequalities for the $\tilde{p}$’s. We always find an infinite number of admissible character solutions, except  for $N = 2, c_{(1)} = 8$ and $N = 4, c_{(1)} = 16$.

Amongst all the admissible characters we have computed in table \ref{t7}, there are only three of them which are known to correspond to actual CFTs. The first one is in the first row, $N = 2, c_{(1)} = 8$, and this corresponds to the $E_{8,1}$ CFT.  The second one is the second row, $N = 4, c_{(1)} = 16$, and this corresponds to two CFTs viz. $E_{8,1}^{\otimes 2}$ and ${\cal E}_1[D_{16,1}]$. The third one is one of the infinite characters of the third row, $N = 6, c_{(1)} = 24$. Reading off from the fourth column of the third row, we have the character $j + 64 \tilde{p}_1$. Among these infinite characters, we can identify the one which corresponds to a CFT by invoking the classification of $c_{(1)} = 24$ CFTs of \cite{Schellekens:1992db}.  In the table at the end of that paper entry no. 42 is the only CFT which is a one-character extension of a tensor product of  copies of $D_{4,1}$\footnote{The reader can find this CFT listed as entry no. 17 of table \ref{t0}.}. This CFT has a character  $j - 576$. Thus amongst the infinite characters $j + 64 \tilde{p}_1$ that we found, only one of them, the one with $\tilde{p}_1 = -9$ corresponds to an actual CFT.

\underline{Lessons learnt and a new method} :  All the results that we have obtained and reported in table \ref{t7} were by running the whole procedure as laid out  in section \ref{2s} . Similar to what we did in section \ref{31s}, we have a simpler  procedure that allows us to write the final answer quickly.  We first note that $P_1$ is a homogenous polynomial of the characters of degree $2$ and that $P_2$ is a homogenous polynomial of the characters of degree $3$. A monomial of the form $P_1^a\,P_2^b$ is a homogenous polynomial of the characters of degree $2a + 3b$. And for this monomial to be present in $P^{D_{4,1}}_{c_{(1)} = 8k}$ it’s degree needs to be equal to $N$. Thus we look for  non-negative integral solutions to 
\be \label{87c}
2a + 3b = N = 2 k, 
\ee
we can obtain all the $\tilde{{\cal Z}}$'s  in $P^{D_{4,1}}_{c_{(1)} = 8k}$.  All except one of the $\tilde{{\cal Z}}$'s  viz. $\tilde{{\cal Z}}_0$ are multiplied by constants the $\tilde{p}$'s.  Which of the solutions of \eqref{87c} corresponds to $\tilde{{\cal Z}}_0$ is easily decided by seeing which of them contains the identity character $\chi_0^N$. Here it is the $a = k, b = 0$ solution. This quick computation of the solutions to \eqref{87c} gives us the same answer as the long procedure.  There is one caveat. We should only consider solutions of \eqref{87c} that give monomials of characters which are commensurate to the identity character. Here, in every solution, we find that $b$ is even which makes all the monomials of characters present in $P_2^b$ to be commensurate to the identity character. It is easy to see that $b$ of every solution of \eqref{87c} is always even when we rewrite as $3b = 2k - 2a$ ; the right hand side is an even number from which it follows that $b$ is even. 

\begin{table}[H]
\begin{center}
\begin{threeparttable}
\rowcolors{2}{Mywhite}{Mygrey}
\resizebox{\textwidth}{!}{
   \renewcommand{\arraystretch}{1.3}
   $\begin{array}{|c|c|c|C{3cm}|C{9cm}|}
\hline
\# &       N & c_{(1)} & P_{c_{(1)} = 8k}^{D_{4,1}} (\chi_0, \chi_1)  &  q\text{-series}  \\
\hline
1. &       \mathcal E_1[D_{4,1}^{~\otimes 2}] & 8 & {\cal \tilde{Z}}_0 = P_1 & j^{\frac{1}{3}}   \\
\hline
2. &    \mathcal E_1[D_{4,1}^{~\otimes 4}] & 16 & {\cal \tilde{Z}}_0 = P_1^2 & j^{\frac{2}{3}}  \\
\hline
3. &   \mathcal E_1[D_{4,1}^{~\otimes 6}] & 24 & {\cal \tilde{Z}}_0 = P_1^3,&  j+N  \\
&& & {\cal \tilde{Z}}_1 = P_2^2 & N_{1}=64 \tilde{p}_1\\
\hline
4. &   \mathcal E_1[D_{4,1}^{~\otimes 8}] & 32 & {\cal \tilde{Z}}_0 = P_1^4,&   j^{\frac{1}{3}}(j+N) \\
&& & {\cal \tilde{Z}}_1 = P_1 P_2^2  & N_{1}=64 \tilde{p}_1\\
\hline
5. &   \mathcal E_1[D_{4,1}^{~\otimes 10}] & 40 & {\cal \tilde{Z}}_0 = P_1^5,&   j^{\frac{2}{3}}(j+N) \\
&& & {\cal \tilde{Z}}_1 = P_1^2 P_2^2  &N_{1}=64 \tilde{p}_1\\
\hline
6. &   \mathcal E_1[D_{4,1}^{~\otimes 12}] & 48 & {\cal \tilde{Z}}_0 =P_1^6,&  j^2+N_1j+N_2  \\
&& & {\cal \tilde{Z}}_1 = P_1^3 P_2^2,& N_{1}=64 \tilde{p}_1,\\
&& & {\cal \tilde{Z}}_2 = P_2^4  & \  N_{2}=4096 \tilde{p}_2\\
\hline
7. &   \mathcal E_1[D_{4,1}^{~\otimes 14}] & 56 & {\cal \tilde{Z}}_0 = P_1^7,&   j^{\frac{1}{3}}(j^2+N_1j+N_2) \\
&& & {\cal \tilde{Z}}_1 =P_1^4 P_2^2,& N_{1}=64 \tilde{p}_1,\\
&& & {\cal \tilde{Z}}_2 = P_1 P_2^4& \  N_{2}=4096 \tilde{p}_2\\
\hline
  8. &  \mathcal E_1[D_{4,1}^{~\otimes 16}] & 64 & {\cal \tilde{Z}}_0 =P_1^8,&   j^{\frac{2}{3}}(j^2+N_1j+N_2) \\
&&& {\cal \tilde{Z}}_1 = P_1^5 P_2^2,&  N_{1}=64 \tilde{p}_1,\\
&&& {\cal \tilde{Z}}_2 = P_1^2 P_2^4 & \  N_{2}=4096 \tilde{p}_2\\
\end{array}$
}
\end{threeparttable}
\end{center}
\end{table}

\begin{table}[H]
\begin{center}
\begin{threeparttable}
\rowcolors{2}{Mywhite}{Mygrey}
\resizebox{\textwidth}{!}{
   \renewcommand{\arraystretch}{1.3}
   $\begin{array}{|c|c|c|C{3cm}|C{7cm}|}
9. &   \mathcal E_1[D_{4,1}^{~\otimes 18}] & 72 & {\cal \tilde{Z}}_0 = P_1^9,&   j^3+N_1j^2+N_2j+N_3 \\
&&& {\cal \tilde{Z}}_1 = P_1^6 P_2^2,& N_{1}=64 \tilde{p}_1,\\
&&& {\cal \tilde{Z}}_2 = P_1^3 P_2^4,&  \  N_{2}=4096 \tilde{p}_2,\\
&&& {\cal \tilde{Z}}_3 = P_2^6  &  \  N_{3}=262144 \tilde{p}_3\\
\hline
10. &    \mathcal E_1[D_{4,1}^{~\otimes 20}] & 80 & {\cal \tilde{Z}}_0 =P_1^{10},&  j^{\frac{1}{3}}(j^3+N_1j^2+N_2j+N_3)  \\
&&& {\cal \tilde{Z}}_1 = P_1^7 P_2^2,&N_{1}=64 \tilde{p}_1,\\
&&& {\cal \tilde{Z}}_2 = P_1^4 P_2^4,&  N_{2}=4096 \tilde{p}_2,\\
&&& {\cal \tilde{Z}}_3 = P_1 P_2^6  &  N_{3}=262144 \tilde{p}_3\\
\hline
11. &    \mathcal E_1[D_{4,1}^{~\otimes 22}] & 88 & {\cal \tilde{Z}}_0 = P_1^{11},&   j^{\frac{2}{3}}(j^3+N_1j^2+N_2j+N_3) \\
&&& {\cal \tilde{Z}}_1 =P_1^8 P_2^2,& N_{1}=64 \tilde{p}_1,\\
&&& {\cal \tilde{Z}}_2 =P_1^5 P_2^4,&   N_{2}=4096 \tilde{p}_2,\\
&&& {\cal \tilde{Z}}_3 = P_1^2 P_2^6 &   N_{3}=262144 \tilde{p}_3\\
\hline
12. &   \mathcal E_1[D_{4,1}^{~\otimes 24}] & 96 & {\cal \tilde{Z}}_0 = P_1^{12},&  j^4+N_1j^3+N_2j^2+N_3j+N_4  \\
&&&  {\cal \tilde{Z}}_1 =P_1^9 P_2^2,&N_{1}=64 \tilde{p}_1,\\
&&&  {\cal \tilde{Z}}_2 =P_1^6 P_2^4,&  N_{2}=4096 \tilde{p}_2,\\
&&& {\cal \tilde{Z}}_3 =P_1^3 P_2^6,&  N_{3}=262144 \tilde{p}_3,\\
&&&   {\cal \tilde{Z}}_4 =P_2^8  &   N_{4}=16777216 \tilde{p}_4\\
\hline
13. &   \mathcal E_1[D_{4,1}^{~\otimes 26}] & 104 & {\cal \tilde{Z}}_0 = P_1^{13},& j^{\frac{1}{3}}(j^4+N_1j^3+N_2j^2+N_3j+N_4)   \\
&&&  {\cal \tilde{Z}}_1 = P_1^{10}P_2^2,& N_{1}=64 \tilde{p}_1,\\
&&& {\cal \tilde{Z}}_2 =P_1^7 P_2^4,&   N_{2}=4096 \tilde{p}_2,\\ 
&&&  {\cal \tilde{Z}}_3 =P_1^4 P_2^6,&   N_{3}=262144 \tilde{p}_3,\\
&&&   {\cal \tilde{Z}}_4 =P_1 P_2^8  &    N_{4}=16777216 \tilde{p}_4\\
\hline
14. &   \mathcal E_1[D_{4,1}^{~\otimes 28}] & 112 & {\cal \tilde{Z}}_0 = P_1^{14},&  j^{\frac{2}{3}}(j^4+N_1j^3+N_2j^2+N_3j+N_4)  \\
&&&  {\cal \tilde{Z}}_1 = P_1^{11}P_2^2,&  N_{1}=64 \tilde{p}_1,\\
&&& {\cal \tilde{Z}}_2 = P_1^8P_2^4,&   N_{2}=4096 \tilde{p}_2,\\
&&&  {\cal \tilde{Z}}_3 =P_1^5 P_2^6,&   N_{3}=262144 \tilde{p}_3,\\
&&&  {\cal \tilde{Z}}_4 =P_1^2 P_2^8  &   N_{4}=16777216 \tilde{p}_4\\
\end{array}$
}
\end{threeparttable}
\end{center}
\end{table}

\begin{table}[H]
\begin{center}
\begin{threeparttable}
\rowcolors{2}{Mywhite}{Mygrey}
\resizebox{\textwidth}{!}{
   \renewcommand{\arraystretch}{1.3}
   $\begin{array}{|c|c|c|C{3cm}|C{7cm}|}
15. &   \mathcal E_1[D_{4,1}^{~\otimes 30}] & 120 &  {\cal \tilde{Z}}_0 =P_1^{15},&  j^5+N_1j^4+N_2j^3+N_3j^2+N_4j+N_5 \\
&&&  {\cal \tilde{Z}}_1 = P_1^{12}P_2^2,&N_{1}=64 \tilde{p}_1,\\
&&&  {\cal \tilde{Z}}_2 =P_1^9 P_2^4,&    N_{2}=4096 \tilde{p}_2,\\
&&&  {\cal \tilde{Z}}_3 =P_1^6 P_2^6,&    N_{3}=262144 \tilde{p}_3,\\ 
&&&  {\cal \tilde{Z}}_4 =P_1^3 P_2^8,&  N_{4}=16777216 \tilde{p}_4,\\
&&&  {\cal \tilde{Z}}_5 =P_2^{10}   &    N_{5}=1073741824 \tilde{p}_5\\
\hline
16. &   \mathcal E_1[D_{4,1}^{~\otimes 32}] & 128 &   {\cal \tilde{Z}}_0 =P_1^{16},&  j^{\frac{1}{3}}(j^5+N_1j^4+N_2j^3+N_3j^2+N_4j+N_5) \\
&&&  {\cal \tilde{Z}}_1 =P_1^{13} P_2^2,& N_{1}=64 \tilde{p}_1,\\
&&&  {\cal \tilde{Z}}_2 =P_1^{10} P_2^4,&    N_{2}=4096 \tilde{p}_2,\\
&&&  {\cal \tilde{Z}}_3 =P_1^7 P_2^6,&    N_{3}=262144 \tilde{p}_3,\\
&&&  {\cal \tilde{Z}}_4 =P_1^4 P_2^8,&    N_{4}=16777216 \tilde{p}_4,\\
&&&   {\cal \tilde{Z}}_5 =P_1 P_2^{10}&   N_{5}=1073741824 \tilde{p}_5\\
\hline
\end{array}$
}
\end{threeparttable}
\end{center}
\caption{Characters of one-character extensions of tensor products of $D_{4,1}$ written in a basis of $S$-invariant polynomials}\label{t7}
\end{table}

To conclude this section which only studied two-character WZW CFTs, we note that in every case, we have two $S$-invariant polynomials (\eqref{34b}, \eqref{38b}, \eqref{41b}, \eqref{44b}, \eqref{47b}, \eqref{50b}, \eqref{53b}). In every case, the existence of these $S$-invariant polynomials  have aided our computations heavily.  Take the character of any of the $\mathcal{E}_1[\mathcal{C}^{\otimes N}]$ in any of the tables of this section \ref{2s}. If the S-invariant polynomials did not exist, we would have them as polynomials of the basic characters $\chi_0, \chi_1$, their degree would be much higher, sometimes as high as $8$ times. 

Apart from contributing to the compactness of expressions, one asks why there are only two $S$-invariant polynomials.  Perhaps there is an answer from polynomial algebra.  Having to do with the fact that the basic characters, which are two in number, form some kind of a basis of generators, not just for the entire polynomial algebra but also for the sub-algebra of $S$-invariant polynomials.  Now, the polynomials $P_1$ and $P_2$, also seem to form a basis of generators for the sub-algebra of $S$-invariant polynomials. And since any two sets of bases should have the same cardinality, we perhaps have an explanation for the existence of two $S$-invariant polynomials $P_1$ and $P_2$ for two-character theories. This kind of argument will perhaps gain more strength after studying three-character CFTs in section \ref{4s}, where we find three $S$-invariant polynomials $P_1$, $P_2$ and $P_3$ for every three-character CFT studied there.

\section{ $\mathbf{[3,0]}$ CFTs : Tensor Products and One-character Extensions \label{4s}}

In this section, we study $\mathbf{[3,0]}$ CFTs. We implement the procedure described in section \ref{2s} for each of the following $\mathbf{[3,0]}$  CFTs. Except for the Ising CFT, all the CFTs we study are WZW CFTs. We start with the $A_{4,1}$ CFT in \ref{41s}. Then we take on the interesting example of the Ising CFT a.k.a ${\cal M}(4,3)$ in section \ref{42s}. Then we study the infinite class of CFTs $D_{r,1}$ in section \ref{43s}. The  $D_{r,1}$ CFTs can be separated into four infinite classes of CFTs. These are (i) when $r$ is a odd number, (ii) $r$ is $2$ times an odd number, (iii) $r$ is a $4$ times an odd number, (iv) $r$ is a multiple of $8$. We study these separately in sections \ref{431s}, \ref{432s}, \ref{433s}, \ref{434s}. All CFTs in  any of these four classes of CFTs  have a common answer which we will explain below. We thus are able to solve for all infinite CFTs $D_{r,1}$. Then in section \ref{44s} we study and solve the infinite class of CFTs $B_{r,1}$.

\subsection{$\mathcal C = A_{4,1}$ \label{41s}}

$A_{4,1}$ is a $\mathbf{[3,0]}$ CFT with central charge $4$. Let us denote it's three characters by $\chi_0, \chi_1, \chi_2$, the latter two  are the non-identity characters with conformal dimensions $h_1 = \frac25, h_2 = \frac35$ respectively. Let us note the $q$-series for the characters :
\begin{align}
     \chi_0& =q^{-\frac{1}{6}}(1+24 q+124 q^2+500 q^3+1625 q^4+4752 q^5+12524 q^6+31000 q^7+72250 q^8+ \ldots) \nonumber \\
     \chi_1& =q^{\frac{7}{30}}(5+50 q+220 q^2+820 q^3+2525 q^4+7070 q^5+18125 q^6+43780 q^7+100030 q^8+\ldots) \nonumber \\
     \chi_2& =q^{\frac{13}{30}}(10+65 q+300 q^2+1025 q^3+3140 q^4+8565 q^5+21760 q^6+51785 q^7+117500 q^8 + \ldots) \label{54b}
\end{align}
and the modular $S$-transformations for the characters :
\begin{equation}
   S(\chi_0)=\frac{\chi_0}{\sqrt{5}}+\frac{2 \chi_1}{\sqrt{5}}+\frac{2 \chi_2}{\sqrt{5}},
   \qquad S(\chi_1)=\frac{\chi_0}{\sqrt{5}}+\frac{\chi_1}{\sqrt{5}\,\varphi}-\frac{\varphi \, \chi_2}{\sqrt{5}},
   \qquad S(\chi_2)=\frac{\chi_0}{\sqrt{5}}-\frac{\varphi\, \chi_1}{ \sqrt{5}}+\frac{ \chi_2}{ \sqrt{5}\,\varphi}. \label{55b}
\end{equation}
where $\varphi = \frac{1+\sqrt{5}}{2}$. The allowed values of $k$ and $N$ can be read from entry 2 of table \ref{t3}. Here, every possible $c_{(1)} = 8k$ occurs and the corresponding number of tensor product factors is  $N = 2k$. 

We follow all the steps laid out in section \ref{2s}.  We solve the set of equations \eqref{7b}, \eqref{8b} and construct the ansatz character \eqref{10b}. We cannot come up with a closed formula for $b$ the number of monomials with $q$-series commensurate to the identity character. But we can readily compute them. And we can set up the overdetermined system of linear equations and find solutions. 

We start with smallest possible $c_{(1)}$. This is the $N = 2,  ~ c_{(1)} = 8$ case, whose solution gives us our first $S$-invariant polynomial $P_1$. This is a non-constant quadratic polynomial. The solution for $N = 4,~  c_{(1)} = 16$ is simply  $P_1^2$. While solving the $N = 6, ~ c_{(1)} = 24$, we discover the second $S$-invariant polynomial $P_2$ which is of degree-$6$  and computes (using \eqref{54b}) to a constant. Then, while solving for  $N = 10, ~ c_{(1)} = 40$, we discover the third $S$-invariant polynomial $P_3$ which is of degree-$10$  and computes (using \eqref{54b}) to a constant.  We collect all the three $S$-invariant polynomials for $A_{4,1}$ below :\\
\begin{small}
\begin{equation}\label{56a}
\begin{array}{lll}
   P_1(\chi_0, \chi_1, \chi_2)&=\chi_0^2 \,+\, 4 \,\chi_1\,\chi_2 ~ = ~j^{\frac{1}{3}}\\
   P_2(\chi_0, \chi_1, \chi_2) &=\chi_0^4 \,\chi_1\, \chi_ 2\,-\,\chi_0^2 \,\chi_1^2 \,\chi_2^2\,-\,\chi_0\, \chi_1^5\,-\,\chi_0\, \chi_2^5\,+\,2\, \chi_1^3\, \chi_2^3 ~ =~ 50\\
   P_3(\chi_0, \chi_1, \chi_2) &=5\, \chi_0^6\, \chi_1^2 \,\chi_2^2 \,-\,4 \,\chi_0^5\, \chi_1^5\,-\,4\, \chi_0^5\, \chi_2^5\,-\,10\, \chi_0^4\, \chi_1^3 \,\chi_2^3\,+\,10 \,\chi_0^3\, \chi_1^6\, \chi_2\,+\,10 \chi_0^3\, \chi_1 \,\chi_2^6\,+\,5\, \chi_0^2\, \chi_1^4 \,\chi_2^4\\& \,-\,10\, \chi_0\, \chi_1^7\, \chi_2^2\,-\,10\, \chi_0\, \chi_1^2\, \chi_2^7\,+\,\chi_1^{10}\,+\,6\, \chi_1^5\, \chi_2^5\,+\,\chi_2^{10}~=~ 0
\end{array}
\end{equation}
\end{small}
We then computed upto $N = 24,~  c_{(1)} = 96$ and we could always express the character $P_{c_{(1)} = 8k}^{A_{4,1}}$ as a polynomial of these three $S$-invariant polynomials. There does not seem to be any more new $S$-invariant polynomials; neither is it  there nor do we need it.  The result for $N = 24, c_{(1)} = 96$ is :
\be \label{91c}
P^{A_{4,1}}_{c_{(1)} = 96}(\chi_0, \chi_1)  = {\cal Z}_0  +  \sum_{i = 1}^{8}\, p_i \,{\cal Z}_i 
\ee
with 
\begin{small}
\begin{align}
     {\cal Z}_0& = P_1^{12}-48 P_1^9 P_2+12 P_1^7 P_3+564 P_1^6 P_2^2-156 P_1^4 P_2 P_3-1252 P_1^3 P_2^3+3 P_1^2 P_3^2+258 P_1 P_2^2 P_3-789 P_2^4,  \nonumber \\
     {\cal Z}_1& = 16 P_1^9 P_2- 4 P_1^7 P_3 - 540 P_1^6 P_2^2 + 228 P_1^4 P_2 P_3 + 2588 P_1^3 P_2^3 - 23 P_1^2 P_3^2 + 134 P_1 P_2^2 P_3 -  4159 P_2^4, \nonumber \\
     {\cal Z}_2& = 16 P_1^6 P_2^2- 8 P_1^4 P_2 P_3- 156 P_1^3 P_2^3 +   P_1^2 P_3^2 + 150 P_1 P_2^2 P_3 - 599 P_2^4,\nonumber \\
     {\cal Z}_3& = -  P_1^7 P_3 + 5 P_1^6 P_2^2 + 23 P_1^4 P_2 P_3 - 95 P_1^3 P_2^3 - 6 P_1^2 P_3^2 + 26 P_1 P_2^2 P_3- 200 P_2^4,\nonumber \\
     {\cal Z}_4& = 4 P_1^3 P_2^3 - 3 P_1 P_2^2 P_3 - 21 P_2^4,  ~~{\cal Z}_5 =  - 4 P_1^4 P_2 P_3 + 20 P_1^3 P_2^3 +  P_1^2 P_3^2 + 30 P_1 P_2^2 P_3 - 95 P_2^4  \nonumber \\
     {\cal Z}_6& = P_2^4, ~~{\cal Z}_7 = 5 P_2^4 -  P_1 P_2^2 P_3, ~~{\cal Z}_8 =  P_1^2 P_3^2 - 10 P_1 P_2^2 P_3 + 25 P_2^4.
     \end{align}
\end{small}
In the following, we are able to rewrite the same in an even more compact manner :
\be \label{93c}
P^{A_{4,1}}_{c_{(1)} = 96}(\chi_0, \chi_1)  = \tilde{{\cal Z}}_0  +  \sum_{i = 1}^{8}\, \tilde{p}_i \,\tilde{{\cal Z}}_i 
\ee
with 
\begin{small}
\begin{align}
     \tilde{{\cal Z}}_0 &= P_1^{12}, \quad \tilde{{\cal Z}}_1 = P_1^9 P_2,  \quad \tilde{{\cal Z}}_2 =  P_1^7 P_3, \quad  \tilde{{\cal Z}}_3 =  P_1^6 P_2^2,  \quad  \tilde{{\cal Z}}_4 =  P_1^4 P_2 P_3, \nonumber \\     \tilde{{\cal Z}}_5 &=  P_1^3 P_2^3,  \quad  \tilde{{\cal Z}}_6 = P_1^2 P_3^2,  \quad  \tilde{{\cal Z}}_7 = P_1 P_2^2 P_3, \quad  \tilde{{\cal Z}}_8 =P_2^4.  
     \end{align}
\end{small}
The relation between the constant coefficients of \eqref{91c} and \eqref{93c} is:
\begin{small}
\begin{align}
         \tilde{p}_1& = - 48 + 16 p_1, \quad  \tilde{p}_2 =  12 - 4 p_1 - p_3, \quad \tilde{p}_3 = 564-540p_1+16p_2+ 5p_3, \nonumber \\
     \tilde{p}_4& = -156+228p_1-8p_2+23p_3-4p_5 , \quad  \tilde{p}_5 = -1252+2588p_1-156p_2-95p_3+4p_4+20p_5 , \nonumber \\
      \tilde{p}_6& = 3-23p_1+p_2-6p_3+p_5+p_8  , \quad  \tilde{p}_7 = 258+134p_1+150p_2+26p_3+30p_5-p_7-10p_8, \nonumber \\
      \tilde{p}_8& =-789-4159p_1-599p_2-200p_3-24p_4-95p_5+p_6+25p_8. 
     \end{align}
\end{small}
Since, the most compact form is that of \eqref{93c}, we will report this form of the results in table \ref{t8}.  The way to read the results is as follows.   The first column is the serial no. which is also the value of $k$. The second column contains the number of tensor product factors $N$ (which equals $2k$). The third column contains the value of $c_{(1)} = 8k$.  The fourth column contains the character $P^{A_{4,1}}_{c_{(1)} = 8k}(\chi_0, \chi_1)$.  The number of summands ranges from $1$ for $k = 1$, to $9$ for $k = 12$. For $k = 12$, the last entry of the table \ref{t8} we have nine $S$-invariant summands $\tilde{{\cal Z}}_0, \tilde{{\cal Z}}_1, \ldots \tilde{{\cal Z}}_8$ (see \eqref{93c}).   The expressions for the $\tilde{{\cal Z}}$'s when written in terms of the characters  $\chi_0, \chi_1$ are very voluminous. Each of them is a  degree $N$ polynomial.  For example for $N= 24$, the last entry of table \ref{t8}, we have  nine homogenous degree-$24$ polynomials. But, when written in terms of the $S$-invariant polynomials \eqref{56a}, they are in fact polynomials of much smaller degree. For $N = 24$, the highest degree term that occurs is $12$.  This is one of the results to report in this paper. Solving the huge linear system of equations that follow from  $S$-invariance,  organising in terms of some number of $S$-invariant summands and writing each of  the summands in terms of $S$-invariant polynomials and thus achieving compact expressions. 

Now we need to impose admissibility. For that, we first need to compute the $q$-series of $P^{A_{4,1}}_{c_{(1)} = 8k}$. We employ the explicit $q$-series of the characters available in \eqref{54b} and give the  $q$-series of $P^{A_{4,1}}_{c_{(1)} = 8k}$ in the  fifth column of table \ref{t8}. Then we need to compare this with the generic admissible character for that particular central charge, given in table \ref{tapp} of the appendix \ref{app1}. The generic admissible character for central charge $8k$ has $[\frac{k}{3}]$ parameters which are denoted by $N_1, N_2, \ldots$. And these $N$’s need to satisfy a set of nested inequalities for admissibility, which are available from table \ref{tapp} of the appendix \ref{app1}.  In the fifth column of table \ref{t8}, we give the relation between the $\tilde{p}$’s and the $N$’s. Then, the nested inequalities for the $N$’s translate to inequalities for the $\tilde{p}$’s. We always find an infinite number of admissible character solutions, except  for $N = 2, c_{(1)} = 8$ and $N = 4, c_{(1)} = 16$.

Amongst all the admissible characters we have computed in table \ref{t8}, there are only three of them which are known to correspond to actual CFTs. The first one is in the first row, $N = 2, c_{(1)} = 8$, and this corresponds to the $E_{8,1}$ CFT.  The second one is the second row, $N = 4, c_{(1)} = 16$, and this corresponds to two CFTs viz. $E_{8,1}^{\otimes 2}$ and ${\cal E}_1[D_{16,1}]$. The third one is one of the infinite characters of the third row, $N = 6, c_{(1)} = 24$.  Reading off from the fifth column of the third row, we have the character $j + 50 \tilde{p}_1$. Among these infinite characters, we can identify the one which corresponds to a CFT by invoking the classification of $c_{(1)} = 24$ CFTs of \cite{Schellekens:1992db}.  In the table at the end of that paper entry no. 37 is the only CFT which is a one-character extension of a tensor product of  copies of $A_{4,1}$\footnote{The reader can find this CFT listed as entry no. 14 of table \ref{t0}.}.  This CFT has a character  $j - 600$. Thus amongst the infinite characters $j + 50 \tilde{p}_1$ that we found, only one of them, the one with $\tilde{p}_1 = -12$ corresponds to an actual CFT.

\underline{Lessons learnt and a new method} :  All the results that we have obtained and reported in table \ref{t8} were by running the whole procedure as laid out  in section \ref{2s} . Similar to what we did in section \ref{3s}, we have a simpler  procedure that allows us to write the final answer quickly.  We first note that $P_1$ is a homogenous polynomial of the characters of degree $2$,   $P_2$ is a homogenous polynomial of the characters of degree $6$ and $P_3$ is a homogenous polynomial of the characters of degree $10$.  A monomial of the form $P_1^a\,P_2^b\,P_3^c$ is a homogenous polynomial of the characters of degree $2a + 6b+10c$. And for this monomial to be present in $P^{A_{4,1}}_{c_{(1)} = 8k}$ it’s degree needs to be equal to $N$. Thus we look for  non-negative integral solutions to 
\be \label{96c}
2a + 6b+10c = N = 2 k, 
\ee
we can obtain all the $\tilde{{\cal Z}}$'s  in $P^{A_{4,1}}_{c_{(1)} = 8k}$.  All except one of the $\tilde{{\cal Z}}$'s  viz. $\tilde{{\cal Z}}_0$ are multiplied by constants the $\tilde{p}$'s.  Which of the solutions of \eqref{96c} corresponds to $\tilde{{\cal Z}}_0$ is easily decided by seeing which of them contains the identity character $\chi_0^N$. Here it is the $a = k, b = 0, c=0$ solution. This quick computation of the solutions to \eqref{96c} gives us the same answer as the long procedure.  There is one caveat. We should only consider solutions of \eqref{96c} that give monomials of characters which are commensurate to the identity character. This caveat does not bother us because each of $P_1$, $P_2$ and $P_3$ only gives monomials of characters which are commensurate with the identity character and hence any solution to \eqref{96c} also gives only monomials of characters which are commensurate to the identity character. 

\begin{table}[H]
\begin{center}
\begin{threeparttable}
\rowcolors{2}{Mywhite}{Mygrey}
\resizebox{\textwidth}{!}{
   \renewcommand{\arraystretch}{1.3}
   $\begin{array}{|c|c|c|C{5cm}|C{5.5cm}|}
\hline
 \#(k) &      N & c_{(1)} & P_{c_{(1)} = 8k}^{A_{4,1}} (\chi_0, \chi_1, \chi_2) &  q\text{-series}  \\
\hline
  1. &     \mathcal E_1[A_{4,1}^{~\otimes 2}] & 8 &  {\cal \tilde{Z}}_0 = P_1 & j^{\frac{1}{3}}   \\
\hline
2. &    \mathcal E_1[A_{4,1}^{~\otimes 4}] &  16 &{\cal \tilde{Z}}_0 = P_1^2 & j^{\frac{2}{3}}  \\
\hline
3. &   \mathcal E_1[A_{4,1}^{~\otimes 6}] &  24 &{\cal \tilde{Z}}_0 =P_1^3,&  j+N  \\
&&& {\cal \tilde{Z}}_1 =P_2 &   N=50 \tilde{p}_1 \\
\hline
4. &   \mathcal E_1[A_{4,1}^{~\otimes 8}] & 32 & {\cal \tilde{Z}}_0 =P_1^4,&   j^{\frac{1}{3}}(j+N) \\
&&& {\cal \tilde{Z}}_1 =P_1 P_2 &  N=50 \tilde{p}_1  \\
\hline
5. &    \mathcal E_1[A_{4,1}^{~\otimes 10}] & 40 & {\cal \tilde{Z}}_0 =P_1^5,&   j^{\frac{2}{3}}(j+N) \\
&&& {\cal \tilde{Z}}_1 =P_1^2 P_2,\ {\cal \tilde{Z}}_2 =P_3  &  N=50 \tilde{p}_1  \\
\hline
6. &   \mathcal E_1[A_{4,1}^{~\otimes 12}] & 48 & {\cal \tilde{Z}}_0 =P_1^6,&  j^2+N_1j+N_2  \\
&&& {\cal \tilde{Z}}_1 =P_1^3 P_2,&   N_1 = 50 \tilde{p}_1, \\
&&& {\cal \tilde{Z}}_2 =P_1 P_3,\ {\cal \tilde{Z}}_3 =P_2^2& N_2 =2500 \tilde{p}_3\\
\hline
7. &   \mathcal E_1[A_{4,1}^{~\otimes 14}] & 56 & {\cal \tilde{Z}}_0 =P_1^7, &   j^{\frac{1}{3}}(j^2+N_1j+N_2) \\
&&& {\cal \tilde{Z}}_1 =P_1^4 P_2,& N_1 = 50 \tilde{p}_1,\\
&&& {\cal \tilde{Z}}_2 =P_1^2 P_3,\ {\cal \tilde{Z}}_3 =P_1 P_2^2& N_2 = 2500 \tilde{p}_3\\
\hline
8. &   \mathcal E_1[A_{4,1}^{~\otimes 16}] & 64 & {\cal \tilde{Z}}_0 =P_1^8, &   j^{\frac{2}{3}}(j^2+N_1j+N_2) \\
&&& {\cal \tilde{Z}}_1 = P_1^5 P_2,\  {\cal \tilde{Z}}_2 = P_1^3 P_3, & N_1 = 50 \tilde{p}_1,\\ 
&&& {\cal \tilde{Z}}_3 = P_1^2 P_2^2,\  {\cal \tilde{Z}}_4 = P_2 P_3 & N_2 = 2500 \tilde{p}_3 \\
\end{array}$
}
\end{threeparttable}
\end{center}
\end{table}

\begin{table}[H]
\begin{center}
\begin{threeparttable}
\rowcolors{2}{Mywhite}{Mygrey}
\resizebox{\textwidth}{!}{
   \renewcommand{\arraystretch}{1.3}
   $\begin{array}{|c|c|c|C{5cm}|C{5.5cm}|}
9. &   \mathcal E_1[A_{4,1}^{~\otimes 18}] & 72 & {\cal \tilde{Z}}_0 =P_1^9, &   j^3+N_1j^2+N_2j+N_3 \\
&&& {\cal \tilde{Z}}_1 =P_1^6 P_2, &  N_1 = 50 \tilde{p}_1,\\
&&& {\cal \tilde{Z}}_2 =P_1^4 P_3,\ {\cal \tilde{Z}}_3 =P_1^3 P_2^2, & N_2 =  2500 \tilde{p}_3, \\
&&& {\cal \tilde{Z}}_4 =P_1 P_2 P_3,\ {\cal \tilde{Z}}_5 =P_2^3 & N_3 = 125000 \tilde{p}_5\\
\hline
10. &   \mathcal E_1[A_{4,1}^{~\otimes 20}] & 80 & {\cal \tilde{Z}}_0 =P_1^{10}, &  j^{\frac{1}{3}}(j^3+N_1j^2+N_2j+N_3)  \\
&&& {\cal \tilde{Z}}_1 =P_1^7 P_2,\ {\cal \tilde{Z}}_2 =P_1^5 P_3, & N_1 = 50 \tilde{p}_1,\\
&&& {\cal \tilde{Z}}_3 =P_1^4 P_2^2,\ {\cal \tilde{Z}}_4 =P_1^2 P_2 P_3, & N_2 =  2500 \tilde{p}_3, \\
&&& {\cal \tilde{Z}}_5 =P_1 P_2^3,\ {\cal \tilde{Z}}_6 =P_3^2 & N_3 = 125000 \tilde{p}_5\\
\hline
11. &   \mathcal E_1[A_{4,1}^{~\otimes 22}] & 88 & {\cal \tilde{Z}}_0 =P_1^{11},\ {\cal \tilde{Z}}_1 = P_1^8 P_2, &   j^{\frac{2}{3}}(j^3+N_1j^2+N_2j+N_3) \\
&&& {\cal \tilde{Z}}_2 = P_1^6 P_3,\ {\cal \tilde{Z}}_3 = P_1^5 P_2^2,& N_1 = 50 \tilde{p}_1,\\
&&& {\cal \tilde{Z}}_4 = P_1^3 P_2 P_3,\ {\cal \tilde{Z}}_5 = P_1^2 P_2^3,& N_2 =  2500 \tilde{p}_3, \\
&&& {\cal \tilde{Z}}_6 = P_1 P_3^2,\  {\cal \tilde{Z}}_7 = P_2^2 P_3& N_3 = 125000 \tilde{p}_5\\
\hline
12. &   \mathcal E_1[A_{4,1}^{~\otimes 24}] & 96 & {\cal \tilde{Z}}_0 =P_1^{12}, &  j^4+N_1j^3+N_2j^2+N_3j+N_4  \\
&&& {\cal \tilde{Z}}_1 =P_1^9 P_2,\ {\cal \tilde{Z}}_2 =P_1^7 P_3,& N_1 = 50 \tilde{p}_1,\\
&&& {\cal \tilde{Z}}_3 =P_1^6 P_2^2,\ {\cal \tilde{Z}}_4 =P_1^4 P_2 P_3,& N_2 =  2500 \tilde{p}_3, \\
&&& {\cal \tilde{Z}}_5 =P_1^3 P_2^3,\ {\cal \tilde{Z}}_6 =P_1^2 P_3^2,& N_3 = 125000 \tilde{p}_5,\\
&&& {\cal \tilde{Z}}_7 =P_1 P_2^2 P_3,\ {\cal \tilde{Z}}_8 =P_2^4& N_4 = 6250000 \tilde{p}_8\\
\hline
\end{array}$
}
\end{threeparttable}
\end{center}
\caption{Characters of one-character extensions of tensor products of $A_{4,1}$ written in a basis of $S$-invariant polynomials }\label{t8}
\end{table}

\subsection{$\mathcal C = {\cal M}(4,3)$ a.k.a the Ising CFT \label{42s}}

The Virasoro minimal model ${\cal M}(4,3)$ is a $\mathbf{[3,0]}$ CFT with central charge $\frac12$. Let us denote it's three characters by $\chi_0, \chi_1, \chi_2$, the latter two  are the non-identity characters with conformal dimensions $h_1 = \frac{1}{16}, h_2 = \frac12$ respectively. Let us note the $q$-series for the characters :
\begin{align}
   \chi_0(q)& = q^{-\frac{1}{48}}(1+q^2+q^3+2 q^4+2 q^5+3 q^6+3 q^7+5 q^8+5 q^9+7 q^{10}+8 q^{11} + \ldots) \nonumber \\
   \chi_1(q)& = q^{\frac{1}{24}}(1+q+q^2+2 q^3+2 q^4+3 q^5+4 q^6+5 q^7+6 q^8+8 q^9+10 q^{10}+ \ldots) \nonumber \\
   \chi_2(q)& = q^{23/48}(1+q+q^2+q^3+2 q^4+2 q^5+3 q^6+4 q^7+5 q^8+6 q^9+8 q^{10}+ \ldots)\label{57a}
\end{align}
and the modular $S$-transformations for the characters :
\begin{align}
   S(\chi_0)&=\frac{\chi_0+\sqrt{2}\chi_1+\chi_2}{2},
   \qquad S(\chi_1)=\frac{\chi_0-\chi_2}{\sqrt{2}}, 
   \qquad S(\chi_2)=\frac{\chi_0-\sqrt{2}\chi_1+\chi_2}{2} \label{58a}
\end{align}

The allowed values of $k$ and $N$ can be read from entry 1 of table \ref{t3}. Here, every possible $c_{(1)} = 8k$ occurs and the corresponding number of tensor product factors is  $N = 16k$. This CFT is one of the hardest in terms of degree of polynomials required as well as the number of monomials one has to work with : for the smallest $c_{(1)} = 8$, we need to work with homogenous polynomials of degree $16$ in $3$ variables, which has $153$ monomials and for the largest $c_{(1)} = 48$ that we consider here, we need to work with homogenous polynomials of degree $96$ in $3$ variables, which has $4753$ monomials.

We follow all the steps laid out in section \ref{2s}.  We solve the set of equations \eqref{7b}, \eqref{8b} and construct the ansatz character \eqref{10b}. We cannot come up with a closed formula for $b$ the number of monomials with $q$-series commensurate to the identity character. But we can readily compute them. And we can set up the overdetermined system of linear equations and find solutions. The answers are tabulated in table \ref{t9b}.

We start with smallest possible $c_{(1)}$. This is the $N = 16,  c_{(1)} = 8$ case. The ansatz character in this case has $13$ arbitrary constants, the $b$’s. The $S$-invariance equations are an overdetermined system of $153$ equations for these $13$ arbitrary variables. It was quite surprising that there are two $S$-invariant polynomials, $P_1$ and  $P_3$, hidden within the answer :
\be \label{99c}
P_{c_{(1)} = 8}^{\mathcal M(4,3)}  = P_1 + \tilde{p}_1\, P_3^2 
\ee
One more source of surprise is the presence of the arbitrary constant $\tilde{p}_1$, given that the character at $c_{(1)} = 8$ viz. $j^{\frac13}$ has no free parameters. But it turns out when one computes the $q$-series for $P_3$ using \eqref{57a}, it computes to zero, thus effectively removing the $\tilde{p}_1$ from the answer in \eqref{99c}.  $P_1$ is a non-constant $S$-invariant polynomial that computes as a $q$-series to $j^{\frac13}$. Thus, the answer in \eqref{99c}, reduces to the expected $j^{\frac13}$. The more important upshot of the computation for $N = 16,  c_{(1)} = 8$ is the discovery of the two $S$-invariant polynomials.

Then, we take on the next case $N = 32,  c_{(1)} = 16$. This is also computationally hard. The ansatz character has $42$ arbitrary constants, the $b$'s.  The $S$-invariance equations are an overdetermined system of $561$ equations for these $42$ arbitrary variables. We managed to find the following solution that has hidden in it a third $S$-invariant polynomial, viz. $P_2$ :
\be \label{100c}
P_{c_{(1)} = 16}^{\mathcal M(4,3)}  = {\cal Z}_0 + p_1\, {\cal Z}_1 + p_2\, {\cal Z}_2 + p_3\, {\cal Z}_3 
\ee
where
\begin{small}
\begin{align}
   {\cal Z}_0& =  P_1^2-480 P_2 P_3+480 P_1 P_3^2-4120 P_3^4, \qquad  {\cal Z}_1 = 2 P_2 P_3-P_1 P_3^2+2 P_3^4 \nonumber \\
   {\cal Z}_2& = P_3^4,  \qquad   {\cal Z}_3 = P_2 P_3-P_1 P_3^2+8 P_3^4. \label{101c}
\end{align}
\end{small}
We can put the above in the following more-compact form :
\be \label{102c}
P_{c_{(1)} = 16}^{\mathcal M(4,3)}  = 
\tilde{{\cal Z}}_0+ \tilde{p}_1\, \tilde{{\cal Z}}_1 + \tilde{p}_2\, \tilde{{\cal Z}}_2 + \tilde{p}_3\, \tilde{{\cal Z}}_3 
\ee
where
\be \label{103c}
\tilde{{\cal Z}}_0 = P_1^2, \quad \tilde{{\cal Z}}_1 = P_1P_3^2, \quad \tilde{{\cal Z}}_2 = P_2P_3, \quad \tilde{{\cal Z}}_3 = P_3^4
\ee
The relation between the constant coefficients of \eqref{102c} and \eqref{100c} are :
\begin{small}
\begin{align}
         \tilde{p}_1& =  480 - p_1 - p_3, \qquad  \tilde{p}_2 =  - 480 +2p_1 + p_3, \qquad \tilde{p}_3 = - 4120 +2p_1 + p_2 + 8 p_3
 \end{align}
\end{small}
Given that the character at $c_{(1)} = 16$ viz. $j^{\frac23}$ has no free parameters, it is initially surprising to find three free parameters in \eqref{102c} but the presence of $P_3$ (which computes to zero as a $q$-series) along with each of these free parameters effectively removes all of them.  And the right hand side of \eqref{102c} reduces to the expected $j^{\frac23}$. The important outcome of the computations so far  is that we have a set of three $S$-invariant polynomials. And we expect these three to form a better basis for $S$-invariant characters than the basis one started with viz. $\chi_0, \chi_1, \chi_2$. We collect all the three $S$-invariant polynomials for ${\cal M}(4,3)$ below :
\begin{small}
\begin{equation}\label{39a}
\begin{array}{lll}
   P_1 &=~\chi_0^{16}\,+ \,140 \,\chi_0^{12} \,\chi_2^4\,+\,448\, \chi_0^{10} \,\chi_2^6\,+\,870 \,\chi_0^8\, \chi_2^8\,+\,240\, \chi_0^7\, \chi_1^8 \,\chi_2\,+\,448\, \chi_0^6 \,\chi_2^{10}\,+\,1680 \,\chi_0^5\, \chi_1^8\, \chi_2^3\,\notag\\&+\,140 \,\chi_0^4\, \chi_2^{12}\,+\,1680 \,\chi_0^3\, \chi_1^8\, \chi_2^5\,+\,240 \chi_0\, \chi_1^8\, \chi_2^7\,+\,8 \,\chi_1^{16}\,+\,\chi_2^{16} ~=j~^{\frac{1}{3}}\\
   P_2 & = ~ \chi_0^{23} \,\chi_2\,-\,\chi_0^{21} \,\chi_2^3\,-\,21\, \chi_0^{19} \,\chi_2^5\,+\,85 \,\chi_0^{17} \,\chi_2^7\,-\,134 \,\chi_0^{15} \,\chi_2^9\,+\,256 \,\chi_0^{14}\, \chi_1^8\, \chi_2^2\,+\,70 \chi_0^{13} \,\chi_2^{11}\,\notag\\&\,+\,3584 \,\chi_0^{12}\, \chi_1^8\, \chi_2^4\,70\, \chi_0^{11} \,\chi_2^{13}\,  +\,16128 \,\chi_0^{10} \,\chi_1^8 \,\chi_2^6\,-\,134 \,\chi_0^9 \,\chi_2^{15}\,+\,25600 \chi_0^8\, \chi_1^8\, \chi_2^8\,-\,256 \,\chi_0^7 \,\chi_1^{16} \,\chi_2\,\notag\\&+\,85 \chi_0^7 \,\chi_2^{17}\,+\,16128\, \chi_0^6\, \chi_1^8 \,\chi_2^{10}\,-\,1792 \,\chi_0^5\, \chi_1^{16} \,\chi_2^3\,-\,21 \,\chi_0^5\, \chi_2^{19}\,+\,3584\, \chi_0^4\, \chi_1^8\, \chi_2^{12}\,-\,1792\, \chi_0^3 \,\chi_1^{16} \,\chi_2^5\, \notag\\&-\,\chi_0^3\, \chi_2^{21}\,+\,256\, \chi_0^2\, \chi_1^8\, \chi_2^{14}\,-\,256 \,\chi_0\, \chi_1^{16}\, \chi_2^7\,+\,\chi_0\, \chi_2^{23} \,=\,1\\
   P_3 &=~\chi_0^7\, \chi_2\,+\,7\, \chi_0^5\, \chi_2^3\,+\,7\, \chi_0^3\, \chi_2^5\,+\,\chi_0\, \chi_2^7\,-\,\chi_1^8 ~=~0
\end{array}
\end{equation}
\end{small}
We have gone ahead and computed for (i) $N = 48,  c_{(1)} = 24$, (ii) $N = 64,  c_{(1)} = 32$, (iii) $N = 80,  c_{(1)} = 40$ and (iv) $N = 96,  c_{(1)} = 48$. The number of arbitrary coefficients, the $b$'s in the ansatz character for these cases are respectively, $87$, $148$, $225$ and $318$. The number of equations that one needs to solve in these cases are respectively, $1225$, $2145$, $3321$ and $4753$. We have tabulated the results in table \ref{t9b}. It should be noted that we give the most compact form of the results. In every case, we give the equivalent of \eqref{102c} rather than \eqref{100c}. The number of $S$-invariant summands, the $\tilde{{\cal Z}}_i$'s in the six cases $N = 16, 32, 48, 64, 80, 96$ are respectively, $2, 4, 7, 10, 14, 19$. We have this information only after a long process of solving a huge number of equations. Below, we will give a quick method to get the number of summands as well as the summands themselves, without having to solve the equations.

The way to read the results from table \eqref{t9b} is as follows.   The first column is the serial no. which is also the value of $k$. The second column contains the number of tensor product factors $N$ (which equals $16k$). The third column contains the value of $c_{(1)} = 8k$.  The fourth column contains the character $P^{\mathcal M(4,3)}_{c_{(1)} = 8k}(\chi_0, \chi_1)$.  

Now we need to impose admissibility. For that, we first need to compute the $q$-series of $P^{\mathcal M(4,3)}_{c_{(1)} = 8k}$. We employ the explicit $q$-series of the characters available in \eqref{57a} and give the  $q$-series of $P^{\mathcal M(4,3)}_{c_{(1)} = 8k}$ in the  fifth column of table \ref{t9b}. Then we need to compare this with the generic admissible character for that particular central charge, given in table \ref{tapp} of the appendix \ref{app1}. The generic admissible character for central charge $8k$ has $[\frac{k}{3}]$ parameters which are denoted by $N_1, N_2, \ldots$. And these $N$’s need to satisfy a set of nested inequalities for admissibility, which are available from table \ref{tapp} of the appendix \ref{app1}.  In the fifth column of table \ref{t9b}, we give the relation between the $\tilde{p}$’s and the $N$’s. Then, the nested inequalities for the $N$’s translate to inequalities for the $\tilde{p}$’s. We always find an infinite number of admissible character solutions, except  for $N = 16, c_{(1)} = 8$ and $N = 32, c_{(1)} = 16$.

It is remarkable that even though the $P^{\mathcal M(4,3)}_{c_{(1)} = 8k}$ has some large number of $S$-invariant summands and hence that many arbitrary constants, the $\tilde{p}$'s, only a small subset of these constants enter into the $q$-series. For example, in the last row of table \ref{t9b}, we find in the fourth column that there are a total of $19$ summands in $P^{\mathcal M(4,3)}_{c_{(1)} = 96}$ and hence $18$ $\tilde{p}$'s. But in the fifth column, we find that only $2$ of them viz. $\tilde{p}_3$ and $\tilde{p}_8$ can be found in the $q$-series. The primary reason for this is that the $S$-invariant polynomial $P_3$ evaluates to zero. We are thus led to believe that the correct way to think of $P^{\mathcal M(4,3)}_{c_{(1)} = 8k}$ is as a polyomial in the quotient ring obtained by quotienting the polynomial ring of characters by the ideal generated by the constant $S$-invariant polynomials. In this case, the ideal is $(P_3, P_2 -1 )$. $P^{\mathcal M(4,3)}_{c_{(1)} = 8k}$, when thought of as an element of this quotient ring, can be written out very compactly.

Amongst all the admissible characters we have computed in table \ref{t9b}, there are only three of them which are known to correspond to actual CFTs. The first one is in the first row, $N = 16, c_{(1)} = 8$, and this corresponds to the $E_{8,1}$ CFT.  The second one is the second row, $N = 32, c_{(1)} = 16$, and this corresponds to two CFTs\footnote{We find it somewhat remarkable that WZW CFTs can be obtained as one-character extensions of tensor products of non-WZW CFTs, here, the Ising CFT.} viz. $E_{8,1}^{\otimes 2}$ and ${\cal E}_1[D_{16,1}]$. The third one is one of the infinite characters of the third row, $N = 48, c_{(1)} = 24$.   Among these infinite characters, we can identify the one which corresponds to a CFT by invoking the classification of $c_{(1)} = 24$ CFTs of \cite{Schellekens:1992db}, which includes the monster CFT.

It is well-known that the character of the monster CFT is a one-character extension of the tensor product of $48$ copies of the Ising model. Here, we can report the character of the monster CFT in terms of the characters of the Ising model. From the third entry of table \ref{t9b}, we have 
\be \label{59a}
P_{c_{(1)} = 24}^{\mathcal M(4,3)} (\chi_0, \chi_1, \chi_2) = \tilde{{\cal Z}}_0 + \tilde{p}_1\,\tilde{{\cal Z}}_1+ \tilde{p}_2\,\tilde{{\cal Z}}_2+ \tilde{p}_3\,\tilde{{\cal Z}}_3+ \tilde{p}_4\,\tilde{{\cal Z}}_4+ \tilde{p}_5\,\tilde{{\cal Z}}_5+ \tilde{p}_6\,\tilde{{\cal Z}}_6.
\ee
There are seven summands in \eqref{59a}. As a $q$-series, this evaluates to $j + \tilde{p}_3 $. Now, it is well-known that the character of the monster CFT is $j - 744$. To obtain the character of the monster CFT from \eqref{59a}, we need to choose $\tilde{p}_3 = -744$. This gives us the character of the monster CFT:
\bea \label{60a}
\chi_{\text{Monster}} &= P_1^3 - 744 P_2^2 \nonumber \\
& = P_1^3 - 744
\eea
The first line is a presentation in which the homogeneity of the Ising characters is clear. The second line is in the simplest form.

\underline{Lessons learnt and a new method} :  All the results that we have obtained and reported in table \ref{t9b} were by running the whole procedure as laid out  in section \ref{2s} . Similar to what we did in section \ref{3s}, we have a simpler  procedure that allows us to write the final answer quickly.  We first note that $P_1$ is a homogenous polynomial of the characters of degree $16$,   $P_2$ is a homogenous polynomial of the characters of degree $24$ and $P_3$ is a homogenous polynomial of the characters of degree $8$.  A monomial of the form $P_1^a\,P_2^b\,P_3^c$ is a homogenous polynomial of the characters of degree $16a + 24b+8c$. And for this monomial to be present in $P^{\mathcal{M}(4,3)}_{c_{(1)} = 8k}$ it’s degree needs to be equal to $N$. Thus we look for  non-negative integral solutions to 
\be \label{107c}
16a + 24b+8c = N = 16 k, 
\ee
we can obtain all the $\tilde{{\cal Z}}$'s  in $P^{\mathcal{M}(4,3)}_{c_{(1)} = 8k}$.  All except one of the $\tilde{{\cal Z}}$'s  viz. $\tilde{{\cal Z}}_0$ are multiplied by constants the $\tilde{p}$'s.  Which of the solutions of \eqref{107c} corresponds to $\tilde{{\cal Z}}_0$ is easily decided by seeing which of them contains the identity character $\chi_0^N$. Here it is the $a = k, b = 0, c=0$ solution. This quick computation of the solutions to \eqref{107c} gives us the same answer as the long procedure.  There is one caveat. We should only consider solutions of \eqref{107c} that give monomials of characters which are commensurate to the identity character. It turns out that every solution of \eqref{107c} is such that  $b$ and $c$ are either both even numbers of both odd numbers; this can be easily seen by rewriting \eqref{107c}  as $c = 2k - 2a - 3b$. It then follows that all monomials of characters present in $P_1^a\,P_2^b\,P_3^c$ are commensurate to the identity character. 
 
\begin{table}[H]
\begin{center}
\begin{threeparttable}
\rowcolors{2}{Mywhite}{Mygrey}
\resizebox{\textwidth}{!}{
   \renewcommand{\arraystretch}{1.3}
   $\begin{array}{|c|c|c|C{8cm}|C{4cm}|}
\hline
 \# (k) &      N & c_{(1)} & P_{c_{(1)} = 8k}^{\mathcal M(4,3)} (\chi_0, \chi_1, \chi_2) &  q\text{-series}  \\
\hline
 1. &  \mathcal E_1[\mathcal M(4,3)^{~\otimes 16}] & 8 & {\cal \tilde{Z}}_{0}=P_{1},& j^{\frac{1}{3}}   \\
      &&& {\cal \tilde{Z}}_{1}=P_{3}^2&\\
\hline
2. &    \mathcal E_1[\mathcal M(4,3)^{~\otimes 32}] &  16 &{\cal \tilde{Z}}_{0}=P_{1}^2,\  {\cal \tilde{Z}}_{1}=P_{1} P_{3}^2,& j^{\frac{2}{3}}  \\
&&& {\cal \tilde{Z}}_{2}=P_{2} P_{3},\  {\cal \tilde{Z}}_{3}=P_{3}^4&\\
\hline
3. &      \mathcal E_1[\mathcal M(4,3)^{~\otimes 48}] &  24 &{\cal \tilde{Z}}_0 =P_1^3,\  {\cal \tilde{Z}}_{1}=P_{1}^2 P_{3}^2,\  {\cal \tilde{Z}}_{2}=P_{1} P_{2} P_{3},&  j+N  \\
&&& {\cal \tilde{Z}}_{3}=P_{2}^2,\  {\cal \tilde{Z}}_{4}=P_{1} P_{3}^4,\  {\cal \tilde{Z}}_{5}=P_{2} P_{3}^3,& N_{1}=\tilde{p}_3\\
&&& {\cal \tilde{Z}}_{6}=P_{3}^6&\\
\hline
4. &      \mathcal E_1[\mathcal M(4,3)^{~\otimes 64}] & 32 & {\cal \tilde{Z}}_0 = P_1^4,\  {\cal \tilde{Z}}_{1}=P_{1}^3 P_{3}^2,\  {\cal \tilde{Z}}_{2}=P_{1}^2 P_{2} P_{3},& j^{\frac{1}{3}}(j+N) \\
&&& {\cal \tilde{Z}}_{3}=P_{1} P_{2}^2,\  {\cal \tilde{Z}}_{4}=P_{1}^2 P_{3}^4,\  {\cal \tilde{Z}}_{5}=P_{1} P_{2} P_{3}^3,& N_{1}=\tilde{p}_3\\
&&& {\cal \tilde{Z}}_{6}=P_{2}^2 P_{3}^2,\  {\cal \tilde{Z}}_{7}=P_{1} P_{3}^6,\  {\cal \tilde{Z}}_{8}=P_{2} P_{3}^5,&\\
&&& {\cal \tilde{Z}}_{9}=P_{3}^8&\\
\end{array}$
}
\end{threeparttable}
\end{center}
\end{table}

\begin{table}[H]
\begin{center}
\begin{threeparttable}
\rowcolors{2}{Mywhite}{Mygrey}
\resizebox{\textwidth}{!}{
   \renewcommand{\arraystretch}{1.3}
   $\begin{array}{|c|c|c|C{8cm}|C{4cm}|}
5. &      \mathcal E_1[\mathcal M(4,3)^{~\otimes 80}] & 40 & {\cal \tilde{Z}}_0 = P_1^5,\  {\cal \tilde{Z}}_{1}=P_{1}^4 P_{3}^2,\  {\cal \tilde{Z}}_{2}=P_{1}^3 P_{2} P_{3},& j^{\frac{2}{3}}(j+N)\\
&&& {\cal \tilde{Z}}_{3}=P_{1}^2 P_{2}^2,\  {\cal \tilde{Z}}_{4}=P_{1}^3 P_{3}^4,\  {\cal \tilde{Z}}_{5}=P_{1}^2 P_{2} P_{3}^3,& N_{1}=\tilde{p}_3\\
&&& {\cal \tilde{Z}}_{6}=P_{1} P_{2}^2 P_{3}^2,\  {\cal \tilde{Z}}_{7}=P_{2}^3 P_{3},\  {\cal \tilde{Z}}_{8}=P_{1}^2 P_{3}^6,&\\
&&& {\cal \tilde{Z}}_{9}=P_{1} P_{2} P_{3}^5,\  {\cal \tilde{Z}}_{10}=P_{2}^2 P_{3}^4,\  {\cal \tilde{Z}}_{11}=P_{1} P_{3}^8,&\\
&&& {\cal \tilde{Z}}_{12}=P_{2} P_{3}^7,\  {\cal \tilde{Z}}_{13}=P_{3}^{10}&\\
\hline
6. &      \mathcal E_1[\mathcal M(4,3)^{~\otimes 96}] & 48 & {\cal \tilde{Z}}_0 = P_1^6,\  {\cal \tilde{Z}}_{1}=P_{1}^5 P_{3}^2,\  {\cal \tilde{Z}}_{2}=P_{1}^4 P_{2} P_{3},& j^2+N_1j+N_2 \\
&&& {\cal \tilde{Z}}_{3}=P_{1}^3 P_{2}^2,\  {\cal \tilde{Z}}_{4}=P_{1}^4 P_{3}^4,\  {\cal \tilde{Z}}_{5}=P_{1}^3 P_{2} P_{3}^3,& N_{1}=\tilde{p}_3,\\
&&& {\cal \tilde{Z}}_{6}=P_{1}^2 P_{2}^2 P_{3}^2,\  {\cal \tilde{Z}}_{7}=P_{1} P_{2}^3 P_{3},\  {\cal \tilde{Z}}_{8}=P_{2}^4,& N_{2}=\tilde{p}_8\\
&&& {\cal \tilde{Z}}_{9}=P_{1}^3 P_{3}^6,\  {\cal \tilde{Z}}_{10}=P_{1}^2 P_{2} P_{3}^5,\  {\cal \tilde{Z}}_{11}=P_{1} P_{2}^2 P_{3}^4,&\\
&&& {\cal \tilde{Z}}_{12}=P_{2}^3 P_{3}^3,\  {\cal \tilde{Z}}_{13}=P_{1}^2 P_{3}^8,\  {\cal \tilde{Z}}_{14}=P_{1} P_{2} P_{3}^7,&\\
&&& {\cal \tilde{Z}}_{15}=P_{2}^2 P_{3}^6,\  {\cal \tilde{Z}}_{16}=P_{1} P_{3}^{10},\  {\cal \tilde{Z}}_{17}=P_{2} P_{3}^9,&\\
&&& {\cal \tilde{Z}}_{18}=P_{3}^{12}&\\
\hline
\end{array}$
}
\end{threeparttable}
\end{center}
\caption{Characters of one-character extensions of tensor products of ${\cal M}(4,3)$ written in a basis of $S$-invariant polynomials}\label{t9b}
\end{table}

\subsection{$\mathcal C = D_{r,1}$ \label{43s}}

In this subsection, we study the infinite class of CFTs $D_{r,1}$ where $r$ is a positive integer and $r \geq 3$ (but $r \neq 4$). Each of these CFTs is a $\mathbf{[3,0]}$ CFT \cite{Das:2020wsi}. A $D_{r,1}$ CFT has a central charge equalling $r$. Let us denote it's three characters by $\chi_0, \chi_1, \chi_2$, the latter two  are the non-identity characters with conformal dimensions $h_1 = \frac{r}{8}, ~h_2 = \frac12$ respectively. Let us note the $q$-series for the characters (these formulae are sourced from chapter 14 of \cite{DiFrancesco:1997nk}):
\begin{align}
   \chi_0(q)& = \frac{\theta_3^r(q)+\theta_4^r(q)}{2\eta^r}, \qquad \chi_1(q)  = \frac{\theta_2^r(q)}{2\eta^r}, \qquad \chi_2(q) = \frac{\theta_3^r(q)-\theta_4^r(q)}{2\eta^r}\label{108c}
\end{align}
where $\theta_2, \theta_3, \theta_4 $ are the $\theta$-constants : $\theta_i(\tau) \equiv \theta_i(0|\tau)$ with $q$-series given by :
\begin{align*}
   \theta_2(q)&=2q^{\frac{1}{8}}\prod_{n=1}^{\infty}(1-q^n)(1+q^n)^2, \qquad 
   \theta_3(q) =\prod_{n=1}^{\infty}(1-q^n)(1+q^{n-\frac{1}{2}})^2\\
   \theta_4(q)&=\prod_{n=1}^{\infty}(1-q^n)(1-q^{n-\frac{1}{2}})^2, \qquad 
   \eta(q) =q^{\frac{1}{24}}\prod_{n=1}^{\infty}(1-q^n).
\end{align*}
The  modular $S$-transformations for the characters is given by :
\begin{small}
\begin{align}
   S(\chi_0)&= \frac{\chi_0+2\chi_1+\chi_2}{2}, \qquad  S(\chi_1)= \frac{\chi_0-\chi_2}{2}, \qquad S(\chi_2)=\frac{\chi_0-2\chi_1+\chi_2}{2}. \label{58ad}
\end{align}
\end{small}
We note that the $S$-transformations do not have a $r$ dependence. This means that we can hope that imposing $S$-invariance can be done independent of $r$. Before we get to imposing $S$-invariance, we need to construct the ansatz character \eqref{10b}. This has a $r$-dependence. We can see this from examining entries 3, 4, 5 and 6 of table \ref{t3}. We find that the number of tensor powers $N$ depends on $r$ and the particular form of the dependence on $r$ makes us divide into four (infinite) sub-classes (i) when $r$ is a odd number, (ii) $r$ is $2$ times an odd number, (iii) $r$ is a $4$ times an odd number, (iv) $r$ is a multiple of $8$.  Within each sub-class, the number of tensor powers $N$, is independent of $r$.  And then it turns out that there is a common solution for all the infinite members of a sub-class. We will take up each sub-class below.

\subsubsection{$\mathcal C = D_{r,1}, \quad  r = 2p +1, \quad p = 1, 2, \ldots$ \label{431s}}

Here, we study all $D_{r,1}$  CFTs with $r = 2p + 1$ : $D_{3,1}, D_{5,1}, D_{7,1}, \ldots$.   From row 3 of table \ref{t3},  it is clear that  all $D_{r,1}$  CFTs with $r = 2p + 1$  have the same $N$ values; $N = 8s$ and the resulting central charge is $c_{(1)} = 8s(2p+1)$. Hence  for all such $r$, when constructing the ansatz character,  we are looking for monomials of the same degree. Then we note that $h_1$ and $h_2$ values for all  $r = 2p + 1$  have the same structure, i.e. denominators are $8$ and $2$.  What this means is that a monomial which solves the equations \eqref{7b} and \eqref{8b} for any odd $r$  also solves the equations for any other odd $r$ (but for different values of $M$).   For example, the following is the ansatz character for $D_{3,1}$ and  $N = 8$ :
\be \label{110c}
\chi_0^8 + b_1\,\chi_0^6  \chi_2^2 + b_2\,\chi_0^3 \chi_1^4 \chi_2 + b_3\,\chi_0^4  \chi_2^4 + b_4\,\chi_1^8  + b_5\,\chi_0 \chi_1^4 \chi_2^3 + b_6\,\chi_0^2  \chi_2^6 + b_7\, \chi_2^8.
\ee
The integer conformal dimensions for the monomials in \eqref{110c} are respectively, $0, 1, 2, 2, 3, 3, 3, 4$. When one computes the ansatz character for $D_{5,1}$ and $N = 8$, one obtains \eqref{110c} again.  But, when thought of as an ansatz character for $D_{5,1}$, the integer conformal dimensions for the monomials in \eqref{110c} are respectively, $0, 1, 3, 2, 5, 4, 3, 4$.  More generally, for $D_{2p+1, 1}$ and $N = 8$, the ansatz character is \eqref{110c} and the integer conformal dimensions for the monomials are respectively, $0, 1, p+1, 2, 2p+1, p+2, 3, 4$. What we need to note here is that the the monomials do not depend on the theory i.e. do not depend on $p$ while the conformal dimensions of the monomials do depend on it. We are able to observe this phenomenon from the computation of the ansatz character for $N = 8$ in \eqref{110c}.

We now give a general proof for this. The monomial $\chi_0^a \chi_1^b \chi_2^c$ is present in the ansatz character for $N = 8s$ if the following equations are satisfied :
\be \label{111d}
a + b + c  = N = 8s, \qquad (2p + 1) b + 4 c = 8 M
\ee
for non-negative integral $M$. We will be able to show that every solution of \eqref{111d} depends on $s$ and not on the theory $p$. We will also be able to show that the solutions will have a conformal dimension $M$ which does depend on $p$. To solve \eqref{111d}, we first note, from the second equation, that $b$ should be a multiple of $4$. Let us designate $b = 4l$. From the fact that $b \leq 8s$, it then follows that $l$ can take values $0, 1, 2, \ldots, 2s-1, 2s$. It then also follows that when $l$ is an odd number, $c$ should also be an odd number, as well as $a$; simillarly when $l$ is an even number, $c$ should also be an even number, as well as $a$. The values $c$ and $a$  take, can be determined and we find two classes of monomial solutions for the ansatz character :
\be
\sum_{l \in L} \left( \sum_{k \in K}~b_{k,l} ~\chi_0^{8s - 4l - k} \chi_1^{4l} \chi_2^k\right) \quad + \quad  \sum_{l' \in L'} \left( \sum_{k' \in K'}~b_{k',l'} ~\chi_0^{8s - 4l' - k'} \chi_1^{4l'} \chi_2^{k'}\right)
\ee \label{112d}
where
\bea
&L = \{1, 3, \ldots, 2s-1 \}, \qquad K = \{1, 3, \ldots, 8s-4l-1\} \nonumber \\
&\text{and} \nonumber \\
&L' = \{0, 2, \ldots, 2s \}, \qquad K' = \{ 0, 2, \ldots, 8s-4l \}. \nonumber
\eea
Note that there is no $p$ dependence in the ansatz character of \eqref{112d}. The $p$ dependence shows up when one computes the conformal dimensions.  The character multiplying $b_{k,l}$ has  a conformal dimension $\frac{(2p+1)l + k}{2}$ and the character multiplying $b_{k',l'}$ has  a conformal dimension $\frac{(2p+1)l' + k'}{2}$.

Now, we proceed to complete the procedure laid out in section \ref{2s}.  We consider the S-invariance equations \eqref{18b}. Since the $S$-matrix for all $D_{r,1}$  CFTs are identical,  we get the same set of linear equations for all odd $r$.  And then we solve the  overdetermined system of linear equations. The answers are tabulated in table \ref{t10b}.

We start with smallest possible $c_{(1)}$. This is the $N = 8,  c_{(1)} = 8(2p + 1)$ case. The ansatz character in this case has $7$ arbitrary constants, the $b$’s, see \eqref{110c}. The $S$-invariance equations are an overdetermined system of $45$ equations for these $7$ arbitrary variables. It was quite surprising that there are two $S$-invariant polynomials, $P_1$ and  $P_2$, hidden within the answer :
\be \label{111c}
P_{c_{(1)} = 8(2p + 1)}^{D_{2p +1,1}}  = P_1 + \tilde{p}_1\, P_2^2 
\ee
Then, we take on the next case $N = 16,  c_{(1)} = 16(2p + 1)$.  The ansatz character \eqref{111d} has $22$ arbitrary constants, the $b$'s.   The $S$-invariance equations are an overdetermined system of $153$ equations for these $22$ arbitrary variables. We managed to find the following solution that has hidden in it a third $S$-invariant polynomial, viz. $P_3$ :
\be \label{112c}
P_{c_{(1)} = 16(2p + 1)}^{D_{2p +1,1}}  = {\cal Z}_0 + p_1\, {\cal Z}_1 + p_2\, {\cal Z}_2 + p_3\, {\cal Z}_3 
\ee
where
\begin{small}
\begin{align}
   {\cal Z}_0& =  P_1^2-28 P_2^4-224 P_2 P_3^4, \qquad  {\cal Z}_1 = P_1 P_2^2-2 P_2^4+4 P_2 P_3^4 \nonumber \\
   {\cal Z}_2& = P_2^4,  \qquad   {\cal Z}_3 = P_2 P_3^4. \label{101c}
\end{align}
\end{small}
We can put the above in the following more-compact form :
\be \label{114c}
P_{c_{(1)} = 16(2p + 1)}^{D_{2p +1,1}}  = 
\tilde{{\cal Z}}_0+ \tilde{p}_1\, \tilde{{\cal Z}}_1 + \tilde{p}_2\, \tilde{{\cal Z}}_2 + \tilde{p}_3\, \tilde{{\cal Z}}_3 
\ee
where
\be \label{115c}
\tilde{{\cal Z}}_0 = P_1^2, \quad \tilde{{\cal Z}}_1 = P_1 P_2^2, \quad \tilde{{\cal Z}}_2 = P_2^4, \quad \tilde{{\cal Z}}_3 = P_2 P_3^4
\ee
The relation between the constant coefficients of \eqref{112c} and \eqref{114c} are :
\begin{small}
\begin{align}
         \tilde{p}_1& =  p_1, \qquad  \tilde{p}_2 =  - 28 -2p_1 + p_2, \qquad \tilde{p}_3 = - 224 +4p_1 + p_3
 \end{align}
\end{small}
The important outcome of the computations so far  is that we have a set of three $S$-invariant polynomials. And we expect these three to form a better basis for $S$-invariant characters than the basis one started with viz. $\chi_0, \chi_1, \chi_2$. We collect all the three $S$-invariant polynomials for $D_{2p+1,1}$ below :
\begin{equation}\label{59ad}
\begin{array}{lll}
   P_1(\chi_0, \chi_1, \chi_2) &=\chi_0^8 \,+\,14 \,\chi_0^4 \,\chi_2^4\,+\,112\, \chi_0^3\, \chi_1^4\, \chi_2 \,+\,112 \,\chi_0\, \chi_1^4\, \chi_2^3\,+\,16\, \chi_1^8\,+\,\chi_2^8\\
   P_2(\chi_0, \chi_1, \chi_2) &=\chi_0^3 \,\chi_2\,+\,\chi_0\, \chi_2^3\,-\,2 \,\chi_1^4\\
   P_3(\chi_0, \chi_1, \chi_2)&= \chi_0^2\,\chi_1 - \chi_1\,\chi_2^2 = 2^{r-1}.
\end{array}
\end{equation}
We note that each of the $S$-invariant polynomials has a $q$-series that depends on $r$.  $P_1$, for all values of $r$ is a non-constant $S$-invariant polynomial. $P_2$ turns out to be constant $S$-invariant polynomial only for $r = 3$, taking the value $6$ and is a non-constant polynomial for $r \geq 5$. $P_3$ is always a constant $S$-invariant polynomial, for all values of $r$, taking the value $2^{r-1}$.

We have gone ahead and computed for $N = 24, 32, 40, 48$.  The number of arbitrary coefficients, the $b$'s in the ansatz character for these cases are respectively, $45$, $76$, $115$ and $162$. The number of equations that one needs to solve in these cases are respectively, $325$, $396$, $861$ and $1225$. We have tabulated the results in table \ref{t10b}. It should be noted that we give the most compact form of the results. In every case, we give the equivalent of \eqref{114c} rather than \eqref{112c}. The number of $S$-invariant summands, the $\tilde{{\cal Z}}_i$'s in the six cases $N = 8, 16, 24,  32, 40, 48$ are respectively, $2, 4, 7, 10, 14, 19$. We have this information only after a long process of solving a huge number of equations. Below, we will give a quick method to get the number of summands as well as the summands themselves, without having to solve the equations. 

The way to read the results from table \eqref{t10b} is as follows.   The first column is the serial no. ($s$). The second column contains the number of tensor product factors $N$ (which equals $8s$). The third column contains the value of $c_{(1)} = 8s(2p+1)$.  The fourth column contains the character $P^{D_{2p+1,1}}_{c_{(1)} = 8s(2p+1)}(\chi_0, \chi_1)$.  We should emphasise that the table \ref{t10b} contains the characters of the one-character extensions of tensor products of an infinite number of CFTs. It is remarkable that we are able to encode an infinite amount of information in such a finite compact form. We do not give the $q$-series of the characters in table \ref{t10b} due to that information being dependent on $r$. 

Now, we should consider admissibility. There are an infinite number of CFTs to study. We only study three cases viz. $r = 3, 5, 7$. We have computed the $q$-series of the characters that can be read off from table \ref{t10b} and we always find infinite numbers of admissible characters at each central charge. 

Amongst all the admissible characters that we have found, only one of them corresponds to a genuine CFT. This happens for $D_{3,1}$ when $N = 8$. The character is given in \eqref{110c}.  Noting that for $r=3$, $P_1 = j-624$ and $P_2 = 6$, we find 
\be \label{118c}
P_{c_{(1)} = 24}^{D_{3,1}} = j - 624 + 36 \tilde{p}_1.
\ee
There are an infinite number of admissible characters in \eqref{118c} at central charge $24$. To answer the question as which of them correspond to a genuine CFT, we once again invoke the classification of one-character $c = 24$ CFTs of \cite{Schellekens:1992db}. In the table at the end of that paper entry no. 30 is the only CFT which is a one-character extension of a tensor product of  copies of $D_{3,1}$\footnote{The reader can find this CFT listed as entry no. 13 of table \ref{t0}. It can be found as $A_{3,1}$.}.  This CFT has a character  $j - 624$. Thus amongst the infinite characters \eqref{118c} that we found, only one of them, the one with $\tilde{p}_1 = 0$ corresponds to an actual CFT. 

Further considerations (beyond those in this paper and) establish \cite{Schellekens:1992db}, in the case of $D_{3,1}$,  that  the single character obtained for $N = 8$ viz. $P_1$ corresponds to a genuine CFT. We suspect that these further considerations would  establish that the same character for all the other $D_{2p+1,1}$'s also  corresponds to genuine CFT(s). We thus are led to the following surmise/conjecture : \emph{For $D_{2p+1,1}$, $p \geq 2$  there exist genuine one-character CFT(s) with central charge $8(2p+1)$ with character $P_1$.}

\underline{Lessons learnt and a new method} :  All the results that we have obtained and reported in table \ref{t10b} were by running the whole procedure as laid out  in section \ref{2s} . Similar to what we did in section \ref{3s}, we have a simpler  procedure that allows us to write the final answer quickly.  We first note that $P_1$ is a homogenous polynomial of the characters of degree $8$,   $P_2$ is a homogenous polynomial of the characters of degree $4$ and $P_3$ is a homogenous polynomial of the characters of degree $3$.  A monomial of the form $P_1^A\,P_2^B\,P_3^C$ is a homogenous polynomial of the characters of degree $8A+4B+3C$. And for this monomial to be present in $P^{D_{2p+1,1}}_{c_{(1)} = 8s(2p+1)}$ it’s degree needs to be equal to $N=8s$. Thus we look for  non-negative integral solutions to 
\be \label{121c}
8A+4B+3C = N = 8s, 
\ee
and we can obtain all the $\tilde{{\cal Z}}$'s  in $P^{D_{2p+1,1}}_{c_{(1)} = 8s(2p+1)}$.  All except one of the $\tilde{{\cal Z}}$'s  viz. $\tilde{{\cal Z}}_0$ are multiplied by constants the $\tilde{p}$'s.  Which of the solutions of \eqref{121c} corresponds to $\tilde{{\cal Z}}_0$ is easily decided by seeing which of them contains the identity character $\chi_0^N$. Here it is the $A = s, B= 0, C=0$ solution i.e. $P_1^s$. This quick computation of the solutions to \eqref{121c} gives us the same answer as the long procedure.  There is one caveat. We should only consider solutions of \eqref{121c} that give monomials of characters which are commensurate to the identity character. It turns out that every solution of \eqref{121c} is such that $B$ is an even number and $C$ is a mutliple of $8$ or $B$ is odd and $C$ is a multiple of $4$; this is transparent when we write \eqref{121c} as $4B + 3C = 8(s - A)$. For both cases, it follows that  all monomials of characters present in $P_1^A\,P_2^B\,P_3^C$ are commensurate to the identity character. Finally, we recognise that  in this short method, we are asked to find solutions to \eqref{121c} which has no dependence on the theory (no dependence on $p$).  Thus the answers we obtain are also $p$ independent. And we have solved for all theories in one go.

\begin{table}[H]
\begin{center}
\begin{threeparttable}
\rowcolors{2}{Mywhite}{Mygrey}
\resizebox{\textwidth}{!}{
   \renewcommand{\arraystretch}{1.3}
   $\begin{array}{|c|c|c|c|}
\hline
 \#(s) &      N & c_{(1)} & P_{c_{(1)}= 8s(2p+1) }^{D_{(2p+1),1}} (\chi_0, \chi_1, \chi_2) \\
\hline
 1. &   \mathcal E_1[D_{(2p+1),1}^{~\otimes 8}] & 8(2p+1) & {\cal \tilde{Z}}_0 =P_1 ,\quad {\cal \tilde{Z}}_1 =P_2^2 \\
\hline
2. &  \mathcal E_1[D_{(2p+1),1}^{~\otimes 16}] & 16(2p+1) & {\cal \tilde{Z}}_0 =P_1^2,\quad {\cal \tilde{Z}}_1 =P_1 P_2^2,\quad {\cal \tilde{Z}}_2 =P_2^4,\quad {\cal \tilde{Z}}_3 =P_2 P_3^4\\
\hline
3. &   \mathcal E_1[D_{(2p+1),1}^{~\otimes 24}] & 24(2p+1) & {\cal \tilde{Z}}_0 =P_1^3,\quad {\cal \tilde{Z}}_1 =P_1^2 P_2^2,\quad {\cal \tilde{Z}}_2 =P_1 P_2^4,\quad {\cal \tilde{Z}}_3 =P_1 P_2 P_3^4,\\
&&& {\cal \tilde{Z}}_4 =P_2^6,\quad {\cal \tilde{Z}}_5 =P_2^3 P_3^4,\quad {\cal \tilde{Z}}_6 =P_3^8\\
\hline
4. &   \mathcal E_1[D_{(2p+1),1}^{~\otimes 32}] & 32(2p+1) & {\cal \tilde{Z}}_0 =P_1^4,\quad {\cal \tilde{Z}}_1=P_1^3 P_2^2,\quad {\cal \tilde{Z}}_2=P_1^2 P_2^4,\quad {\cal \tilde{Z}}_3=P_1^2 P_2 P_3^4,\\
&&& {\cal \tilde{Z}}_4=P_1 P_2^6,\quad {\cal \tilde{Z}}_5=P_1 P_2^3 P_3^4,\quad {\cal \tilde{Z}}_6=P_1 P_3^8,\quad {\cal \tilde{Z}}_7=P_2^8,\\
&&& {\cal \tilde{Z}}_8=P_2^5 P_3^4,\quad {\cal \tilde{Z}}_9=P_2^2 P_3^8\\
\hline
5. &      \mathcal E_1[D_{(2p+1),1}^{~\otimes 40}] & 40(2p+1) & {\cal \tilde{Z}}_0 =P_1^5,\quad {\cal \tilde{Z}}_1=P_1^4 P_2^2,\quad {\cal \tilde{Z}}_2=P_1^3 P_2^4,\quad {\cal \tilde{Z}}_3=P_1^3 P_2 P_3^4,\\
&&& {\cal \tilde{Z}}_4=P_1^2 P_2^6,\quad {\cal \tilde{Z}}_5=P_1^2 P_2^3 P_3^4,\quad {\cal \tilde{Z}}_6=P_1^2 P_3^8,\quad {\cal \tilde{Z}}_7=P_1 P_2^8,\\
&&& {\cal \tilde{Z}}_8=P_1 P_2^5 P_3^4,\quad {\cal \tilde{Z}}_9=P_1 P_2^2 P_3^8,\quad {\cal \tilde{Z}}_{10}=P_2^{10},\quad {\cal \tilde{Z}}_{11}=P_2^7 P_3^4,\\
&&& {\cal \tilde{Z}}_{12}=P_2^4 P_3^8,\quad {\cal \tilde{Z}}_{13}=P_2 P_3^{12}\\
\hline
6. &   \mathcal E_1[D_{(2p+1),1}^{~\otimes 48}] & 48(2p+1) & {\cal \tilde{Z}}_0 =P_1^6,\quad {\cal \tilde{Z}}_1=P_1^5 P_2^2,\quad {\cal \tilde{Z}}_2=P_1^4 P_2^4,\quad {\cal \tilde{Z}}_3=P_1^4 P_2 P_3^4,\\
&&& {\cal \tilde{Z}}_4=P_1^3 P_2^6,\quad {\cal \tilde{Z}}_5=P_1^3 P_2^3 P_3^4,\quad {\cal \tilde{Z}}_6=P_1^3 P_3^8,\quad {\cal \tilde{Z}}_7=P_1^2 P_2^8,\\
&&& {\cal \tilde{Z}}_8=P_1^2 P_2^5 P_3^4,\quad {\cal \tilde{Z}}_9=P_1^2 P_2^2 P_3^8,\quad {\cal \tilde{Z}}_{10}=P_1 P_2^{10},\quad {\cal \tilde{Z}}_{11}=P_1 P_2^7 P_3^4,\\
&&& {\cal \tilde{Z}}_{12}=P_1 P_2^4 P_3^8,\quad {\cal \tilde{Z}}_{13}=P_1 P_2 P_3^{12},\quad {\cal \tilde{Z}}_{14}=P_2^{12},\quad {\cal \tilde{Z}}_{15}=P_2^9 P_3^4,\\
&&& {\cal \tilde{Z}}_{16}=P_2^6 P_3^8,\quad {\cal \tilde{Z}}_{17}=P_2^3 P_3^{12},\quad {\cal \tilde{Z}}_{18}=P_3^{16}\\
\hline
\end{array}$
}
\end{threeparttable}
\end{center}
\caption{Characters of one-character extensions of tensor products of $D_{2p+1,1}$ written in a basis of $S$-invariant polynomials }\label{t10b}
\end{table}

\subsubsection{$\mathcal C = D_{r,1}, \quad  r = 2(2p +1), \quad p = 1, 2, \ldots $ \label{432s}}

Here, we study all $D_{r,1}$  CFTs with $r = 2 (2p + 1)$ : $D_{6,1}, D_{10,1}, D_{14,1}, \ldots$.   From row 4 of table \ref{t3},  it is clear that  all $D_{r,1}$  CFTs with $r = 2(2p + 1)$  have the same $N$ values; $N = 4s$ and the resulting central charge is $c_{(1)} = 8s(2p+1)$. Hence  for all such $r$, when constructing the ansatz character,  we are looking for monomials of the same degree. Then we note that $h_1$ and $h_2$ values for all  $r = 2(2p + 1)$  have the same structure, i.e. denominators are $4$ and $2$.  What this means is that a monomial which solves the equations \eqref{7b} and \eqref{8b} for any $r= 2(2p + 1)$  also solves the equations for any other such $r$ (but for different values of $M$).   For example, the following is the ansatz character for $D_{6,1}$ and  $N = 4$ :
\bea \label{122c}
\chi_0^4 + b_1\,\chi_0^2  \chi_2^2  + b_2\,\chi_0 \chi_1^2 \chi_2 + b_3\, \chi_2^4 + b_4\, \chi_1^4.
\eea
The integer conformal dimensions for the monomials in \eqref{110c} are respectively, $0, 1, 2, 2, 3$. When one computes the ansatz character for $D_{10,1}$ and $N = 4$, one obtains \eqref{122c} again.  But, when thought of as an ansatz character for $D_{10,1}$, the integer conformal dimensions for the monomials in \eqref{122c} are respectively, $0, 1, 3, 2, 5$.  More generally, for $D_{2(2p+1), 8}$ and $N = 4$, the ansatz character is \eqref{122c} and the integer conformal dimensions for the monomials are respectively, $0, 1, p+1, 2, 2p+1$. What we need to note here is that the the monomials do not depend on the theory i.e. do not depend on $p$ while the conformal dimensions of the monomials do depend on it. We are able to observe this phenomenon from the computation of the ansatz character for $N = 4$ in \eqref{122c}.

We now give a general proof for this. The monomial $\chi_0^a \chi_1^b \chi_2^c$ is present in the ansatz character for $N = 4s$ if the following equations are satisfied :
\be \label{123c}
a + b + c  = N = 4s, \qquad (2p + 1) b + 2 c = 4 M
\ee
for non-negative integral $M$. We will be able to show that every solution of \eqref{123c} depends on $s$ and not on the theory $p$. We will also be able to show that the solutions will have a conformal dimension $M$ which does depend on $p$. To solve \eqref{123c}, we first note, from the second equation, that $b$ should be a multiple of $2$. Let us designate $b = 2l$. From the fact that $b \leq 4s$, it then follows that $l$ can take values $0, 1, 2, \ldots, 2s-1, 2s$. It then also follows that when $l$ is an odd number, $c$ should also be an odd number, as well as $a$; simillarly when $l$ is an even number, $c$ should also be an even number, as well as $a$. The values $c$ and $a$  take, can be determined and we find two classes of monomial solutions for the ansatz character :
\be \label{124c}
\sum_{l \in L} \left( \sum_{k \in K}~b_{k,l} ~\chi_0^{4s - 2l - k} \chi_1^{2l} \chi_2^k\right) \quad + \quad  \sum_{l' \in L'} \left( \sum_{k' \in K'}~b_{k',l'} ~\chi_0^{4s - 2l' - k'} \chi_1^{2l'} \chi_2^{k'}\right)
\ee
where
\bea
&L = \{1, 3, \ldots, 2s-1 \}, \qquad K = \{1, 3, \ldots, 4s-2l-1\} \nonumber \\
&\text{and} \nonumber \\
&L' = \{0, 2, \ldots, 2s \}, \qquad K' = \{ 0, 2, \ldots, 4s-2l \}. \nonumber
\eea
Note that there is no $p$ dependence in the ansatz character of \eqref{124c}. The $p$ dependence shows up when one computes the conformal dimensions.  The character multiplying $b_{k,l}$ has  a conformal dimension $\frac{(2p+1)l + k}{2}$ and the character multiplying $b_{k',l'}$ has  a conformal dimension $\frac{(2p+1)l' + k'}{2}$.

Now, we proceed to complete the procedure laid out in section \ref{2s}.  We consider the S-invariance equations \eqref{18b}. Since the $S$-matrix for all $D_{r,1}$  CFTs are identical,  we get the same set of linear equations for all $r = 2(2p+1)$.  And then we solve the  overdetermined system of linear equations. The answers are tabulated in table \ref{t11b}.

We start with smallest possible $c_{(1)}$. This is the $N = 4,  c_{(1)} = 8(2p + 1)$ case. The ansatz character in this case has $4$ arbitrary constants, the $b$’s, see \eqref{122c}. The $S$-invariance equations are an overdetermined system of $15$ equations for these $4$ arbitrary variables. It was quite surprising that there are two $S$-invariant polynomials, $P_1$ and  $P_3$, hidden within the answer :
\be \label{125c}
P_{c_{(1)} = 8(2p + 1)}^{D_{2(2p +1),1}}  = P_1 + \tilde{p}_1\, P_3^2 
\ee
Then, we take on the next case $N = 8,  c_{(1)} = 16(2p + 1)$.  The ansatz character \eqref{124c} has $12$ arbitrary constants, the $b$'s.   The $S$-invariance equations are an overdetermined system of $45$ equations for these $12$ arbitrary variables. We managed to find the following solution that has hidden in it a third $S$-invariant polynomial, viz. $P_2$ :
\be \label{126c}
P_{c_{(1)} = 16(2p + 1)}^{D_{2(2p +1),1}}  = {\cal Z}_0 + p_1\, {\cal Z}_1 + p_2\, {\cal Z}_2 + p_3\, {\cal Z}_3 
\ee
where
\begin{small}
\begin{align}
   {\cal Z}_0& =  P_1^2-24 P_2^2 P_3-2 P_3^4, \qquad  {\cal Z}_1 = P_1 P_3^2+2 P_2^2 P_3 \nonumber \\
   {\cal Z}_2& = P_2^2 P_3,  \qquad   {\cal Z}_3 = P_3^4. \label{127c}
\end{align}
\end{small}
We can put the above in the following more-compact form :
\be \label{128c}
P_{c_{(1)} = 16(2p + 1)}^{D_{2(2p +1),1}}  = 
\tilde{{\cal Z}}_0+ \tilde{p}_1\, \tilde{{\cal Z}}_1 + \tilde{p}_2\, \tilde{{\cal Z}}_2 + \tilde{p}_3\, \tilde{{\cal Z}}_3 
\ee
where
\be \label{129c}
\tilde{{\cal Z}}_0 = P_1^2, \quad \tilde{{\cal Z}}_1 = P_1 P_3^2, \quad \tilde{{\cal Z}}_2 = P_2^2 P_3, \quad \tilde{{\cal Z}}_3 = P_3^4
\ee
The relation between the constant coefficients of \eqref{126c} and \eqref{128c} are :
\begin{small}
\begin{align}
         \tilde{p}_1& =  p_1, \qquad  \tilde{p}_2 =  - 24 + 2p_1 + p_2, \qquad \tilde{p}_3 = - 2 + p_3
 \end{align}
\end{small}
The important outcome of the computations so far  is that we have a set of three $S$-invariant polynomials. And we expect these three to form a better basis for $S$-invariant characters than the basis one started with viz. $\chi_0, \chi_1, \chi_2$. We collect all the three $S$-invariant polynomials for $D_{2(2p+1),1}$ below :
\begin{equation}\label{60ad}
\begin{array}{lll}
   P_1(\chi_0, \chi_1, \chi_2) &=\chi_0^4\,+\,12\, \chi_0\, \chi_1^2\, \chi_2\,+\,2 \,\chi_1^4\,+\,\chi_2^4\\
   P_2(\chi_0, \chi_1, \chi_2) &=\chi_0^2\,\chi_1 - \chi_1\,\chi_2^2 = 2^{r-1}\\
   P_3(\chi_0, \chi_1, \chi_2) &=\chi_0 \chi_2-\chi_1^2
\end{array}
\end{equation}
We note that each of the $S$-invariant polynomials has a $q$-series that depends on $r$.  $P_1$, for all values of $r$ is a non-constant $S$-invariant polynomial. $P_2$ is always a constant $S$-invariant polynomial, for all values of $r$, taking the value $2^{r-1}$. $P_3$ turns out to be constant $S$-invariant polynomial only for $r = 6$, taking the value  $12$ and is a non-constant polynomial for $r \geq 10$. 

We have gone ahead and computed for $N = 12, 16, 20, 24$.  The number of arbitrary coefficients, the $b$'s in the ansatz character for these cases are respectively, $24$, $40$, $60$ and $84$. The number of equations that one needs to solve in these cases are respectively, $91$, $153$, $231$ and $325$. We have tabulated the results in table \ref{t11b}. It should be noted that we give the most compact form of the results. In every case, we give the equivalent of \eqref{128c} rather than \eqref{126c}. The number of $S$-invariant summands, the $\tilde{{\cal Z}}_i$'s in the six cases $N = 4, 8, 12,  16, 20, 24$ are respectively, $2, 4, 7, 10, 14, 19$. We have this information only after a long process of solving a huge number of equations. Below, we will give a quick method to get the number of summands as well as the summands themselves, without having to solve the equations. 

The way to read the results from table \eqref{t11b} is as follows.   The first column is the serial no. ($s$). The second column contains the number of tensor product factors $N$ (which equals $4s$). The third column contains the value of $c_{(1)} = 8s(2p+1)$.  The fourth column contains the character $P^{D_{2(2p+1),1}}_{c_{(1)} = 8s(2p+1)}(\chi_0, \chi_1)$.  We should emphasise that the table \ref{t11b} contains the characters of the one-character extensions of tensor products of an infinite number of CFTs. It is remarkable that we are able to encode an infinite amount of information in such a finite compact form. We do not give the $q$-series of the characters in table \ref{t11b} due to that information being dependent on $r$. 

Now, we should consider admissibility. There are an infinite number of CFTs to study. We only study three cases viz. $r = 6, 10, 14$. We have computed the $q$-series of the characters that can be read off from table \ref{t11b} and we always find infinite numbers of admissible characters at each central charge. 

Amongst all the admissible characters that we have found, only one of them corresponds to a genuine CFT. This happens for $D_{6,1}$ when $N = 4$. The character is given in \eqref{125c}.  Noting that for $r=6$, $P_1 = j-480$ and $P_3 = 12$, we find 
\be \label{132c}
P_{c_{(1)} = 24}^{D_{6,1}} = j - 480 + 144 \tilde{p}_1.
\ee
There are an infinite number of admissible characters in \eqref{132c} at central charge $24$. To answer the question as which of them correspond to a genuine CFT, we once again invoke the classification of one-character $c = 24$ CFTs of \cite{Schellekens:1992db}. In the table at the end of that paper entry no. 54 is the only CFT which is a one-character extension of a tensor product of  copies of $D_{6,1}$\footnote{The reader can find this CFT listed as entry no. 20 of table \ref{t0}.}.  This CFT has a character  $j - 480$. Thus amongst the infinite characters \eqref{132c} that we found, only one of them, the one with $\tilde{p}_1 = 0$ corresponds to an actual CFT.

Further considerations (beyond those in this paper and) establish \cite{Schellekens:1992db}, in the case of $D_{6,1}$,  that  the single character obtained for $N = 4$ viz. $P_1$ corresponds to a genuine CFT. We suspect that these further considerations would  establish that the same character for all the other $D_{2(2p+1),1}$'s also  corresponds to genuine CFT(s). We thus are led to the following surmise/conjecture : \emph{For $D_{2(2p+1),1}$, $p \geq 2$  there exist genuine one-character CFT(s) with central charge $8(2p+1)$ with character $P_1$.}

\underline{Lessons learnt and a new method} :  All the results that we have obtained and reported in table \ref{t11b} were by running the whole procedure as laid out  in section \ref{2s} . Similar to what we did in section \ref{3s}, we have a simpler  procedure that allows us to write the final answer quickly.  We first note that $P_1$ is a homogenous polynomial of the characters of degree $4$,   $P_2$ is a homogenous polynomial of the characters of degree $3$ and $P_3$ is a homogenous polynomial of the characters of degree $2$.  A monomial of the form $P_1^A\,P_2^B\,P_3^C$ is a homogenous polynomial of the characters of degree $4A+3B+2C$. And for this monomial to be present in $P^{D_{2(2p+1),1}}_{c_{(1)} = 8s(2p+1)}$ it’s degree needs to be equal to $N=4s$. Thus we look for  non-negative integral solutions to 
\be \label{133c}
4A+3B+2C = N = 4s, 
\ee
and we can obtain all the $\tilde{{\cal Z}}$'s  in $P^{D_{2(2p+1),1}}_{c_{(1)} = 8s(2p+1)}$.  All except one of the $\tilde{{\cal Z}}$'s  viz. $\tilde{{\cal Z}}_0$ are multiplied by constants the $\tilde{p}$'s.  Which of the solutions of \eqref{133c} corresponds to $\tilde{{\cal Z}}_0$ is easily decided by seeing which of them contains the identity character $\chi_0^N$. Here it is the $A = s, B= 0, C=0$ solution i.e. $P_1^s$. This quick computation of the solutions to \eqref{133c} gives us the same answer as the long procedure.  There is one caveat. We should only consider solutions of \eqref{133c} that give monomials of characters which are commensurate to the identity character.

It turns out that every solution of \eqref{133c} is such that $C$ is an even number and $B$ is a mutliple of $4$ or $C$ is odd and $B$ is a multiple of $2$; this is transparent when we write \eqref{133c} as $3B + 2C = 4(s - A)$. For both cases, it follows that  all monomials of characters present in $P_1^A\,P_2^B\,P_3^C$ are commensurate to the identity character. Finally, we recognise that  in this short method, we are asked to find solutions to \eqref{133c} which has no dependence on the theory (no dependence on $p$).  Thus the answers we obtain are also $p$ independent. And we have solved for all theories in one go.  

\begin{table}[H]
\begin{center}
\begin{threeparttable}
\rowcolors{2}{Mywhite}{Mygrey}
\resizebox{\textwidth}{!}{
   \renewcommand{\arraystretch}{1.3}
   $\begin{array}{|c|c|c|C{9cm}|}
\hline
 \#(s) &      N & c_{(1)} & P_{c_{(1)} = 8s(2p+1) }^{D_{2(2p+1),1}} (\chi_0, \chi_1, \chi_2) \\
\hline
1. &  \mathcal E_1[D_{2(2p+1),1}^{~\otimes 4}] & 8(2p+1) & {\cal \tilde{Z}}_0 =P_1 ,\  {\cal \tilde{Z}}_1 =P_2^2 \\
\hline
2. &   \mathcal E_1[D_{2(2p+1),1}^{~\otimes 8}] & 16(2p+1) &  {\cal \tilde{Z}}_{0}=P_1^2,\   {\cal \tilde{Z}}_{1}=P_1 P_2^2,\   {\cal \tilde{Z}}_{2}=P_2 P_3^2,\   {\cal \tilde{Z}}_{3}=P_2^4 \\
\hline
3. &   \mathcal E_1[D_{2(2p+1),1}^{~\otimes 12}] & 24(2p+1) &  {\cal \tilde{Z}}_{0}=P_1^3,\   {\cal \tilde{Z}}_{1}=P_1^2 P_2^2,\  {\cal \tilde{Z}}_{2}=P_1 P_2 P_3^2,\  {\cal \tilde{Z}}_{3}=P_1 P_2^4,\\
&&& {\cal \tilde{Z}}_{4}=P_3^4,\  {\cal \tilde{Z}}_{5}=P_2^3 P_3^2,\  {\cal \tilde{Z}}_{6}=P_2^6\\
\hline
4. &   \mathcal E_1[D_{2(2p+1),1}^{~\otimes 16}] & 32(2p+1) & {\cal \tilde{Z}}_{0}=P_1^4,\   {\cal \tilde{Z}}_{1}=P_1^3 P_2^2,\   {\cal \tilde{Z}}_{2}=P_1^2 P_2 P_3^2,\   {\cal \tilde{Z}}_{3}=P_1^2 P_2^4,\\
&&& {\cal \tilde{Z}}_{4}=P_1 P_3^4,\   {\cal \tilde{Z}}_{5}=P_1 P_2^3 P_3^2,\   {\cal \tilde{Z}}_{6}=P_1 P_2^6,\   {\cal \tilde{Z}}_{7}=P_2^2 P_3^4,\\
&&& {\cal \tilde{Z}}_{8}=P_2^5 P_3^2,\   {\cal \tilde{Z}}_{9}=P_2^8\\
\end{array}$
}
\end{threeparttable}
\end{center}
\end{table}

\begin{table}[H]
\begin{center}
\begin{threeparttable}
\rowcolors{2}{Mywhite}{Mygrey}
\resizebox{\textwidth}{!}{
   \renewcommand{\arraystretch}{1.3}
   $\begin{array}{|c|c|c|C{10cm}|}
5. &   \mathcal E_1[D_{2(2p+1),1}^{~\otimes 20}] & 40(2p+1) & {\cal \tilde{Z}}_{0}=P_1^5,\   {\cal \tilde{Z}}_{1}=P_1^4 P_2^2,\   {\cal \tilde{Z}}_{2}=P_1^3 P_2 P_3^2,\   {\cal \tilde{Z}}_{3}=P_1^3 P_2^4,\\
&&& {\cal \tilde{Z}}_{4}=P_1^2 P_3^4,\   {\cal \tilde{Z}}_{5}=P_1^2 P_2^3 P_3^2,\   {\cal \tilde{Z}}_{6}=P_1^2 P_2^6,\   {\cal \tilde{Z}}_{7}=P_1 P_2^2 P_3^4,\\
&&& {\cal \tilde{Z}}_{8}=P_1 P_2^5 P_3^2,\   {\cal \tilde{Z}}_{9}=P_1 P_2^8,\   {\cal \tilde{Z}}_{10}=P_2 P_3^6,\   {\cal \tilde{Z}}_{11}=P_2^4 P_3^4,\\
&&& {\cal \tilde{Z}}_{12}=P_2^7 P_3^2,\   {\cal \tilde{Z}}_{13}=P_2^{10}\\
\hline
6. &   \mathcal E_1[D_{2(2p+1),1}^{~\otimes 24}] & 48(2p+1) & {\cal \tilde{Z}}_{0}=P_1^6,\   {\cal \tilde{Z}}_{1}=P_1^5 P_2^2,\   {\cal \tilde{Z}}_{2}=P_1^4 P_2 P_3^2,\   {\cal \tilde{Z}}_{3}=P_1^4 P_2^4,\\
&&& {\cal \tilde{Z}}_{4}=P_1^3 P_3^4,\   {\cal \tilde{Z}}_{5}=P_1^3 P_2^3 P_3^2,\   {\cal \tilde{Z}}_{6}=P_1^3 P_2^6,\   {\cal \tilde{Z}}_{7}=P_1^2 P_2^2 P_3^4,\\
&&& {\cal \tilde{Z}}_{8}=P_1^2 P_2^5 P_3^2,\   {\cal \tilde{Z}}_{9}=P_1^2 P_2^8,\   {\cal \tilde{Z}}_{10}=P_1 P_2 P_3^6,\   {\cal \tilde{Z}}_{11}=P_1 P_2^4 P_3^4,\\
&&& {\cal \tilde{Z}}_{12}=P_1 P_2^7 P_3^2,\   {\cal \tilde{Z}}_{13}=P_1 P_2^{10},\   {\cal \tilde{Z}}_{14}=P_3^8,\   {\cal \tilde{Z}}_{15}=P_2^3 P_3^6,\\
&&& {\cal \tilde{Z}}_{16}=P_2^6 P_3^4,\   {\cal \tilde{Z}}_{17}=P_2^9 P_3^2,\   {\cal \tilde{Z}}_{18}=P_2^{12}\\
\hline
\end{array}$
}
\end{threeparttable}
\end{center}
\caption{Characters of one-character extensions of tensor products of $D_{2(2p+1),1}$ written in a basis of $S$-invariant polynomials}\label{t11b}
\end{table}

\subsubsection{$\mathcal C = D_{r,1}, \quad  r = 4(2p +1), \quad p = 1, 2, \ldots$ \label{433s}}

Here, we study all $D_{r,1}$  CFTs with $r = 4(2p + 1)$ : $D_{12,1}, D_{20,1}, D_{28,1}, \ldots$.   From row 5 of table \ref{t3},  it is clear that  all $D_{r,1}$  CFTs with $r = 4(2p + 1)$  have the same $N$ values; $N = 2s$ and the resulting central charge is $c_{(1)} = 8s(2p+1)$. Hence  for all such $r$, when constructing the ansatz character,  we are looking for monomials of the same degree. Then we note that $h_1$ and $h_2$ values for all  $r = 4(2p + 1)$  have the same structure, i.e. denominators are $2$ and $2$.  What this means is that a monomial which solves the equations \eqref{7b} and \eqref{8b} for any odd $r$  also solves the equations for any other odd $r$ (but for different values of $M$).   For example, the following is the ansatz character for $D_{12,1}$ and  $N = 2$ :
\be \label{134c}
\chi_0^2 + b_1\,\chi_2^2 + b_2\,\chi_1 \chi_2 + b_3\,\chi_1^2.
\ee
The integer conformal dimensions for the monomials in \eqref{134c} are respectively, $0, 1, 2, 3$. When one computes the ansatz character for $D_{20,1}$ and $N = 2$, one obtains \eqref{134c} again.  But, when thought of as an ansatz character for $D_{20,1}$, the integer conformal dimensions for the monomials in \eqref{134c} are respectively, $0, 1, 3, 5$.  More generally, for $D_{2p+1, 8}$ and $N = 4$, the ansatz character is \eqref{134c} and the integer conformal dimensions for the monomials are respectively, $0, 1, p+1, 2p+1$. What we need to note here is that the the monomials do not depend on the theory i.e. do not depend on $p$ while the conformal dimensions of the monomials do depend on it. We are able to observe this phenomenon from the computation of the ansatz character for $N = 2$ in \eqref{134c}.

 We now give a general proof for this. The monomial $\chi_0^a \chi_1^b \chi_2^c$ is present in the ansatz character for $N = 2s$ if the following equations are satisfied :
\be \label{135c}
a + b + c  = N = 2s, \qquad (2p + 1) b +  c = 2 M
\ee
for non-negative integral $M$. We will be able to show that every solution of \eqref{135c} depends on $s$ and not on the theory $p$. We will also be able to show that the solutions will have a conformal dimension $M$ which does depend on $p$. To solve \eqref{135c}, we first note, from the second equation, that either both $b$ and $c$ are odd or both $b$ and $c$ are even.  Let us designate $b = l$ (partly as a continuity of notation from sections \ref{431s} and \ref{432s}). From the fact that $b \leq 2s$, we have that  $l$ takes values $0, 1, 2, \ldots, 2s-1, 2s$.  The values $c$ and $a$  take, can be determined and we find two classes of monomial solutions for the ansatz character :
\be
\sum_{l \in L} \left( \sum_{k \in K}~b_{k,l} ~\chi_0^{2s - l - k} \chi_1^{l} \chi_2^k\right) \quad + \quad  \sum_{l' \in L'} \left( \sum_{k' \in K'}~b_{k',l'} ~\chi_0^{2s - l' - k'} \chi_1^{l'} \chi_2^{k'}\right)
\ee \label{136c}
where
\bea
&L = \{1, 3, \ldots, 2s-1 \}, \qquad K = \{1, 3, \ldots, 2s-l\} \nonumber \\
&\text{and} \nonumber \\
&L' = \{0, 2, \ldots, 2s \}, \qquad K' = \{ 0, 2, \ldots, 2s-l \}. \nonumber
\eea
Note that there is no $p$ dependence in the ansatz character of \eqref{136c}. The $p$ dependence shows up when one computes the conformal dimensions.  The character multiplying $b_{k,l}$ has  a conformal dimension $\frac{(2p+1)l + k}{2}$ and the character multiplying $b_{k',l'}$ has  a conformal dimension $\frac{(2p+1)l' + k'}{2}$.

Now, we proceed to complete the procedure laid out in section \ref{2s}.  We consider the S-invariance equations \eqref{18b}. Since the $S$-matrix for all $D_{r,1}$  CFTs are identical,  we get the same set of linear equations for all  $r=4(2p+1)$.  And then we solve the  overdetermined system of linear equations. The answers are tabulated in table \ref{t12b}.

We start with smallest possible $c_{(1)}$. This is the $N = 2,  c_{(1)} = 8(2p + 1)$ case. The ansatz character in this case has $3$ arbitrary constants, the $b$’s, see \eqref{134c}. The $S$-invariance equations are an overdetermined system of $6$ equations for these $3$ arbitrary variables. It was quite surprising that there are two $S$-invariant polynomials, $P_1$ and  $P_3$, hidden within the answer :
\be \label{137c}
P_{c_{(1)} = 8(2p + 1)}^{D_{4(2p +1),1}}  = P_1 + \tilde{p}_1\, P_3^2 
\ee
Then, we take on the next case $N = 4,  c_{(1)} = 16(2p + 1)$.  The ansatz character \eqref{136c} has $8$ arbitrary constants, the $b$'s.   The $S$-invariance equations are an overdetermined system of $15$ equations for these $8$ arbitrary variables. We managed to find the following solution that has hidden in it a third $S$-invariant polynomial, viz. $P_2$ :
\be \label{138c}
P_{c_{(1)} = 16(2p + 1)}^{D_{4(2p +1),1}}  = {\cal Z}_0 + p_1\, {\cal Z}_1 + p_2\, {\cal Z}_2 + p_3\, {\cal Z}_3 
\ee
where
\begin{small}
\begin{align}
   {\cal Z}_0& =  P_1^2+8 P_2 P_3+8 P_1 P_3^2, \qquad  {\cal Z}_1 = -2 P_2 P_3-P_1 P_3^2 \nonumber \\
   {\cal Z}_2& = -P_2 P_3-P_1 P_3^2,  \qquad   {\cal Z}_3 = P_3^4. \label{139c}
\end{align}
\end{small}
We can put the above in the following more-compact form :
\be \label{140c}
P_{c_{(1)} = 16(2p + 1)}^{D_{4(2p +1),1}}  = 
\tilde{{\cal Z}}_0+ \tilde{p}_1\, \tilde{{\cal Z}}_1 + \tilde{p}_2\, \tilde{{\cal Z}}_2 + \tilde{p}_3\, \tilde{{\cal Z}}_3 
\ee
where
\be \label{141c}
\tilde{{\cal Z}}_0 = P_1^2, \quad \tilde{{\cal Z}}_1 = P_2 P_3, \quad \tilde{{\cal Z}}_2 = P_1 P_3^2, \quad \tilde{{\cal Z}}_3 = P_3^4
\ee
The relation between the constant coefficients of \eqref{138c} and \eqref{140c} are :
\begin{small}
\begin{align}
         \tilde{p}_1& =  8 -2p_1 - p_2, \qquad  \tilde{p}_2 =  8 -p_1 - p_2, \qquad \tilde{p}_3 =  p_3
 \end{align}
\end{small}
The important outcome of the computations so far  is that we have a set of three $S$-invariant polynomials. And we expect these three to form a better basis for $S$-invariant characters than the basis one started with viz. $\chi_0, \chi_1, \chi_2$. We collect all the three $S$-invariant polynomials for $D_{4(2p+1),1}$ below :
\begin{equation}\label{61a}
\begin{array}{lll}
   P_1(\chi_0, \chi_1, \chi_2) &=\chi_0^2\,+\,4\, \chi_1\, \chi_2\,-\,\chi_2^2 \qquad \qquad \\
   P_2(\chi_0, \chi_1, \chi_2) &=\chi_0^2\,\chi_1 - \chi_1\,\chi_2^2 = 2^{r-1} \\
   P_3(\chi_0, \chi_1, \chi_2) &=\chi_2-\chi_1
\end{array}
\end{equation}
We note that each of the $S$-invariant polynomials has a $q$-series that depends on $r$.  $P_1$, for all values of $r$ is a non-constant $S$-invariant polynomial. $P_2$ is always a constant $S$-invariant polynomial, for all values of $r$, taking the value $2^{r-1}$. $P_3$ turns out to be constant $S$-invariant polynomial only for $r = 12$, taking the value $24$ and is a non-constant polynomial for $r \geq 20$.

We have gone ahead and computed for $N = 6, 8, 10, 12$.  The number of arbitrary coefficients, the $b$'s in the ansatz character for these cases are respectively, $15$, $24$, $35$ and $48$. The number of equations that one needs to solve in these cases are respectively, $28$, $45$, $66$ and $91$. We have tabulated the results in table \ref{t12b}. It should be noted that we give the most compact form of the results. In every case, we give the equivalent of \eqref{140c} rather than \eqref{138c}. The number of $S$-invariant summands, the $\tilde{{\cal Z}}_i$'s in the six cases $N = 2, 4, 6,  8, 10, 12$ are respectively, $2, 4, 7, 10, 14, 19$. We have this information only after a long process of solving a huge number of equations. Below, we will give a quick method to get the number of summands as well as the summands themselves, without having to solve the equations. 

The way to read the results from table \eqref{t12b} is as follows.   The first column is the serial no. ($s$). The second column contains the number of tensor product factors $N$ (which equals $2s$). The third column contains the value of $c_{(1)} = 8s(2p+1)$.  The fourth column contains the character $P^{D_{4(2p+1),1}}_{c_{(1)} = 8s(2p+1)}(\chi_0, \chi_1)$.  We should emphasise that the table \ref{t12b} contains the characters of the one-character extensions of tensor products of an infinite number of CFTs. It is remarkable that we are able to encode an infinite amount of information in such a finite compact form. We do not give the $q$-series of the characters in table \ref{t12b} due to that information being dependent on $r$. 

Now, we should consider admissibility. There are an infinite number of CFTs to study. We only study three cases viz. $r = 12, 20, 28$. We have computed the $q$-series of the characters that can be read off from table \ref{t12b} and we always find infinite numbers of admissible characters at each central charge. 

Amongst all the admissible characters that we have found, only one of them corresponds to a genuine CFT. This happens for $D_{12,1}$ when $N = 2$. The character is given in \eqref{137c}.  Noting that for $r=12$, $P_1 = j-768$ and $P_3 = 24$, we find 
\be \label{145c}
P_{c_{(1)} = 24}^{D_{12,1}} = j - 768 + 576 \tilde{p}_1.
\ee
There are an infinite number of admissible characters in \eqref{145c} at central charge $24$. To answer the question as which of them correspond to a genuine CFT, we once again invoke the classification of one-character $c = 24$ CFTs of \cite{Schellekens:1992db}. In the table at the end of that paper entry no. 66 is the only CFT which is a one-character extension of a tensor product of  copies of $D_{12,1}$\footnote{The reader can find this CFT listed as entry no. 25 of table \ref{t0}.}.  This CFT has a character  $j - 192$. Thus amongst the infinite characters \eqref{145c} that we found, only one of them, the one with $\tilde{p}_1 = 1$ corresponds to an actual CFT. 

Further considerations (beyond those in this paper and) establish \cite{Schellekens:1992db}, in the case of $D_{12,1}$,  that  the single character obtained for $N = 2$ viz. $P_1 +  P_3^2$ corresponds to a genuine CFT. We suspect that these further considerations would  establish that the same character for all the other $D_{4(2p+1),1}$'s also  corresponds to genuine CFT(s). We thus are led to the following surmise/conjecture : \emph{For $D_{4(2p+1),1}$, $p \geq 2$  there exist genuine one-character CFT(s) with central charge $8(2p+1)$ with character $P_1 +  P_3^2$.}

\underline{Lessons learnt and a new method} :  All the results that we have obtained and reported in table \ref{t12b} were by running the whole procedure as laid out  in section \ref{2s} . Similar to what we did in section \ref{3s}, we have a simpler  procedure that allows us to write the final answer quickly.  We first note that $P_1$ is a homogenous polynomial of the characters of degree $2$,   $P_2$ is a homogenous polynomial of the characters of degree $3$ and $P_3$ is a homogenous polynomial of the characters of degree $1$.  A monomial of the form $P_1^A\,P_2^B\,P_3^C$ is a homogenous polynomial of the characters of degree $2A+3B+C$. And for this monomial to be present in $P^{D_{2p+1,1}}_{c_{(1)} = 8s(2p+1)}$ it’s degree needs to be equal to $N=2s$. Thus we look for  non-negative integral solutions to 
\be \label{146c}
2A+3B+C = N = 2s, 
\ee
and we can obtain all the $\tilde{{\cal Z}}$'s  in $P^{D_{4(2p+1),1}}_{c_{(1)} = 8s(2p+1)}$.  All except one of the $\tilde{{\cal Z}}$'s  viz. $\tilde{{\cal Z}}_0$ are multiplied by constants the $\tilde{p}$'s.  Which of the solutions of \eqref{146c} corresponds to $\tilde{{\cal Z}}_0$ is easily decided by seeing which of them contains the identity character $\chi_0^N$. Here it is the $A = s, B= 0, C=0$ solution i.e. $P_1^s$. This quick computation of the solutions to \eqref{146c} gives us the same answer as the long procedure.  There is one caveat.  We should only consider solutions of \eqref{146c} that give monomials of characters which are commensurate to the identity character.  It turns out that every solution of \eqref{146c} is such that both $B$ and $C$ are even numbers or both are odd numbers; this is transparent when we write \eqref{146c} as $3B + C = 2(s - A)$. For both cases, it follows that  all monomials of characters present in $P_1^A\,P_2^B\,P_3^C$ are commensurate to the identity character. Finally, we recognise that  in this short method, we are asked to find solutions to \eqref{146c} which has no dependence on the theory (no dependence on $p$).  Thus the answers we obtain are also $p$ independent. And we have solved for all theories in one go.

\begin{table}[H]
\begin{center}
\begin{threeparttable}
\rowcolors{2}{Mywhite}{Mygrey}
\resizebox{\textwidth}{!}{
   \renewcommand{\arraystretch}{1.3}
   $\begin{array}{|c|c|c|c|}
\hline
 \#(s) &      N & c_{(1)} & P_{c_{(1) = 8s(2p+1)}}^{D_{4(2p+1),1}} (\chi_0, \chi_1, \chi_2)  \\
\hline
1. &   \mathcal E_1[D_{4(2p+1),1}^{~\otimes 2}] & 8(2p+1) & {\cal \tilde{Z}}_{0}=P_1,\   {\cal \tilde{Z}}_{1}=P_2^2\\
\hline
2. &   \mathcal E_1[D_{4(2p+1),1}^{~\otimes 4}] & 16(2p+1) & {\cal \tilde{Z}}_{0}=P_1^2,\   {\cal \tilde{Z}}_{1}=P_1 P_2^2,\   {\cal \tilde{Z}}_{2}=P_2 P_3,\   {\cal \tilde{Z}}_{3}=P_2^4\\
\hline
3. &   \mathcal E_1[D_{4(2p+1),1}^{~\otimes 6}] & 24(2p+1) & {\cal \tilde{Z}}_{0}=P_1^3,\   {\cal \tilde{Z}}_{1}=P_1^2 P_2^2,\   {\cal \tilde{Z}}_{2}=P_1 P_2 P_3,\   {\cal \tilde{Z}}_{3}=P_1 P_2^4,\\
&&& {\cal \tilde{Z}}_{4}=P_3^2,\   {\cal \tilde{Z}}_{5}=P_2^3 P_3,\   {\cal \tilde{Z}}_{6}=P_2^6\\
\hline
4. &   \mathcal E_1[D_{4(2p+1),1}^{~\otimes 8}] & 32(2p+1) & {\cal \tilde{Z}}_{0}=P_1^4,\   {\cal \tilde{Z}}_{1}=P_1^3 P_2^2,\   {\cal \tilde{Z}}_{2}=P_1^2 P_2 P_3,\   {\cal \tilde{Z}}_{3}=P_1^2 P_2^4,\\
&&& {\cal \tilde{Z}}_{4}=P_1 P_3^2,\   {\cal \tilde{Z}}_{5}=P_1 P_2^3 P_3,\   {\cal \tilde{Z}}_{6}=P_1 P_2^6,\   {\cal \tilde{Z}}_{7}=P_2^2 P_3^2,\\
&&& {\cal \tilde{Z}}_{8}=P_2^5 P_3,\   {\cal \tilde{Z}}_{9}=P_2^8\\
\hline
5. &   \mathcal E_1[D_{4(2p+1),1}^{~\otimes 10}] & 40(2p+1) & {\cal \tilde{Z}}_{0}=P_1^5,\   {\cal \tilde{Z}}_{1}=P_1^4 P_2^2,\   {\cal \tilde{Z}}_{2}=P_1^3 P_2 P_3,\   {\cal \tilde{Z}}_{3}=P_1^3
      P_2^4\\ &&& {\cal \tilde{Z}}_{4}=P_1^2 P_3^2,\   {\cal \tilde{Z}}_{5}=P_1^2 P_2^3 P_3,\   {\cal \tilde{Z}}_{6}=P_1^2 P_2^6,\   {\cal \tilde{Z}}_{7}=P_1 P_2^2 P_3^2,\\
&&& {\cal \tilde{Z}}_{8}=P_1 P_2^5 P_3,\   {\cal \tilde{Z}}_{9}=P_1 P_2^8,\   {\cal \tilde{Z}}_{10}=P_2 P_3^3,\   {\cal \tilde{Z}}_{11}=P_2^4 P_3^2,\\
&&& {\cal \tilde{Z}}_{12}=P_2^7 P_3,\   {\cal \tilde{Z}}_{13}=P_2^{10}\\
\hline
6. &   \mathcal E_1[D_{4(2p+1),1}^{~\otimes 12}] & 48(2p+1) & {\cal \tilde{Z}}_{0}=P_1^6,\   {\cal \tilde{Z}}_{1}=P_1^5 P_2^2,\   {\cal \tilde{Z}}_{2}=P_1^4 P_2 P_3,\   {\cal \tilde{Z}}_{3}=P_1^4 P_2^4,\\
&&& {\cal \tilde{Z}}_{4}=P_1^3 P_3^2,\   {\cal \tilde{Z}}_{5}=P_1^3 P_2^3 P_3,\   {\cal \tilde{Z}}_{6}=P_1^3 P_2^6,\   {\cal \tilde{Z}}_{7}=P_1^2 P_2^2 P_3^2,\\
&&& {\cal \tilde{Z}}_{8}=P_1^2 P_2^5 P_3,\   {\cal \tilde{Z}}_{9}=P_1^2 P_2^8,\   {\cal \tilde{Z}}_{10}=P_1 P_2 P_3^3,\   {\cal \tilde{Z}}_{11}=P_1 P_2^4 P_3^2,\\
&&& {\cal \tilde{Z}}_{12}=P_1 P_2^7 P_3,\   {\cal \tilde{Z}}_{13}=P_1 P_2^{10},\   {\cal \tilde{Z}}_{14}=P_3^4,\   {\cal \tilde{Z}}_{15}=P_2^3 P_3^3,\\
&&& {\cal \tilde{Z}}_{16}=P_2^6 P_3^2,\   {\cal \tilde{Z}}_{17}=P_2^9 P_3,\   {\cal \tilde{Z}}_{18}=P_2^{12}\\
\hline
\end{array}$
}
\end{threeparttable}
\end{center}
\caption{Characters of one-character extensions of tensor products of $D_{4(2p+1),1}$ written in a basis of $S$-invariant polynomials}\label{t12b}
\end{table}

\subsubsection{$\mathcal C = D_{r,1}, \quad  r = 8p, \quad p = 1, 2, \ldots$ \label{434s}}

Here, we study all $D_{r,1}$  CFTs with $r = 8p$ : $D_{8,1}, D_{16,1}, D_{24,1}, \ldots$.   From row 6 of table \ref{t3},  it is clear that  all $D_{r,1}$  CFTs with $r = 8p$  have the same $N$ values; $N = s$ and the resulting central charge is $c_{(1)} = 8sp$. Hence  for all such $r$, when constructing the ansatz character,  we are looking for monomials of the same degree. Then we note that $h_1$ and $h_2$ values for all  $r = 8p$  have the same structure, i.e.  denominators are $2$ and $1$.  What this means is that a monomial which solves the equations \eqref{7b} and \eqref{8b} for any  $r = 8p$  also solves the equations for any other such $r$ (but for different values of $M$).   For example, the following is the ansatz character for $D_{8,1}$ and  $N = 1$ :
\be \label{147c}
\chi_0 + b_1\,\chi_1.
\ee
The integer conformal dimensions for the monomials in \eqref{147c} are respectively, $0, 1$. When one computes the ansatz character for $D_{16,1}$ and $N = 1$, one obtains \eqref{147c} again.  But, when thought of as an ansatz character for $D_{16,1}$, the integer conformal dimensions for the monomials in \eqref{147c} are respectively, $0, 2$.  More generally, for $D_{8p, 8}$ and $N = 1$, the ansatz character is \eqref{147c} and the integer conformal dimensions for the monomials are respectively, $0, p$. What we need to note here is that the the monomials do not depend on the theory i.e. do not depend on $p$ while the conformal dimensions of the monomials do depend on it. We are able to observe this phenomenon from the computation of the ansatz character for $N = 1$ in \eqref{147c}.

We now give a general proof for this. The monomial $\chi_0^a \chi_1^b \chi_2^c$ is present in the ansatz character for $N = s$ if the following equations are satisfied :
\be \label{148c}
a + b + c  = N = s, \qquad 2 p b + c = 2 M
\ee
for non-negative integral $M$. We will be able to show that every solution of \eqref{148c} depends on $s$ and not on the theory $p$. We will also be able to show that the solutions will have a conformal dimension $M$ which does depend on $p$. To solve \eqref{148c}, we first note, from the second equation, that $c$ should be a multiple of $2$. Let us designate $c = 2k$. The values $a$ and $b$  take, can be determined and we find the following monomial solutions for the ansatz character :
\be
\sum_{k \in K} \left( \sum_{l \in L}~b_{k,l} ~\chi_0^{s - l - 2k} \chi_1^{l} \chi_2^{2k}\right) \quad 
\ee \label{149c}
where
\bea
&L = \{0, 1, \ldots, s-2k \}, \qquad K = \{2k~|~0 \leq 2k \leq s\} \nonumber 
\eea
Note that there is no $p$ dependence in the ansatz character of \eqref{149c}. The $p$ dependence shows up when one computes the conformal dimensions.  The character multiplying $b_{k,l}$ has  a conformal dimension $p l+k $.

Now, we proceed to complete the procedure laid out in section \ref{2s}.  We consider the S-invariance equations \eqref{18b}. Since the $S$-matrix for all $D_{r,1}$  CFTs are identical,  we get the same set of linear equations for all $r = 8p$.  And then we solve the  overdetermined system of linear equations. The answers are tabulated in table \ref{t13b}.

We start with smallest possible $c_{(1)}$. This is the $N = 1,  c_{(1)} = 8p$ case. The ansatz character in this case has $1$ arbitrary constant, see \eqref{147c}. The $S$-invariance equations are a system of $3$ equations for  $1$ variable. The solution gives us the first $S$-invariant polynomial :
\be \label{150c}
P_{c_{(1)} = 8p}^{D_{8p,1}}  = P_1  
\ee
In the next case, i.e. for $N = 1,  c_{(1)} = 8p$, we discover the second $S$-invariant polynomial viz. $P_2$. Then, we take on the  case $N = 3,  c_{(1)} = 24p$.  The ansatz character \eqref{149c} has $5$ arbitrary constants, the $b$'s.   The $S$-invariance equations are an overdetermined system of $10$ equations for these $5$  variables. We managed to find the following solution that has hidden in it a third $S$-invariant polynomial, viz. $P_3$ :
\be \label{151c}
P_{c_{(1)} = 24 p}^{D_{8p,1}}  = {\cal Z}_0 + p_1\, {\cal Z}_1 + p_2\, {\cal Z}_2 
\ee
where
\begin{small}
\begin{align}
   {\cal Z}_0 =  P_1^3+3 P_3+6 P_1 P_2, \qquad  {\cal Z}_1 = P_3,  \qquad   {\cal Z}_2 = -2 P_3 - 2 P_1 P_2 . \label{152c}
\end{align}
\end{small}
We can put the above in the following more-compact form :
\be \label{153c}
P_{c_{(1)} = 24 p}^{D_{8p,1}}   = 
\tilde{{\cal Z}}_0+ \tilde{p}_1\, \tilde{{\cal Z}}_1 + \tilde{p}_2\, \tilde{{\cal Z}}_2 
\ee
where
\be \label{154c}
\tilde{{\cal Z}}_0 = P_1^3, \quad \tilde{{\cal Z}}_1 = P_3, \quad \tilde{{\cal Z}}_2 = P_1 P_2.
\ee
The relation between the constant coefficients of \eqref{151c} and \eqref{153c} are :
\begin{small}
\begin{align}
         \tilde{p}_1& =  3 + p_1 - 2p_2, \qquad  \tilde{p}_2 =  6 -2p_2.
 \end{align}
\end{small}
The important outcome of the computations so far  is that we have a set of three $S$-invariant polynomials. And we expect these three to form a better basis for $S$-invariant characters than the basis one started with viz. $\chi_0, \chi_1, \chi_2$. We collect all the three $S$-invariant polynomials for $D_{8p,1}$ below :
\begin{equation}\label{62a}
\begin{array}{lll}
   P_1(\chi_0, \chi_1, \chi_2) &=\chi_0+\chi_1\\
   P_2(\chi_0, \chi_1, \chi_2) &=\chi_1^2\,+\,\chi_2^2\,-\,2\,\chi_0 \,\chi_1\\
   P_3(\chi_0, \chi_1, \chi_2) &=\chi_0^2\,\chi_1 - \chi_1\,\chi_2^2 = 2^{r-1}
\end{array}
\end{equation}
We note that each of the $S$-invariant polynomials has a $q$-series that depends on $r$.  $P_1$, for all values of $r$ is a non-constant $S$-invariant polynomial. $P_2$ turns out to be constant $S$-invariant polynomial only for $r = 8$, taking the value $0$ and is a non-constant polynomial for $r \geq 16$. $P_3$ is always a constant $S$-invariant polynomial, for all values of $r$, taking the value $2^{r-1}$.

We have gone ahead and computed for $N = 4, 5, 6$.  The number of arbitrary coefficients, the $b$'s in the ansatz character for these cases are respectively, $8$, $11$ and $15$. The number of equations that one needs to solve in these cases are respectively, $15$, $21$  and $28$. We have tabulated the results in table \ref{t13b}. It should be noted that we give the most compact form of the results. In every case, we give the equivalent of \eqref{153c} rather than \eqref{151c}. The number of $S$-invariant summands, the $\tilde{{\cal Z}}_i$'s in the six cases $N = 1, 2, 3,  4, 5, 6$ are respectively, $1, 2, 3, 4, 5, 7$. We have this information only after a long process of solving a huge number of equations. Below, we will give a quick method to get the number of summands as well as the summands themselves, without having to solve the equations. 

The way to read the results from table \eqref{t13b} is as follows.   The first column is the serial no. ($s$). The second column contains the number of tensor product factors $N$ (which equals $s$). The third column contains the value of $c_{(1)} = 8sp$.  The fourth column contains the character $P^{D_{8p,1}}_{c_{(1)} = 8s p}(\chi_0, \chi_1)$.  We should emphasise that the table \ref{t13b} contains the characters of the one-character extensions of tensor products of an infinite number of CFTs. It is remarkable that we are able to encode an infinite amount of information in such a finite compact form. We do not give the $q$-series of the characters in table \ref{t13b} due to that information being dependent on $r$. 

Now, we should consider admissibility. There are an infinite number of CFTs to study. We only study three cases viz. $r = 8, 16, 24$. We have computed the $q$-series of the characters that can be read off from table \ref{t13b} and we always find infinite numbers of admissible characters at each central charge. 

Amongst all the admissible characters that we have found, only one of them corresponds to a genuine CFT. This happens for $D_{8,1}$ when $N = 3$. The character is given in \eqref{153c}.  Noting that for $r=8$, $P_1 = j^{\frac13}$, $P_2 = 0$ and $P_3 = 128$, we find 
\be \label{157c}
P_{c_{(1)} = 24}^{D_{8,1}} = j + 128 \tilde{p}_1.
\ee
There are an infinite number of admissible characters in \eqref{157c} at central charge $24$. To answer the question as which of them correspond to a genuine CFT, we once again invoke the classification of one-character $c = 24$ CFTs of \cite{Schellekens:1992db}. In the table at the end of that paper entry no. 61 is the only CFT which is a one-character extension of a tensor product of  copies of $D_{8,1}$\footnote{The reader can find this CFT listed as entry no. 24 of table \ref{t0}.}.  This CFT has a character  $j - 384$. Thus amongst the infinite characters \eqref{157c} that we found, only one of them, the one with $\tilde{p}_1 = -3$ corresponds to an actual CFT. 

Further considerations (beyond those in this paper and) establish \cite{Schellekens:1992db}, in the case of $D_{8,1}$,  that  the single character obtained for $N = 3$ viz. $P_1^3 - 3 P_3$ corresponds to a genuine CFT. We suspect that these further considerations would  establish that the same character for all the other $D_{8p,1}$'s also  corresponds to genuine CFT(s). We thus are led to the following surmise/conjecture : \emph{For $D_{8p,1}$, $p \geq 2$  there exist genuine one-character CFT(s) with central charge $24 p$ with character $P_1^3 - 3 P_3$.}

\underline{Lessons learnt and a new method} :  All the results that we have obtained and reported in table \ref{t13b} were by running the whole procedure as laid out  in section \ref{2s} . Similar to what we did in section \ref{3s}, we have a simpler  procedure that allows us to write the final answer quickly.  We first note that $P_1$ is a homogenous polynomial of the characters of degree $1$,   $P_2$ is a homogenous polynomial of the characters of degree $2$ and $P_3$ is a homogenous polynomial of the characters of degree $3$.  A monomial of the form $P_1^A\,P_2^B\,P_3^C$ is a homogenous polynomial of the characters of degree $A+2B+3C$. And for this monomial to be present in $P^{D_{8p,1}}_{c_{(1)} = 8 s p}$ it’s degree needs to be equal to $N=s$. Thus we look for  non-negative integral solutions to 
\be \label{158c}
A+2B+3C = N = s, 
\ee
and we can obtain all the $\tilde{{\cal Z}}$'s  in $P^{D_{8p,1}}_{c_{(1)} = 8sp}$.  All except one of the $\tilde{{\cal Z}}$'s  viz. $\tilde{{\cal Z}}_0$ are multiplied by constants the $\tilde{p}$'s.  Which of the solutions of \eqref{158c} corresponds to $\tilde{{\cal Z}}_0$ is easily decided by seeing which of them contains the identity character $\chi_0^N$. Here it is the $A = s, B= 0, C=0$ solution i.e. $P_1^s$. This quick computation of the solutions to \eqref{158c} gives us the same answer as the long procedure.  There is one caveat.  We should only consider solutions of \eqref{158c} that give monomials of characters which are commensurate to the identity character.  It turns out that for every solution of \eqref{158c}, all monomials of characters present in $P_1^A\,P_2^B\,P_3^C$ are commensurate to the identity character. This is simply because every monomial present in each of the $S$-invariant polynomials are commensurate to the identity character.  Finally, we recognise that  in this short method, we are asked to find solutions to \eqref{158c} which has no dependence on the theory (no dependence on $p$).  Thus the answers we obtain are also $p$ independent. And we have solved for all theories in one go.

\begin{table}[H]
\begin{center}
\begin{threeparttable}
\rowcolors{2}{Mywhite}{Mygrey}
\resizebox{\textwidth}{!}{
   \renewcommand{\arraystretch}{1.3}
   $\begin{array}{|c|C{2cm}|c|C{10cm}|}
\hline
 \#(s) &      N & c_{(1)} & P_{c_{(1)}= 8sp}^{D_{8 p,1}} (\chi_0, \chi_1, \chi_2)  \\
\hline
1. &    \mathcal E_1[D_{8 p,1}] & 8p & {\cal \tilde{Z}}_0 =P_1\\
\hline
2. &   \mathcal E_1[D_{8 p,1}^{~\otimes 2}] & 16 p & {\cal \tilde{Z}}_{0}=P_1^2,\   {\cal \tilde{Z}}_{1}=P_2\\
\hline
3. &   \mathcal E_1[D_{8 p,1}^{~\otimes 3}] & 24 p & {\cal \tilde{Z}}_{0}=P_1^3,\   {\cal \tilde{Z}}_{1}=P_3,\   {\cal \tilde{Z}}_{2}=P_1 P_2\\
\hline
4. &   \mathcal E_1[D_{8 p,1}^{~\otimes 4}] & 32 p &  {\cal \tilde{Z}}_{0}=P_1^4,\   {\cal \tilde{Z}}_{1}=P_1 P_3,\   {\cal \tilde{Z}}_{2}=P_1^2 P_2,\   {\cal \tilde{Z}}_{3}=P_2^2\\
\hline
5. &      \mathcal E_1[D_{8 p,1}^{~\otimes 5}] & 40 p &  {\cal \tilde{Z}}_{0}=P_1^5,\   {\cal \tilde{Z}}_{1}=P_1^2 P_3,\   {\cal \tilde{Z}}_{2}=P_1^3 P_2,\ {\cal \tilde{Z}}_{3}=P_2 P_3,\\
&&& {\cal \tilde{Z}}_{4}=P_1 P_2^2\\
\hline
6. &      \mathcal E_1[D_{8 p,1}^{~\otimes 6}] & 48 p & {\cal \tilde{Z}}_{0}=P_1^6,\   {\cal \tilde{Z}}_{1}=P_1^3 P_3,\   {\cal \tilde{Z}}_{2}=P_1^4 P_2,\   {\cal \tilde{Z}}_{3}=P_3^2,\\

&&& {\cal \tilde{Z}}_{4}=P_1 P_2 P_3,\   {\cal \tilde{Z}}_{5}=P_1^2 P_2^2,\   {\cal \tilde{Z}}_{6}=P_2^3\\
\hline
\end{array}$
}
\end{threeparttable}
\end{center}
\caption{Characters of one-character extensions of tensor products of $D_{8p,1}$ written in a basis of $S$-invariant polynomials }\label{t13b}
\end{table}

\subsection{$\mathcal C = B_{r,1}$ \label{44s}}

In this subsection, we study the infinite class of CFTs :  $B_{r,1}$ where $r$ is a positive integer and $r \geq 2$. Each of these CFTs is a $\mathbf{[3,0]}$ CFT \cite{Das:2020wsi}. A $B_{r,1}$ CFT has a central charge equalling $\frac{2 r+ 1}{2}$. Let us denote it's three characters by $\chi_0, \chi_1, \chi_2$, the latter two  are the non-identity characters with conformal dimensions $h_1 = \frac{2 r+ 1}{16}, ~h_2 = \frac12$ respectively. Let us note the $q$-series for the characters (these formulae are sourced from chapter 14 of \cite{DiFrancesco:1997nk}):
\begin{align}
   \chi_0(q)& = \frac{\theta_3^{r+\frac{1}{2}}(q)+\theta_4^{r+\frac{1}{2}}(q)}{2\eta^{r+\frac{1}{2}}}, \qquad \chi_1(q)  = \frac{\theta_2^{r+\frac{1}{2}}(q)}{\sqrt{2}\eta^{r+\frac{1}{2}}}, \qquad \chi_2(q) = \frac{\theta_3^{r+\frac{1}{2}}(q)-\theta_4^{r+\frac{1}{2}}(q)}{2\eta^{r+\frac{1}{2}}}\label{159a}
\end{align}
where $\theta_2, \theta_3, \theta_4$ are the $\theta$-constants : $\theta_i(\tau) \equiv \theta_i(0|\tau)$ with $q$-series given by :
\begin{align*}
   \theta_2(q)&=2q^{\frac{1}{8}}\prod_{n=1}^{\infty}(1-q^n)(1+q^n)^2, \qquad 
   \theta_3(q) =\prod_{n=1}^{\infty}(1-q^n)(1+q^{n-\frac{1}{2}})^2\\
   \theta_4(q)&=\prod_{n=1}^{\infty}(1-q^n)(1-q^{n-\frac{1}{2}})^2, \qquad 
   \eta(q) =q^{\frac{1}{24}}\prod_{n=1}^{\infty}(1-q^n).
\end{align*}
The  modular $S$-transformations for the characters is given by :
\begin{small}
\begin{align}
   S(\chi_0)&=\frac{\chi_0+\sqrt{2}\chi_1+\chi_2}{2},
   \qquad S(\chi_1)=\frac{\chi_0-\chi_2}{\sqrt{2}}, 
   \qquad S(\chi_2)=\frac{\chi_0-\sqrt{2}\chi_1+\chi_2}{2} \label{160a} 
\end{align}
\end{small}
We note that the $S$-transformations do not have a $r$ dependence. This means that we can hope that imposing $S$-invariance can be done independent of $r$. Before we get to imposing $S$-invariance, we need to construct the ansatz character \eqref{10b}. This has a $r$-dependence. From row 7 of table \ref{t3},  it is clear that  all $B_{r,1}$  CFTs  have the same $N$ values; $N = 16s$ and the resulting central charge is $c_{(1)} = 8s(2r+1)$. Hence  for all  $r$, when constructing the ansatz character,  we are looking for monomials of the same degree. Then we note that $h_1$ and $h_2$ values for all  $r$  have the same structure, i.e. denominators are $16$ and $2$.  What this means is that a monomial which solves the equations \eqref{7b} and \eqref{8b} for any $r$  also solves the equations for any other $r$ (but for different values of $M$).  For example, the following is the ansatz character for $B_{2,1}$ and  $N = 16$ :
\bea \label{161a}
\chi_0^{16} + b_1\,\chi_0^{14} \, \chi_2^2 + b_2\,\chi_0^{12} \, \chi_2^4 + b_3\,\chi_0^{10} \, \chi_2^6 + b_4\,\chi_0^7\, \chi_1^8 \,\chi_2 + b_5\,\chi_0^{8} \, \chi_2^8 + b_6\,\chi_0^5\, \chi_1^8 \,\chi_2^3 + b_7\,\chi_0^{6} \, \chi_2^{10} + b_8\,\chi_0^3\, \chi_1^8 \,\chi_2^5  \nonumber \\
+ b_9 \chi_1^{16} + b_{10}\,\chi_0^{4} \, \chi_2^{12} + b_{11}\,\chi_0\, \chi_1^8 \,\chi_2^7 + b_{12}\,\chi_0^2\,\chi_2^{14} + b_{13} \chi_2^{16} \quad
\eea
The integer conformal dimensions for the monomials in \eqref{161a} are respectively, $0, 1, 2, 3, 3, 4, 4, 5, 5, 5,\\ 6, 6, 7, 8$. When one computes the ansatz character for $B_{3,1}$ and $N = 16$, one obtains \eqref{161a} again.  But, when thought of as an ansatz character for $B_{3,1}$, the integer conformal dimensions for the monomials in \eqref{161a} are respectively, $0, 1, 2, 3, 4, 4, 5, 5, 6, 7, 6, 7, 7, 8$.  More generally, for $B_{r, 1}$ and $N = 16$, the ansatz character is \eqref{161a} and the integer conformal dimensions for the monomials are respectively, $0, 1, 2, 3, r+1, 4, r+2, 5, r+3, 2r+1, 6, r+4, 7, 8$ . What we need to note here is that the the monomials do not depend on the theory i.e. do not depend on $r$ while the conformal dimensions of the monomials do depend on it. We are able to observe this phenomenon from the computation of the ansatz character for $N = 8$ in \eqref{161a}.

We now give a general proof for this. The monomial $\chi_0^a \chi_1^b \chi_2^c$ is present in the ansatz character for $N = 16s$ if the following equations are satisfied :
\be \label{162a}
a + b + c  = N = 16s, \qquad (2r + 1) b + 8 c = 16 M
\ee
for non-negative integral $M$. We will be able to show that every solution of \eqref{162a} depends on $s$ and not on the theory $r$. We will also be able to show that the solutions will have a conformal dimension $M$ which does depend on $r$. To solve \eqref{162a}, we first note, from the second equation, that $b$ should be a multiple of $8$. Let us designate $b = 8l$. From the fact that $b \leq 16s$, it then follows that $l$ can take values $0, 1, 2, \ldots, 2s-1, 2s$. It then also follows that when $l$ is an odd number, $c$ should also be an odd number, as well as $a$; simillarly when $l$ is an even number, $c$ should also be an even number, as well as $a$. The values $c$ and $a$  take, can be determined and we find two classes of monomial solutions for the ansatz character :
\be
\sum_{l \in L} \left( \sum_{k \in K}~b_{k,l} ~\chi_0^{16s - 8l - k} \chi_1^{8l} \chi_2^k\right) \quad + \quad  \sum_{l' \in L'} \left( \sum_{k' \in K'}~b_{k',l'} ~\chi_0^{16s - 8l' - k'} \chi_1^{8l'} \chi_2^{k'}\right)
\ee \label{163a}
where
\bea
&L = \{1, 3, \ldots, 2s-1 \}, \qquad K = \{1, 3, \ldots, 16s-8l-1\} \nonumber \\
&\text{and} \nonumber \\
&L' = \{0, 2, \ldots, 2s \}, \qquad K' = \{ 0, 2, \ldots, 16s-8l \}. \nonumber
\eea
Note that there is no $r$ dependence in the ansatz character of \eqref{163a}. The $r$ dependence shows up when one computes the conformal dimensions.  The character multiplying $b_{k,l}$ has  a conformal dimension $\frac{(2r+1)l + k}{2}$ and the character multiplying $b_{k',l'}$ has  a conformal dimension $\frac{(2r+1)l' + k'}{2}$.

Now, we proceed to complete the procedure laid out in section \ref{2s}.  We consider the S-invariance equations \eqref{18b}. Since the $S$-matrix for all $B_{r,1}$  CFTs are identical,  we get the same set of linear equations for all  $r$.  And then we solve the  overdetermined system of linear equations. The answers are tabulated in table \ref{t14b}.

We start with smallest possible $c_{(1)}$. This is the $N = 16,  c_{(1)} = 8(2r + 1)$ case. The ansatz character in this case has $13$ arbitrary constants, the $b$’s, see \eqref{161a}. The $S$-invariance equations are an overdetermined system of $153$ equations for these $13$  variables. It was quite surprising that there are two $S$-invariant polynomials, $P_1$ and  $P_3$, hidden within the answer :
\be \label{164a}
P_{c_{(1)} = 8(2r + 1)}^{B_{r,1}}  = P_1 + \tilde{p}_1\, P_3^2 
\ee
Then, we take on the next case $N = 32,  c_{(1)} = 16(2r + 1)$.  The ansatz character \eqref{163a} has $42$ arbitrary constants, the $b$'s.   The $S$-invariance equations are an overdetermined system of $561$ equations for these $42$ arbitrary variables. We managed to find the following solution that has hidden in it a third $S$-invariant polynomial, viz. $P_2$ :
\be \label{165a}
P_{c_{(1)} = 16(2r + 1)}^{B_{r,1}}  = P_{1}^2 + \tilde{p}_1\,  P_{1} P_{3}^2 + \tilde{p}_2 \, P_{2} P_{3} + \tilde{p}_3 \, P_{3}^4.
\ee
The important outcome of the computations so far  is that we have a set of three $S$-invariant polynomials. And we expect these three to form a better basis for $S$-invariant characters than the basis one started with viz. $\chi_0, \chi_1, \chi_2$. We collect all the three $S$-invariant polynomials for $B_{2r+1,1}$ below :
\begin{equation}\label{165a}
\begin{array}{lll}
   P_1 &=~\chi_0^{16}\,+ \,140 \,\chi_0^{12} \,\chi_2^4\,+\,448\, \chi_0^{10} \,\chi_2^6\,+\,870 \,\chi_0^8\, \chi_2^8\,+\,240\, \chi_0^7\, \chi_1^8 \,\chi_2\,+\,448\, \chi_0^6 \,\chi_2^{10}\,+\,1680 \,\chi_0^5\, \chi_1^8\, \chi_2^3\,\notag\\&+\,140 \,\chi_0^4\, \chi_2^{12}\,+\,1680 \,\chi_0^3\, \chi_1^8\, \chi_2^5\,+\,240 \chi_0\, \chi_1^8\, \chi_2^7\,+\,8 \,\chi_1^{16}\,+\,\chi_2^{16} \\
   P_2 & = ~ \chi_0^{23} \,\chi_2\,-\,\chi_0^{21} \,\chi_2^3\,-\,21\, \chi_0^{19} \,\chi_2^5\,+\,85 \,\chi_0^{17} \,\chi_2^7\,-\,134 \,\chi_0^{15} \,\chi_2^9\,+\,256 \,\chi_0^{14}\, \chi_1^8\,
   \chi_2^2\,+\,70 \chi_0^{13} \,\chi_2^{11}\,\notag\\&\,+\,3584 \,\chi_0^{12}\, \chi_1^8\, \chi_2^4\,+\,70\, \chi_0^{11} \,\chi_2^{13}\,  +\,16128 \,\chi_0^{10} \,\chi_1^8 \,\chi_2^6\,-\,134 \,\chi_0^9 \,\chi_2^{15}\,+\,25600 \chi_0^8\, \chi_1^8\, \chi_2^8\,-\,256 \,\chi_0^7 \,\chi_1^{16} \,\chi_2\,\notag\\&+\,85 \chi_0^7 \,\chi_2^{17}\,+\,16128\, \chi_0^6\, \chi_1^8 \,\chi_2^{10}\,-\,1792 \,\chi_0^5\, \chi_1^{16} \,\chi_2^3\,-\,21 \,\chi_0^5\, \chi_2^{19}\,+\,3584\, \chi_0^4\, \chi_1^8\, \chi_2^{12}\,-\,1792\, \chi_0^3 \,\chi_1^{16} \,\chi_2^5\, \notag\\&-\,\chi_0^3\, \chi_2^{21}\,+\,256\, \chi_0^2\, \chi_1^8\, \chi_2^{14}\,-\,256 \,\chi_0\, \chi_1^{16}\, \chi_2^7\,+\,\chi_0\, \chi_2^{23}\\
   P_3 &=~\chi_0^7\, \chi_2\,+\,7\, \chi_0^5\, \chi_2^3\,+\,7\, \chi_0^3\, \chi_2^5\,+\,\chi_0\, \chi_2^7\,-\,\chi_1^8.
\end{array}
\end{equation}
We note that each of the $S$-invariant polynomials has a $q$-series that depends on $r$.  Each of $P_1$, $P_2$ and $P_3$, for all values of $r$ is a non-constant $S$-invariant polynomial. 

We have gone ahead and computed for $N = 48, 64, 80, 96$.  The number of arbitrary coefficients, the $b$'s in the ansatz character for these cases are respectively, $87$, $148$, $225$ and $318$. The number of equations that one needs to solve in these cases are respectively, $1225$, $2016$, $3321$ and $4753$. We have tabulated the results in table \ref{t14b}. It should be noted that we give the most compact form of the results. The number of $S$-invariant summands, the $\tilde{{\cal Z}}_i$'s in the six cases $N = 16, 32, 48, 64, 80, 96$ are respectively, $2, 4, 7, 10, 14, 19$. We have this information only after a long process of solving a huge number of equations. Below, we will give a quick method to get the number of summands as well as the summands themselves, without having to solve the equations. 

The way to read the results from table \eqref{t14b} is as follows.   The first column is the serial no. ($s$). The second column contains the number of tensor product factors $N$ (which equals $16s$). The third column contains the value of $c_{(1)} = 16 s (2r + 1)$.  The fourth column contains the character $P^{B_{r,1}}_{c_{(1)} = 16 s (2r+1)}(\chi_0, \chi_1)$.  We should emphasise that the table \ref{t14b} contains the characters of the one-character extensions of tensor products of an infinite number of CFTs. It is remarkable that we are able to encode an infinite amount of information in such a finite compact form. We do not give the $q$-series of the characters in table \ref{t14b} due to that information being dependent on $r$. 

Now, we should consider admissibility. There are an infinite number of CFTs to study. We only study three cases viz. $r = 2, 3, 4$. We have computed the $q$-series of the characters that can be read off from table \ref{t14b} and we always find infinite numbers of admissible characters at each central charge. 

Amongst all the admissible characters that we have found, none of them correspond to a genuine CFT.   

\underline{Lessons learnt and a new method} :  All the results that we have obtained and reported in table \ref{t14b} were by running the whole procedure as laid out  in section \ref{2s} . Similar to what we did in section \ref{3s}, we have a simpler  procedure that allows us to write the final answer quickly.  We first note that $P_1$ is a homogenous polynomial of the characters of degree $16$,   $P_2$ is a homogenous polynomial of the characters of degree $24$ and $P_3$ is a homogenous polynomial of the characters of degree $8$.  A monomial of the form $P_1^a\,P_2^b\,P_3^c$ is a homogenous polynomial of the characters of degree $16a + 24b+8c$. And for this monomial to be present in $P^{B_{r,1}}_{c_{(1)} = 16 s (2r+1)}$ it’s degree needs to be equal to $N$. Thus we look for  non-negative integral solutions to 
\be \label{166a}
16a + 24b+8c = N = 16 s, 
\ee
we can obtain all the $\tilde{{\cal Z}}$'s  in $P^{B_{r,1}}_{c_{(1)} = 16 s (2r+1)}$.  All except one of the $\tilde{{\cal Z}}$'s  viz. $\tilde{{\cal Z}}_0$ are multiplied by constants the $\tilde{p}$'s.  Which of the solutions of \eqref{166a} corresponds to $\tilde{{\cal Z}}_0$ is easily decided by seeing which of them contains the identity character $\chi_0^N$. Here it is the $a = s, b = 0, c=0$ solution. This quick computation of the solutions to \eqref{166a} gives us the same answer as the long procedure.  There is one caveat. We should only consider solutions of \eqref{166a} that give monomials of characters which are commensurate to the identity character. It turns out that every solution of \eqref{166a} is such that  $b$ and $c$ are either both even numbers of both odd numbers; this can be easily seen by rewriting \eqref{166a}  as $c = 2s - 2a - 3b$. It then follows that all monomials of characters present in $P_1^a\,P_2^b\,P_3^c$ are commensurate to the identity character.

\begin{table}[H]
\begin{center}
\begin{threeparttable}
\rowcolors{2}{Mywhite}{Mygrey}
\resizebox{\textwidth}{!}{
   \renewcommand{\arraystretch}{1.3}
   $\begin{array}{|c|C{2cm}|c|C{12.9cm}|}
\hline
 \#(s) &      N & c_{(1)} & P_{c_{(1)} = 8 s (2r + 1)}^{B_{r,1}} (\chi_0, \chi_1, \chi_2) \\
\hline
1. & \mathcal E_1[B_{r,1}^{~\otimes 16}] & 8(2r+1) & {\cal \tilde{Z}}_{0}=P_{1},\  {\cal \tilde{Z}}_{1}=P_{3}^2\\
\hline
2. &   \mathcal E_1[B_{r,1}^{~\otimes 32}] & 16(2r+1)  &{\cal \tilde{Z}}_{0}=P_{1}^2,\  {\cal \tilde{Z}}_{1}=P_{1} P_{3}^2,\ {\cal \tilde{Z}}_{2}=P_{2} P_{3},\  {\cal \tilde{Z}}_{3}=P_{3}^4\\
\hline
3. &  \mathcal E_1[B_{r,1}^{~\otimes 48}] & 24(2r+1) &{\cal \tilde{Z}}_0 =P_1^3,\  {\cal \tilde{Z}}_{1}=P_{1}^2 P_{3}^2,\  {\cal \tilde{Z}}_{2}=P_{1} P_{2} P_{3},\ {\cal \tilde{Z}}_{3}=P_{2}^2,\\
&&& {\cal \tilde{Z}}_{4}=P_{1} P_{3}^4,\  {\cal \tilde{Z}}_{5}=P_{2} P_{3}^3,\ {\cal \tilde{Z}}_{6}=P_{3}^6\\
\hline
4. &  \mathcal E_1[B_{r,1}^{~\otimes 64}] & 32(2r+1)  & {\cal \tilde{Z}}_0 = P_1^4,\  {\cal \tilde{Z}}_{1}=P_{1}^3 P_{3}^2,\  {\cal \tilde{Z}}_{2}=P_{1}^2 P_{2} P_{3},\ {\cal \tilde{Z}}_{3}=P_{1} P_{2}^2,\\
&&& {\cal \tilde{Z}}_{4}=P_{1}^2 P_{3}^4,\  {\cal \tilde{Z}}_{5}=P_{1} P_{2} P_{3}^3,\ {\cal \tilde{Z}}_{6}=P_{2}^2 P_{3}^2,\ {\cal \tilde{Z}}_{7}=P_{1} P_{3}^6,\\
&&& {\cal \tilde{Z}}_{8}=P_{2} P_{3}^5,\ {\cal \tilde{Z}}_{9}=P_{3}^8\\
\hline
5. &  \mathcal E_1[B_{r,1}^{~\otimes 80}] & 40(2r+1) & {\cal \tilde{Z}}_0 = P_1^5,\  {\cal \tilde{Z}}_{1}=P_{1}^4 P_{3}^2,\  {\cal \tilde{Z}}_{2}=P_{1}^3 P_{2} P_{3},\ {\cal \tilde{Z}}_{3}=P_{1}^2 P_{2}^2,\\
&&&  {\cal \tilde{Z}}_{4}=P_{1}^3 P_{3}^4,\  {\cal \tilde{Z}}_{5}=P_{1}^2 P_{2} P_{3}^3,\ {\cal \tilde{Z}}_{6}=P_{1} P_{2}^2 P_{3}^2,\ {\cal \tilde{Z}}_{7}=P_{2}^3 P_{3},\\
&&&   {\cal \tilde{Z}}_{8}=P_{1}^2 P_{3}^6,\ {\cal \tilde{Z}}_{9}=P_{1} P_{2} P_{3}^5,\  {\cal \tilde{Z}}_{10}=P_{2}^2 P_{3}^4,\  {\cal \tilde{Z}}_{11}=P_{1} P_{3}^8,\\
&&& {\cal \tilde{Z}}_{12}=P_{2} P_{3}^7,\ {\cal \tilde{Z}}_{13}=P_{3}^{10}\\
\end{array}$
}
\end{threeparttable}
\end{center}
\end{table}

\begin{table}[H]
\begin{center}
\begin{threeparttable}
\rowcolors{2}{Mywhite}{Mygrey}
\resizebox{\textwidth}{!}{
   \renewcommand{\arraystretch}{1.3}
   $\begin{array}{|c|C{2cm}|c|C{12.9cm}|}
6. &  \mathcal E_1[B_{r,1}^{~\otimes 96}] & 48(2r+1)  & {\cal \tilde{Z}}_0 = P_1^6,\  {\cal \tilde{Z}}_{1}=P_{1}^5 P_{3}^2,\  {\cal \tilde{Z}}_{2}=P_{1}^4 P_{2} P_{3},\ {\cal \tilde{Z}}_{3}=P_{1}^3 P_{2}^2,\\
&&&  {\cal \tilde{Z}}_{4}=P_{1}^4 P_{3}^4,\  {\cal \tilde{Z}}_{5}=P_{1}^3 P_{2} P_{3}^3,\ {\cal \tilde{Z}}_{6}=P_{1}^2 P_{2}^2 P_{3}^2,\  {\cal \tilde{Z}}_{7}=P_{1} P_{2}^3 P_{3},\\
&&&  {\cal \tilde{Z}}_{8}=P_{2}^4,\ {\cal \tilde{Z}}_{9}=P_{1}^3 P_{3}^6,\  {\cal \tilde{Z}}_{10}=P_{1}^2 P_{2} P_{3}^5,\  {\cal \tilde{Z}}_{11}=P_{1} P_{2}^2 P_{3}^4,\\
&&& {\cal \tilde{Z}}_{12}=P_{2}^3 P_{3}^3,\  {\cal \tilde{Z}}_{13}=P_{1}^2 P_{3}^8,\  {\cal \tilde{Z}}_{14}=P_{1} P_{2} P_{3}^7,\ {\cal \tilde{Z}}_{15}=P_{2}^2 P_{3}^6,\\
&&&  {\cal \tilde{Z}}_{16}=P_{1} P_{3}^{10},\  {\cal \tilde{Z}}_{17}=P_{2} P_{3}^9,\ {\cal \tilde{Z}}_{18}=P_{3}^{12}\\
\hline
\end{array}$
}
\end{threeparttable}
\end{center}
   \caption{Characters of one-character extensions of tensor products of $B_{r,1}$ written in a basis of $S$-invariant polynomials}\label{t14b}
\end{table}

\section{ $S$-invariant Polynomials\label{5s}}

$S$-invariant polynomials have played a crucial role in this paper. They have provided an alternate and more useful basis for homogenous polynomials of characters, allowing us to obtain characters of one-character extensions in a compact form. But, they seem to occur in other parts of mathematical physics. We try to collect these connections here in this section. 

\underline{Constant $S$-invariant Polynomials} :
In this section, we will  collect all the constant $S$-invariant polynomials for various CFTs that we have discovered in this paper. 

\begin{small}
\begin{eqnarray}
&A_{1,1}:&\qquad \qquad  \chi_0^5\,\chi_1 -  \chi_0\,\chi_1^5 = 2 \nonumber \\
&A_{2,1}:&\qquad \qquad \chi_0^3 \chi_1-\chi_1^4 = 3 \nonumber \\
&D_{4,1}:&\qquad \qquad \chi_0^2 \chi_1-\chi_1^3 = 8 \nonumber \\
&A_{4,1}:&\qquad \qquad (\text{i}) ~\chi_0^4 \,\chi_1\, \chi_ 2\,-\,\chi_0^2 \,\chi_1^2 \,\chi_2^2\,-\,\chi_0\, \chi_1^5\,-\,\chi_0\, \chi_2^5\,+\,2\, \chi_1^3\, \chi_2^3 ~ =~ 50 \nonumber \\
&&\qquad \qquad (\text{ii}) ~ 5\, \chi_0^6\, \chi_1^2 \,\chi_2^2 \,-\,4 \,\chi_0^5\, \chi_1^5\,-\,4\, \chi_0^5\, \chi_2^5\,-\,10\, \chi_0^4\, \chi_1^3 \,\chi_2^3\,+\,10 \,\chi_0^3\, \chi_1^6\, \chi_2\, \nonumber \\
&& \qquad \qquad +\,10 \chi_0^3\, \chi_1 \,\chi_2^6\,+\,5\, \chi_0^2\, \chi_1^4 \,\chi_2^4 \,-\,10\, \chi_0\, \chi_1^7\, \chi_2^2\,-\,10\, \chi_0\, \chi_1^2\, \chi_2^7\, \nonumber \\
&& \qquad \qquad +\,\chi_1^{10}\,+\,6\, \chi_1^5\, \chi_2^5\,+\,\chi_2^{10}~=~ 0 \nonumber \\
&{\cal M}(4,3):&\qquad \qquad (\text{i}) ~\chi_0^7\, \chi_2\,+\,7\, \chi_0^5\, \chi_2^3\,+\,7\, \chi_0^3\, \chi_2^5\,+\,\chi_0\, \chi_2^7\,-\,\chi_1^8 ~=~0 \nonumber \\
&&\qquad \qquad (\text{ii}) ~\chi_0^{23} \,\chi_2\,-\,\chi_0^{21} \,\chi_2^3\,-\,21\, \chi_0^{19} \,\chi_2^5\,+\,85 \,\chi_0^{17} \,\chi_2^7\,-\,134 \,\chi_0^{15} \,\chi_2^9 \nonumber \\
&&\qquad \qquad + \,256 \,\chi_0^{14}\, \chi_1^8\, \chi_2^2\,+\,70 \chi_0^{13} \,\chi_2^{11}\,+\,3584 \,\chi_0^{12}\, \chi_1^8\, \chi_2^4\,+ \,70\, \chi_0^{11} \,\chi_2^{13}\, \nonumber \\
&&\qquad \qquad +\,16128 \,\chi_0^{10} \,\chi_1^8 \,\chi_2^6\,-\,134 \,\chi_0^9 \,\chi_2^{15}\,+\,25600 \chi_0^8\, \chi_1^8\, \chi_2^8\,-\,256 \,\chi_0^7 \,\chi_1^{16} \,\chi_2\, \nonumber \\
&&\qquad \qquad +\,85 \chi_0^7 \,\chi_2^{17}\,+\,16128\, \chi_0^6\, \chi_1^8 \,\chi_2^{10}\,-\,1792 \,\chi_0^5\, \chi_1^{16} \,\chi_2^3\,-\,21 \,\chi_0^5\, \chi_2^{19}\, \nonumber \\
&&\qquad \qquad +\,3584\, \chi_0^4\, \chi_1^8\, \chi_2^{12}\,-\,1792\, \chi_0^3 \,\chi_1^{16} \,\chi_2^5\, -\,\chi_0^3\, \chi_2^{21}\,+\,256\, \chi_0^2\, \chi_1^8\, \chi_2^{14}\, \nonumber \\
&&\qquad \qquad -\,256 \,\chi_0\, \chi_1^{16}\, \chi_2^7\,+\,\chi_0\, \chi_2^{23} \,=\,1 \nonumber \\
&D_{r,1}&\qquad \qquad \chi_0^2\,\chi_1 - \chi_1\,\chi_2^2 = 2^{r-1} \label{67a}
\end{eqnarray}
\end{small}

Also, consider the following constant $S$-invariant polynomial of the ${\cal M}(5,2)$ a.k.a Lee-Yang CFT.  
\bea \label{68a}
{\cal M}(5,2): \qquad \qquad \chi_0^{11} \,\chi_1\,+\,11\, \chi_0^6\, \chi_1^6 \,-\, \chi_0\, \chi_1^{11}~= 1
\eea
The power-series part of the characters of the Lee-Yang CFT, $\chi_0(q)$ and $\chi_1(q)$, are represented in the math literature by $G(q)$ and $H(q)$. $G(q)$ is the common generating function for two different combinatorial problems and the commonality is referred to as the first Rogers-Ramanujan identity. Similarly $H(q)$ is the common generating function for two other combinatorial problems and the commonality is referred to as the second Rogers-Ramanujan identity. Thus, from \eqref{68a}, we learn that there is a polynomial relation between the two Rogers-Ramanujan identities. Hence we can refer to this \eqref{68a} as the Rogers-Ramanujan polynomial relation. But for us it is simply the constant $S$-invariant polynomial of the Lee-Yang CFT.

Hence, we can think of what we have obtained in this paper, collected in \eqref{67a}, as generalizations of Rogers-Ramanujan-like polynomial relations from the Lee-Yang CFT to other CFTs. We expect this kind of generalization to exist for other CFTs as well, which we hope to investigate.

Another lesson to learn from the above polynomial relations among characters \eqref{67a} is the following. When we start our analysis, the character of a one-character extension of tensor product of $N$ copies of a CFT is an element of the polynomial ring generated by the characters of the CFT. In fact it is a homogenous polynomial of degree $N$. In CFTs which have constant $S$-invariant polynomials,  their existence means that the expressions for the character of the one-character extension can be simplified and they may not look homogenous any more. More importantly, the correct way to account for the existence of these polynomial relations is that the ideal generated by them can be used to quotient the polynomial ring of characters and the character of the one-character extension is  properly an element of this quotient ring. 

\underline{$S$-invariant Polynomials and Theta-function Identities} :

The characters of the WZW CFTs based on the orthogonal Lie-algebras at level one viz. the $D_{r,1}$ and $B_{r,1}$ are given by theta-constants \eqref{108c}, \eqref{159a}. Hence it should not come as a surprise that the various $S$-invariant polynomials that we have discovered in this paper are a consequence of well-known mathematical identities involving the theta functions. We will give only a sample of them here.

First, we consider the constant $S$-invariant polynomial : $\chi_0^2 \chi_1 - \chi_1 \chi_2^2$ which computes to $2^{r-1}$. This happens for all $D_{r,1}$ (see $P_3$ in \eqref{59ad}, $P_2$ in \eqref{60ad}, $P_2$ in \eqref{61a}, $P_3$ in \eqref{62a}). We find that the following standard theta-function identity is the reason for the existence of this constant $S$-invariant polynomial :
\bea 
\theta_2(q)\,\theta_3(q)\,\theta_4(q) = 2 \eta^3(q) \qquad \Longrightarrow \qquad \chi_0^2 \chi_1 - \chi_1 \chi_2^2 = 2^{r-1}.
\eea

Many of the $S$-invariant polynomials, even though look different on the face of it, simplify to similar expressions. Let us define the following combination of the theta-functions :
\be \label{169a}
Q(a) \equiv \frac{\theta_3^{4a} - \theta_4^{4a} - \theta_2^{4a}}{2 \eta^{4a}}.
\ee 
Then, it turns out that (i) the $S$-invariant polynomial $P_2 \equiv \chi_0^3 \,\chi_2\,+\,\chi_0\, \chi_2^3\,-\,2 \,\chi_1^4$ of \eqref{59ad} simplifies to $\frac{Q(r)}{4}$, (ii) the $S$-invariant polynomial $P_3 \equiv \chi_0 \chi_2-\chi_1^2$ of \eqref{60ad} simplifies to $\frac{Q(\frac{r}{2})}{2}$ and (iii) the $S$-invariant polynomial $P_3 \equiv \chi_2 - \chi_1$ of \eqref{61a} simplifies to $Q(\frac{r}{4})$. These are all non-constant $S$-invariant polynomials and they can be obtained in one shot, by computing the $q$-series for \eqref{169a}:
\bea \label{170a}
Q(a) = 8 a \,j^{\frac{a}{6} - \frac12} + C_2\,j^{\frac{a}{6} - \frac32} + \ldots + C_{[\frac{a}{6}+\frac12]}\,j^{\frac{a}{6} - \frac12-[\frac{a}{6} - \frac12]},\quad a=3,5,7,\ldots
\eea
Here the $C$'s are constants that can be obtained from comparing the $q$-series of the right hand sides of equations \eqref{169a} and \eqref{170a}. There seem to be many more Theta-function identities that are relevant in our computations of the various $S$-invariant polynomials. We will postpone to a future study a more thorough investigation of this connection.

\section{Conclusions and Future Directions \label{6s}}

First, we will summarise the work in this paper.  In section \ref{2s} we set up an operational procedure to construct characters of one-character extensions of tensor products of a single CFT ${\cal C}$. The technical elements involved in the procedure was very basic and algebraic : polynomial algebra and solving systems of linear equations. 

\underline{One-character extensions of tensor products of $\mathbf{[2,0]}$ CFTs} :  We implemented this procedure in section \ref{3s} for the $\mathbf{[2,0]}$ CFTs. For each of $A_{1,1}$, $A_{2,1}$ and $D_{4,1}$ we could construct characters of possible one-character extensions with central charges $c_{(1)} = 8, 16, \ldots 128$. For $G_{2,1}$ we could construct  characters of possible one-character extensions with central charges $c_{(1)} = 56, 112, 168,  \ldots 448$. For $F_{4,1}$ we could construct characters of possible one-character extensions with central charges $c_{(1)} = 104, 208, 312, \ldots 832$. For $E_{6,1}$ we could construct characters  of possible one-character extensions with central charges $c_{(1)} = 24, 48, 96, \ldots 384$. For $E_{7,1}$, we could construct characters of possible one-character extensions with central charges $c_{(1)} = 56, 112, 168,  \ldots 896$.

To start with, the venue of the problem is polynomial rings. The statement of the problem is given in terms of  a natural basis of generators of the ring i.e. the characters of ${\cal C}$. The answers we are seeking i.e. characters of one-character extensions are homogenous polynomials of the generators. But more importantly, as we discovered, they are $S$-invariant. We were able to discover another basis of generators which have the distinction from the earlier basis used in that they are $S$-invariant.  We discovered such a basis with two $S$-invariant polynomials for each of the $\mathbf{[2,0]}$ CFTs.  Apart from it being aesthetically pleasing to work with a $S$-invariant basis to obtain answers that are $S$-invariant, we are rewarded with brevity. The characters when expressed in terms of the $S$-invariant polynomial basis gave compact forms and this is one of the reasons we could go to such high central charges in this paper. 

\underline{One-character extensions of tensor products of $\mathbf{[3,0]}$ CFTs} :  In section \ref{3s}, we have studied some $\mathbf{[3,0]}$ WZW CFTs and the Ising CFT which is not a WZW CFT but  a  $\mathbf{[3,0]}$ CFT.  For $A_{4,1}$ we could construct characters of possible one-character extensions with central charges $c_{(1)} = 8, 16, 24, \ldots 96$. For the Ising CFT we could construct characters of possible one-character extensions  with central charges $c_{(1)} = 8, 16, \ldots 48$. Then we solved the problem we set for ourselves for an infinite number of CFTs the $D_{r,1}$ CFTs. We had to divide the problem into four categories, each with an infinite number of CFTs. We could give a common solution for all the infinite number of CFTs in each of the four categories.

Here again, for each of the $\mathbf{[3,0]}$ CFTs, we constructed a new basis of generators for the homogenous polynomials, which contained only $S$-invariant polynomials. If we had not found this $S$-invariant basis, we could not probably have written this paper. We could provide, for example, the explicit degree-$48$ polynomial of characters of the Ising CFT that is the character of the monster CFT. 

\underline{New Method} : Having understood the importance of the existence of the $S$-invariant polynomial basis for the recasting of the answers in compact forms, a natural question presented itself to us. is there a way to set up and solve the problem of finding characters of one-character extensions of tensor product CFTs with the $S$-invariant basis being incorporated at a much earlier stage of the analysis, perhaps from the very beginning? 

We did find that method. It is described in the text of the article under the heading “Lessons learnt and a new method.” This new method comprises the following steps : 

(i) First obtain the $S$-invariant polynomials, $P_1, P_2, P_3 \ldots$. This can be achieved by imposing $S$-invariance on homogenous polynomials of characters. One starts with the smallest degree and works upwards in degree and one stops till one obtains as many $S$-invariant polynomials as needed (i.e. as many as the number of characters of the ${\mathcal C}$ CFT). We note the degrees of these : $\text{deg}(P_i)$.

(ii) Then one considers monomials of the $S$-invariant polynomials $P_1^{a_1}\,P_2^{a_2}\,P_3^{a_3}\ldots$, which when considered as a polynomial of the characters of ${\mathcal C}$ has a degree given by $\sum_i a_i\,\text{deg}(P_i)$.  We then need to find all monomials of $S$-invariant polynomials which have the degree required for the particular one-character extension under study i.e. the $N$ of tables \ref{t2} and \ref{t3}. That is, we need to find all non-negative integral solutions (while solving for the $a_i$'s) to 
\be \label{171a}
\sum_i a_i\,\text{deg}(P_i) = N.
\ee

(iii) In the final step, we write down the answer for the character of the one-character extension of the tensor product CFT under study, as follows. It is just the sum of all the monomials of the $S$-invariant polynomials obtained from the solutions to \eqref{171a} with arbitrary coefficients. One of the monomials contains the identity character (of the tensor product CFT) and this monomial comes with coefficient $1$. We have described how to identify the monomial that contains the identity character in the previous sections. We have also described some caveats that instruct to eliminate some kind of solutions to \eqref{171a} in certain situations. 

We thus hope to have convinced the reader that this new method is even simpler than the procedure laid out in section \ref{2s} : we only need to find $S$-invariant polynomials which involve solving much smaller systems of linear equations (than the ones that one encounters while implementing the procedure of section \ref{2s}) and solving \eqref{171a}.

\underline{Constant $S$-invariant polynomials} :  Another important result/aspect of our work is the content of section \ref{5s}. Some of the basis $S$-invariant polynomials of characters, in some (not all) CFTs, evaluate magically (after using the $q$-series of the characters) to a constant. We use “magically” because each character is an infinite series, has an infinite amount of information and there are an infinite number of cancellations that happen when it all reduces to one term, the constant term of the series. We called such $S$-invariant polynomials that evaluate to a constant as a constant $S$-invariant polynomial. Historically, the constant $S$-invariant polynomial associated to the Lee-Yang CFT (reproduced here in \eqref{68a}),  appeared as the  polynomial relation between the two Rogers-Ramanujan identities.
 We seem to have stumbled upon such objects for many CFTs. See \eqref{67a}. We wonder if there are more such constant  $S$-invariant polynomials of characters for other CFTs. This is an avenue of investigation for the future.

Finally, we noted  that the constant  $S$-invariant polynomial relations can generate ideals in the polynomial ring of characters. And one should properly work in the corresponding quotient ring. Also, the character of a one-character extension of a tensor product CFT properly belongs to this quotient ring. 

\subsection{Genuine CFTs,  impossible CFTs and conjectures for both}

\underline{Genuine CFTs} : We wished to study the CFTs that are one-character extensions of tensor product CFTs. We have only constructed the characters for them and more analysis is needed to decide which of these characters correspond to genuine CFTs.  However, we are confident that  some of them correspond to genuine CFTs. Our confidence comes from the partial classification of one-character CFTs that are available in the literature. Our primary sources are \cite{Schellekens:1992db} and \cite{Das:2022slz}. Here we list all the genuine CFTs that are one-character extensions of tensor product CFTs that have come about in this paper. 

\begin{small}
\begin{eqnarray*}
&&\mathcal E_1[A_{1,1}^{~\otimes 8}],  \qquad \mathcal E_1[A_{1,1}^{~\otimes 16}],  \qquad \mathcal E_1[A_{1,1}^{~\otimes 24}], \nonumber \\
&& \mathcal E_1[A_{2,1}^{~\otimes 4}],  \qquad \mathcal E_1[A_{2,1}^{~\otimes 8}],  \qquad \mathcal E_1[A_{2,1}^{~\otimes 12}],  \qquad \mathcal E_1[A_{2,1}^{~\otimes 20}], \nonumber \\
\end{eqnarray*}
\end{small}
\bea 
&& \mathcal E_1[D_{4,1}^{~\otimes 2}],  \qquad \mathcal E_1[D_{4,1}^{~\otimes 4}],  \qquad \mathcal E_1[D_{4,1}^{~\otimes 6}], \nonumber \\
&& \mathcal E_1[E_{6,1}^{~\otimes 4}], \nonumber \\
&&\mathcal E_1[A_{4,1}^{~\otimes 2}],  \qquad \mathcal E_1[A_{4,1}^{~\otimes 4}],  \qquad \mathcal E_1[A_{4,1}^{~\otimes 6}],  \qquad \mathcal E_1[A_{4,1}^{~\otimes 10}], \nonumber \\
&& \mathcal E_1[\mathcal{M}(4,3)^{\otimes 48}], \nonumber \\
&& \mathcal E_1[D_{3,1}^{~\otimes 8}], \nonumber \\
&& \mathcal E_1[D_{6,1}^{~\otimes 4}], \nonumber \\
&& \mathcal E_1[D_{12,1}^{~\otimes 2}], \qquad   \mathcal E_1[D_{20,1}^{~\otimes 2}],  \nonumber \\
&& \mathcal E_1[D_{8,1}^{~\otimes 3}], \qquad   \mathcal E_1[D_{24,1}],  \label{172a}
\eea

Among the above CFTS,  $\mathcal E_1[A_{2,1}^{~\otimes 20}]$, $ \mathcal E_1[A_{4,1}^{~\otimes 10}]$ and $\mathcal E_1[D_{20,1}^{~\otimes 2}]$ can be found in table 6 of \cite{Das:2022slz}. The rest are mostly from \cite{Schellekens:1992db}.

We have limited ourselves in this paper to constructing characters of possible CFTs. We have not taken up the task of further analysing these characters to answer the question of which of them are characters of genuine CFTs. We hope we can take up this study in the future. But, it turns out that the methods of this paper, while being too primitive to be able to conclude for the existence of CFTs, are good enough to conclude for the non-existence of CFTs. In certain situations we find that we cannot simultaneously fulfil the admissibility conditions and the modular data of one-character CFTs. In such situations, we can conclude that one-character CFTs which are one-character extensions of tensor products do not exist.

\underline{Impossible CFTs} : We have limited ourselves in this paper to constructing characters of possible CFTs. We have not taken up the task of further analysing these characters to answer the question of which of them are characters of genuine CFTs. We hope we can take up this study in the future. But, it turns out that the methods of this paper, while being too primitive to be able to conclude for the existence of CFTs, are good enough to conclude for the non-existence of CFTs. In certain situations we find that we cannot simultaneously fulfil the admissibility conditions and the modular data of one-character CFTs. In such situations, we can conclude that one-character CFTs which are one-character extensions of tensor products do not exist and this constitutes an impossible CFT. All \cancel{the} impossible CFTs we found in this paper are summarised thus :
\bea 
&&\cancel{\mathcal E_1[G_{2,1}^{~\otimes 20}]},  \quad \cancel{\mathcal E_1[G_{2,1}^{~\otimes 40}]}, \quad \cancel{\mathcal E_1[G_{2,1}^{~\otimes 80}]},  \quad \cancel{\mathcal E_1[G_{2,1}^{~\otimes 100}]}, \quad \cancel{\mathcal E_1[G_{2,1}^{~\otimes 140}]},  \quad \cancel{\mathcal E_1[G_{2,1}^{~\otimes 160}]}, \nonumber \\
&&\cancel{\mathcal E_1[F_{4,1}^{~\otimes 20}]},  \quad \cancel{\mathcal E_1[F_{4,1}^{~\otimes 40}]}, \quad \cancel{\mathcal E_1[F_{4,1}^{~\otimes 80}]},  \quad \cancel{\mathcal E_1[F_{4,1}^{~\otimes 100}]}, \quad \cancel{\mathcal E_1[F_{4,1}^{~\otimes 140}]},  \quad \cancel{\mathcal E_1[F_{4,1}^{~\otimes 160}]}. 
\eea
Thus, there are no one-character CFTs that are one-character extensions of tensor products of the  $G_{2,1}$ CFT with central charges $56, 112, 224, 280, 392, 448.$ Similarly, there are no one-character CFTs that are one-character extensions of tensor products of the  $F_{4,1}$ CFT with central charges $104, 208, 416, 520, 728, 832.$

\underline{Conjectures for genuine CFTs} : The following situation occurred many times in this paper. There are two or more CFTs which  have a common modular $S$-matrix and common set of $S$-invariant polynomials : (i) $A_{1,1}$ and $E_{7,1}$, (ii) $A_{2,1}$ and $E_{6,1}$, (Iii) $G_{2,1}$ and $F_{4,1}$, (iv) all the $D_{2p+1,1}$ CFTs, (v) all the $D_{2(2p+1),1}$ CFTs, (vi) all the $D_{4(2p+1),1}$ CFTs, (vii) all the $D_{8p ,1}$ CFTs ($p$ is a positive integer). Let us denote two such CFTs by $\cal C$ and $\cal C’$ and let us assume that there exists a character obtained as a  one-character extension of the tensor product of  $N$ copies of the  CFT $\cal C$ which is a genuine CFT.  Then we conjecture that the same character when interpreted as a character of the one-character extension of the tensor product of  $N$ copies of the  CFT $\cal C'$ is a character of a genuine CFT. We will denote this conjecture as follows : $\mathcal E_1
[{\cal C}^{\otimes N}] \Longrightarrow \mathcal E_1[{\cal C}'^{\otimes N}]$. With this notation, we note all the conjectured genuine CFTs that we have found in this paper :
\begin{normalsize}
\bea
&& \mathcal E_1[A_{1,1}^{~\otimes 8}] \quad \Longrightarrow \quad \mathcal E_1[E_{7,1}^{~\otimes 8}] \label{174b} \\
&& \mathcal E_1[A_{1,1}^{~\otimes 16}] \quad \Longrightarrow \quad \mathcal E_1[E_{7,1}^{~\otimes 16}] \label{175b} \\
&& \mathcal E_1[A_{1,1}^{~\otimes 24}] \quad \Longrightarrow \quad \mathcal E_1[E_{7,1}^{~\otimes 24}] \label{176b} \\
&& \mathcal E_1[A_{2,1}^{~\otimes 8}] \quad \Longrightarrow \quad \mathcal E_1[E_{6,1}^{~\otimes 8}] \label{177b} \\
&& \mathcal E_1[A_{2,1}^{~\otimes 12}] \quad \Longrightarrow \quad \mathcal E_1[E_{6,1}^{~\otimes 12}] \label{178b} \\
&& \mathcal E_1[A_{2,1}^{~\otimes 20}] \quad \Longrightarrow \quad \mathcal E_1[E_{6,1}^{~\otimes 20}] \label{179b} \\
&& \mathcal E_1[D_{3,1}^{~\otimes 8}] \quad \Longrightarrow \quad \mathcal E_1[D_{2p+1,1}^{~\otimes 8}], \quad p = 1, 2, \ldots \label{180b} \\
&& \mathcal E_1[D_{6,1}^{~\otimes 4}] \quad \Longrightarrow \quad \mathcal E_1[D_{2(2p+1),1}^{~\otimes 4}], \quad p = 1, 2, \ldots \label{181b} \\
&& \mathcal E_1[D_{12,1}^{~\otimes 2}] \quad \Longrightarrow \quad \mathcal E_1[D_{4(2p+1),1}^{~\otimes 2}], \quad p = 1, 2, \ldots \label{182b} \\
&& \mathcal E_1[D_{8,1}^{~\otimes 3}] \quad \Longrightarrow \quad \mathcal E_1[D_{8p,1}^{~\otimes 3}], \quad p = 1, 2, \ldots \label{183b} 
\eea
\end{normalsize}
  The conjectured one-character extension CFTs of \ref{174b}, \ref{175b}, \ref{176b}, \ref{177b}, \ref{178b} and \ref{179b} have central charges of $56, 112, 168, 48, 72$ and $120$ respectively. From \ref{180b}, we have an infinite series of conjectured CFTs with central charges $8(2p+1)$, $p = 2, 3, \ldots$.  From \ref{181b}, we have an infinite series of conjectured CFTs with central charges $8(2p+1)$, $p = 2, 3, \ldots$. From \ref{182b}, we have an infinite series of conjectured CFTs with central charges $8(2p+1)$, $p = 2, 3, \ldots$. From \ref{183b}, we have an infinite series of conjectured CFTs with central charges $24p$, $p = 2, 3, \ldots$. The characters of each of the above conjectured CFTs can be had from the earlier sections. 

We thus have a discrete set of $6$ (\ref{174b}, \ref{175b}, \ref{176b}, \ref{177b}, \ref{178b}, \ref{179b}) and $4$ infinite series (\ref{180b}, \ref{181b}, \ref{182b}, \ref{183b}) of conjectured CFTs.

\underline{Conjectures for impossible CFTs} : Based on our experience with computations for $G_{2,1}$ CFTs, we can make the following surmise : there are no genuine CFTs for any of the one-character extensions of tensor product of $N$ copies of $G_{2,1}$ CFTs with $N$ a multiple of $20$ and not a multiple of $60$. Similarly we make the following surmise for $F_{4,1}$ : there are no genuine CFTs for any of the one-character extensions of tensor product of $N$ copies of $F_{4,1}$ CFTs with $N$ a multiple of $20$ and not a multiple of $60$. 

\newpage
\begin{center}
\textbf{Acknowledgments}
\end{center}
CNG thanks Prof. Sunil Mukhi for raising a question during the course of the collaboration \cite{Das:2022uoe}, which triggered the investigations of this paper.  CNG thanks him in general for general inspiration to pursue this subject of CFTs, MLDEs etc.  CNG also thanks  (i) the CERN theory group,  (ii) Prof. Sudipta Mukherji and IOP, Bhubaneshwar, for their warm hospitality, during the course of this work.  JS would like to acknowledge the warm hospitality of The Abdus Salam International Centre for Theoretical Physics, Trieste where part of the work had been done. JS is indebted to Suresh  Govindarajan for helpful discussions. JS would also like to acknowledge the warm hospitality of IIT Ropar where he met one of the authors and discussed this project physically.

\appendix

\section{ Admissible Characters of One-character MLDEs\label{app1}}

Here, we give the solutions to one-character MLDE’s. These exist for central charge, $c_{(1)}$ being a multiple of $8$. There are three cases, (i) when $c_{(1)} = 24k +8$, (ii) when  $c_{(1)} = 24k + 16$ and (iii)  when $c_{(1)} = 24k$. For $c_{(1)} = 24k$, the solution to the MLDE is a degree-$k$ monic polynomial in the Klein $j$-function. For $c_{(1)} = 24k + 8$, the solution to the MLDE is a degree-$k$ monic polynomial in the Klein $j$-function multiplied by $j^{\frac13}$. For $c_{(1)} = 24k + 16$, the solution to the MLDE is a degree-$k$ monic polynomial in the Klein $j$-function multiplied by $j^{\frac23}$. In all three cases, we have $k$ arbitrary parameters in the solution, which we denote by $N_1, N_2, \ldots, N_k$. For arbitrary solutions, i.e. arbitrary values of $N_1, N_2, \ldots, N_k$, there is no admissibility. One has to tune their values to make the solutions/characters admissible. In this table, we give the nested inequalities for the parameters $N_1, N_2, \ldots, N_k$ that need to be solved to achieve admissible characters.

\begin{table}[H]
\begin{center}
\begin{threeparttable}
\rowcolors{2}{Mywhite}{Mygrey}
\resizebox{\textwidth}{!}{
   \renewcommand{\arraystretch}{1.3}
   $\begin{array}{|C{1cm}|C{1cm}|C{15cm}|}
\hline
 \# & c_{(1)} & q\text{-series}  \\
\hline
1. & 8 & j^{\frac{1}{3}}\\
\hline
2. & 16 & j^{\frac{2}{3}}\\
\hline
3. & 24 &j + N_1\\
&& N_1 \geq -744\\
\hline
4. & 32 &j^{\frac13}(j + N_1)\\
&& N_1 \geq -992\\
\hline
5. & 40 &j^{\frac23}(j + N_1)\\
&& N_1 \geq -1240\\
\hline
6. & 48 &(j^2 + N_1 j + N_2)\\
&&N_1 \geq -1488,\\&& N_2 \geq -744 N_1 - 947304\\
\end{array}$
}
\end{threeparttable}
\end{center}
\end{table}

\begin{table}[H]
\begin{center}
\begin{threeparttable}
\rowcolors{2}{Mywhite}{Mygrey}
\resizebox{\textwidth}{!}{
   \renewcommand{\arraystretch}{1.3}
   $\begin{array}{|C{1cm}|C{1cm}|C{15cm}|}
7. & 56 &j^{\frac13} (j^2 + N_1 j + N_2)\\
&&N_1 \geq -1736,\\&&N_2 \geq -992 N_1 - 1320452\\
\hline
8. & 64 &j^{\frac23} (j^2 + N_1 j + N_2),\\&&N_1 \geq -1984,\\&&N_2 \geq -1240 N_1 - 1755104\\
\hline
9. & 72 &(j^3 + N_1 j^2 + N_2 j + N_3), \nonumber \\
&&N_1 \geq -2232,\\&& N_2 \geq -1488 N_1 - 2251260,\\ &&N_3 \geq -744 N_2 - 947304 N_1 - 1355202240\\
\hline
10. & 80 &j^{\frac13}(j^3 + N_1 j^2 + N_2 j + N_3)\\
&&N_1 \geq -2480,\\ &&N_2 \geq -1736 N_1 - 2808920,\\  &&N_3 \geq -992 N_2 - 1320452 N_1 - 1922754240\\
\hline
11. & 88 &j^{\frac23}(j^3 + N_1 j^2 + N_2 j + N_3)\\
&&N_1 \geq -2728,\\ &&N_2 \geq -1984 N_1 - 3428084 ,\\ &&N_3 \geq -1240 N_2 - 1755104 N_1 - 2629628672\\
\hline
12. & 96 &(j^4 + N_1 j^3 + N_2 j^2 + N_3 j + N_4)  \\
&&N_1 \geq -2976,\\ &&N_2 \geq -2232 N_1 - 4108752,\\ &&N_3 \geq -1488 N_2 - 2251260 N_1 - 3491078528  ,\\ && N_4 \geq -744 N_3 - 947304 N_2 - 1355202240 N_1 - 2042124031080\\
\hline
13. & 104 &j^{\frac13}(j^4 + N_1 j^3 + N_2 j^2 + N_3 j + N_4)  \\
&&N_1 \geq -3224,\\ &&N_2 \geq -2480 N_1 - 4850924,\\ &&N_3 \geq -1736 N_2 - 2808920 N_1 - 4522356800  ,\\ && N_4 \geq -992 N_3 - 1320452 N_2 - 1922754240 N_1 - 2924959634350\\
\hline
14. & 112 &j^{\frac23}(j^4 + N_1 j^3 + N_2 j^2 + N_3 j + N_4)  \\
&&N_1 \geq -3472,\\ &&N_2 \geq -2728 N_1 - 5654600,\\ &&N_3 \geq -1984 N_2 - 3428084 N_1 - 5738716480,  \\ && N_4 \geq -1240 N_3 - 1755104 N_2 - 2629628672 N_1 - 4066621584900\\
\end{array}$
}
\end{threeparttable}
\end{center}
\end{table}

\begin{table}[H]
\begin{center}
\begin{threeparttable}
\rowcolors{2}{Mywhite}{Mygrey}
\resizebox{\textwidth}{!}{
   \renewcommand{\arraystretch}{1.3}
   $\begin{array}{|C{1cm}|C{1cm}|C{15cm}|}
15. & 120 &(j^5 + N_1 j^4 + N_2 j^3 + N_3 j^2 + N_4 j + N_5)  \\
&&N_1 \geq -3720,  \\ &&N_2 \geq -2976 N_1 - 6519780,  \\ &&N_3 \geq -2232 N_2 - 4108752 N_1 - 7155410560,  \\
&& N_4 \geq -1488 N_3 - 2251260 N_2 - 3491078528 N_1 - 5513263714410,  \\
&& N_5 \geq -744 N_4 - 947304 N_3 - 1355202240 N_2 - 2042124031080 N_1 - 3169342733223744\\
\hline
16. & 128 &j^{\frac13}(j^5 + N_1 j^4 + N_2 j^3 + N_3 j^2 + N_4 j + N_5)  \\
&&N_1 \geq -3968,  \\ &&N_2 \geq -3224 N_1 - 7446464,  \\ &&N_3 \geq -2480 N_2 - 4850924 N_1 - 8787692032  ,\\
&& N_4 \geq -1736 N_3 - 2808920 N_2 - 4522356800 N_1 - 7314822596576 , \\
&& N_5 \geq -992 N_4 - 1320452 N_3 - 1922754240 N_2 - 2924959634350 N_1 - 4566368416827648\\
\hline
\end{array}$
}
\end{threeparttable}
\end{center}
\caption{Admissible characters of one-character MLDEs}\label{tapp}
\end{table}

\end{document}